\documentclass[11pt]{emulateapj}
\pdfoutput=1
\usepackage{natbib}
\usepackage{graphicx}
\usepackage{amsmath}
\usepackage{amssymb}
\def\chisq{$\chi^2$ }
\def\chisqx{$\chi^2$}
\def\chf{CH$_4$ }

\def\Wm2{W/m$^2$}
\def\Wpm2sr{Wm$^{-2}sr^{-1}$}
\def\deg{$^\circ$ }
\def\degx{$^\circ$}

\def\jgr{J. Geophys. Res. }

\def\mum{$\mu$m }
\def\mumx{$\mu$m}

\begin{document}
\title{High S/N Keck and Gemini AO imaging of
    Uranus during 2012-2014: new cloud patterns, increasing activity, and improved wind measurements.}
\author{L.A. Sromovsky\altaffilmark{1}, I. de Pater\altaffilmark{2}, P.\ M. Fry\altaffilmark{1},
  H.\ B. Hammel\altaffilmark{3}, P. Marcus\altaffilmark{2}}
\altaffiltext{1}{University of Wisconsin - Madison, Madison WI 53706, USA}
\altaffiltext{2}{University of California, Berkeley, CA 94720, USA}
\altaffiltext{3}{Space Science Institute, Boulder, CO 80303, USA}

\slugcomment{Journal reference: Icarus (2015), http://dx.doi.org/10.1016/j.icarus.2015.05.029.}

\begin{abstract}

We imaged Uranus in the near infrared from 2012 into 2014, using the
Keck/NIRC2 camera and Gemini/NIRI camera, both with adaptive
optics. We obtained exceptional signal to noise ratios by averaging
8-16 individual exposures in a planet-fixed coordinate system.  These
noise-reduced images revealed many low-contrast discrete features and
large scale cloud patterns not seen before, including scalloped
waveforms just south of the equator, and an associated transverse
ribbon wave near 6\degx S. In all three years numerous small (600-700
km wide) and mainly bright discrete features were seen within the
north polar region (north of about 55\degx N).  Two small dark spots
with bright companions were seen at middle latitudes. Over 850 wind
measurements were made, the vast majority of which were in the
northern hemisphere.  Winds at high latitudes were measured with great
precision, revealing an extended region of solid body rotation between
62\degx N and at least 83\degx N, at a rate of 4.08$\pm0.015$\degx/h
westward relative to the planet's interior (radio) rotation of 20.88\degx/h
westward. Near-equatorial speeds measured with high accuracy
give different results for waves and small discrete features, with
eastward drift rates of 0.4\degx/h and 0.1\degx/h respectively.  The
region of polar solid body rotation is a close match to the region of
small-scale polar cloud features, suggesting a dynamical relationship.
The winds from prior years and those from 2012-2014 are consistent
with a mainly symmetric wind profile up to middle latitudes, with a
small asymmetric component of $\sim$0.09\degx /h peaking near
$\pm$30\degx, and about 60\% greater amplitude if only prior years are
included, suggesting a declining mid-latitude asymmetry. While
winds at high southern latitudes (50\degx S - 90\degx S) are
unconstrained by groundbased observations, a recent reanalysis of 1986
Voyager 2 observations by Karkoschka (2015, Icarus 250, 294-307) has
revealed an extremely large north-south asymmetry in this region,
which might be seasonal.  Greatly increased activity was seen in 2014,
including the brightest ever feature seen in K$'$ images (de Pater et
al. 2015, Icarus 252, 121-128), as well as other significant features,
some of which had long lives. Over the 2012-2014 period we identified
six persistent discrete features.  Three were tracked for more than
two years, two more for more than one year, and one for at least 5
months and continuing.  Several drifted in latitude towards the
equator, and others appeared to exhibit latitudinal oscillations with
long periods. We found two pairs of long-lived features that survived
multiple passages within their own diameters of each other. Zonally
averaged cloud patterns were found to persist over 2012-2014. When
averaged over longitude, there is a brightness variation with latitude
from 55\degx N to the pole that is similar to effective methane mixing
ratio variations with latitude derived from 2012 STIS observations
(Sromovsky et al. 2014, Icarus 238, 137-155).
\end{abstract}

\maketitle
\shortauthors{Sromovsky et al.} 
\shorttitle{High signal-to-noise imaging of Uranus.}


\newpage

\section{Introduction}

Visual observers reported detection of zonal bands and occasional
spots on Uranus as early as 1870 \citep{Alexander1965}, but a reliable measure
of atmospheric motions on Uranus had to wait until 1986, when close-up Voyager-2
images revealed eight discrete cloud features between planetocentric
latitudes of 35\deg S and
70\deg S \citep{SmithBA1986}. Tracking these features, combined with
one radio occultation wind measurement at 5\deg S, yielded a crude
zonal wind profile \citep{Allison1991uranbook}, the main feature of
which appeared to be a high-speed ($\sim$250 m/s) prograde jet near
60\deg S and a weaker broad retrograde equatorial jet.  Voyager did
not provide wind information in the northern hemisphere, which was
near its winter solstice and dark at the time of the encounter.
Eleven years later, Hubble Space Telescope (HST) near-IR images of Uranus
revealed many more discrete cloud features, enabling the extension of
wind
 measurements into the northern hemisphere
\citep{Kark1998Sci}, confirming an approximate N-S symmetry in the
wind profile.  Groundbased observations from
the Keck II telescope, which combined a large aperture, near-IR
wavelengths, and adaptive optics, produced a bounty of cloud features
far beyond the Hubble results \citep{Hammel2001Icar,
 Hammel2005winds, Sro2005dyn, Sro2007bright,Hammel2009Icar,
 Sro2009eqdyn}.

The wind profile of Uranus was last updated by \cite{Sro2012polar},
combining wind observations from the above references with new
measurements in 2009-2011 from Gemini, Keck, and HST
observatories. Here we use as a reference their 13-term asymmetric Legendre fit given
in their Table 6 and plotted in their Fig. 11.  This will be hereafter
referred to as Model {\bf S13A}.
The main results enabled by their new measurements were:
(1) clear definition of the magnitude and latitude of the northern jet
peak; (2) discovery that motions north of the peak were consistent
with solid body rotation; (3) discovery of a new class of cloud
features in the polar regions that resemble cumulus cloud fields and
bear similarities to the cloud structure in the polar regions of
Saturn; and (4) characterization of the large morphological asymmetry
between southern summer polar latitudes and northern spring polar
latitudes. The updated profile still suffered from a lack of targets
in low latitudes (20\deg S - 20\deg N) and in the south polar region,
where, save for one UV feature identified by Voyager, no discrete
clouds had ever been seen.

\begin{figure}[!h]\centering
\includegraphics[width=3.3in]{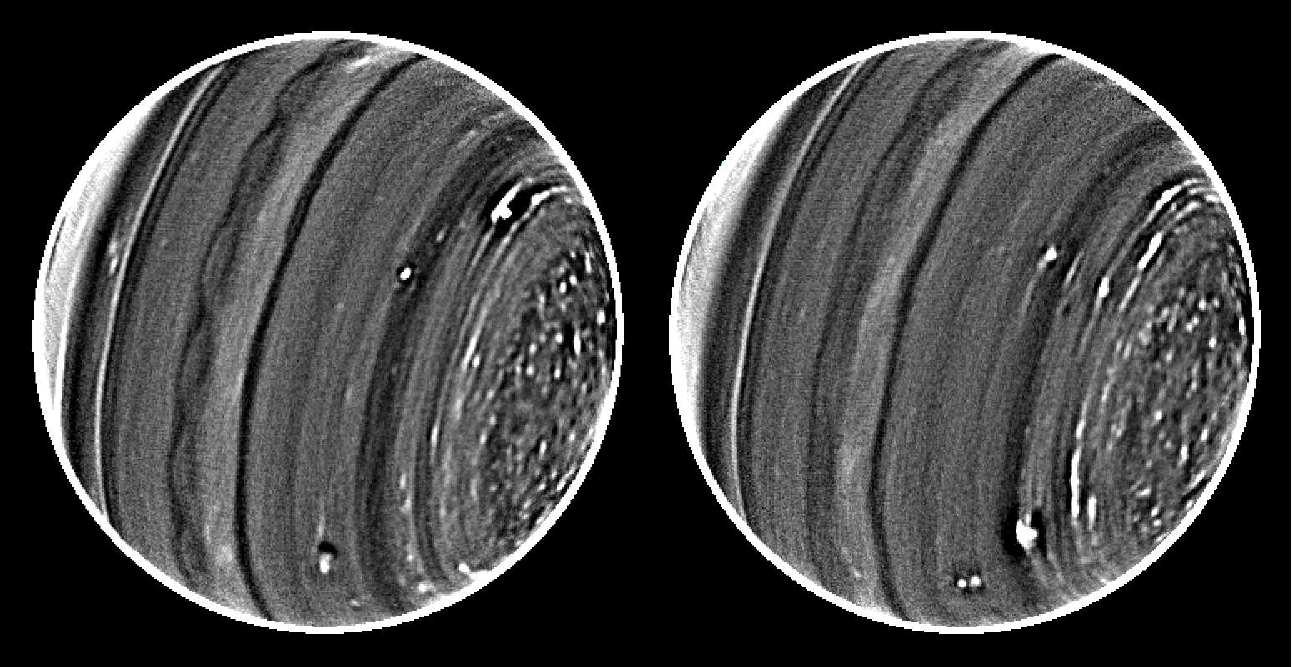}
\caption{Keck II de-rotated, averaged, and high-pass filtered H-filter images of
Uranus from 25 July (L) and 26 July(R), 2012. The north pole is at
the right (about 100\deg clockwise from up), and the dark band just
right of the middle of each image is at $\approx$8.5\degx N (planetographic),
the scalloped waveform centered near 4.5\degx S, and the ribbon wave
near 6\degx S. Note the narrow bright latitude bands spaced about
2.5\deg apart in the 48\degx-53\degx range, and the numerous small bright spots
north of 55\degx N. The bright narrow feature in the southern hemisphere
is the epsilon ring of Uranus.}\label{Fig:bestim}
\end{figure}

The \cite{Sro2012polar} work was advanced by the use of high S/N
techniques to improve detectability and facilitate tracking of low
contrast cloud features.  The basic idea was to use long exposures to
grow the signal linearly in exposure time while random noise grew as
the square root, thus increasing S/N as the square root of the
exposure time.  To avoid the smearing of long exposures by planetary
rotation, we took multiple short exposures, rotated the planet images
to the same central meridian longitude, then averaged the de-rotated
images.  \cite{Fry2012} verified this process using Keck imagery and
applied it to specially designed HST imaging to provide new wind
measurements.  \cite{Sro2012polar} applied it to Gemini imaging in
2011, and to a limited degree to Keck imaging in 2011, though highly
variable seeing limited the benefits that were possible.  Much
improved results were obtained from subsequent observations with
better seeing that resulted in better AO performance.  The 2012
results shown in Fig.\ \ref{Fig:bestim}, first presented by
\cite{Sro2012dps} and \cite{Fry2012DPS}, were the most detailed images
of Uranus ever obtained up to that time.  Features revealed by those
images are discussed in Section\ \ref{Sec:patterns}.

Since 2007, our view of the north polar region of Uranus improved
significantly: at opposition, the the sub-solar latitude reached
19.5\degx N in 2012, 23.5\degx N in 2013, and 27.6\degx N in 2014.
This resulted in observations well suited to further refinement of the
zonal wind profile in the northern hemisphere and better
characterization of its polar cloud features.  It also helped to fill
in poorly sampled regions, mainly at low latitudes. Fortunate periods
of high quality seeing also allowed us to apply high S/N techniques
more extensively and thereby detect and track more subtle features.
In the following we describe the new observations, the measurement
results from each data set, new views of the north polar region of
Uranus and the different styles of discrete cloud features located
there, new features discovered at low latitudes, and finally,
approximate altitude constraints on the cloud features.

\section{Observations, image processing, and navigation}

Table\ \ref{Tbl:obslist} summarizes Keck and Gemini imaging
observations of Uranus acquired from 2012 through 2014.  The camera
characteristics for each groundbased observing configuration are given
in Table \ref{Tbl:cameras}, in which the pixel scale of
0.02138$\pm$0.0005 arcsec/pixel listed for the Gemini NIRI camera was
derived from measurements of Uranus and its satellites in comparison
with HORIZONS ephemeris positions, as described by \cite{Sro2012polar}.
We used the Keck II/NIRC2 pixel scale provided by the Keck observatory
web site, which we found to be consistent with measurements of the
Uranian ring system.
For Gemini AO operation, we used a bright Uranian satellite (usually
Ariel) as the wavefront reference, while for Keck II AO operation we
were able to use Uranus itself, providing a substantial gain in reference object
brightness (VMAG=5.6 vs.  VMAG$\approx$14) and thus in image quality for
similar seeing conditions.  Except for the image S/N enhancement, described in more
detail by \cite{Fry2012}, image processing and navigation followed the
same procedures described by \cite{Sro2009eqdyn}. To aid in long-term tracking of a long
lived feature that was first observed in August 2014 (identified as F
in a later section), we also used non-proprietary images from HST
Program 13712 (Target of Opportunity Observation of an Episodic Storm
on Uranus, PI K. Sayanagi) and HST Program 13937 (Hubble 2020: Outer
Planet Atmospheres Legacy (OPAL) Program, PI A. Simon).

\begin{table*}\centering
\caption{Imaging observations used to track and characterize discrete cloud features.}
\vspace{0.15in}
\begin{tabular}{c c c c c c c l}
\hline\\[-0.1in] 
 Date & Time Range & Telescope/Camera/Program & PI & Filters (images) \\  
\hline\\[-0.1in] 
2012/07/25 & 09:59-12:29  & Keck/NIRC2/2012A-N145N2 & LAS & H(93), Hc(24), K$'$(1) \\
2012/07/26 & 10:04-15:29  & Keck/NIRC2/2012A-N145N2 & LAS & H(88), Hc(32) \\
2012/08/16 & 10:31-15:41  & Keck/NIRC2/2012B-N125N2 & LAS & H(91), Hc(56) \\
2012/09/28 & 08:20-12:24 & Gemini-N/NIRI/GN-2012B-Q-121 & LAS & H(11), Hc(10)\\
2012/09/30 & 08:15-12:09 & Gemini-N/NIRI/GN-2012B-Q-121 & LAS & H(8), Hc(9)\\
2012/10/04 & 08:58-12:50 & Gemini-N/NIRI/GN-2012B-Q-121 & LAS & H(11), Hc(12)\\
2012/10/09 & 06:17-10:10 & Gemini-N/NIRI/GN-2012B-Q-121 & LAS & H(11), Hc(11)\\
2012/11/04 & 04:17-09:34 & Keck/NIRC2/2012B-U011N2 & IDP & H(120), Hc(3), K$'$(1) \\
2012/11/05 & 04:11-09:17 & Keck/NIRC2/2012B-U011N2 & IDP & H(42), Hc(76), K$'$(2) \\
2013/08/15 & 10:35-14:43 & Keck/NIRC2/2013B-U009N2 & IDP & H(74), CH4S(10), K$'$(3)\\
2013/08/16 & 10:34-15:24 & Keck/NIRC2/2013B-U009N2 & IDP & H(84),  CH4S(14), K$'$(10) \\
2014/08/05 & 11:34-15:36 & Keck/NIRC2/2014B-U037N2 & IDP & H(64), K$'$(5), CH4S(8)\\
2014/08/06 & 11:24-13:44 & Keck/NIRC2/2014B-U037N2 & IDP & H(78),J(1),K$'$(9),CH4S(10)\\
2014/08/20 & 13:38-13:44 & Keck/NIRC2/2014B-U014N2 & IDP & H(2), K$'$(2)\\
2014/10/30 & 08:57-09:03 & Gemini-N/NIRI/GN-2014B-DD-5 & LAS & H(4)\\
2014/11/09 & 09:19-09:25 & Gemini-N/NIRI/GN-2014B-DD-5 & LAS & H(4)\\
2014/11/26 & 07:14-07:48 & Gemini-N/NIRI/GN-2014B-DD-5 & LAS & H(4), K$'$(4), Hc(8)\\
2015/01/08 & 07:05-07:24 & Gemini-N/NIRI/GN-2014B-DD-5 & LAS & H(4), K$'$(4), Hc(8)\\
\hline\\[-0.1in]
\end{tabular}
\vspace{0.2in}
\parbox{5.5in}{NOTE: Times are UTC. PI codes are LAS for Sromovsky,
  IDP for de Pater.  Filter bands are: H(1.48-1.78 \mumx),
  Hcont(1.57-1.59 \mumx), CH4S(1.53-1.66 \mum), and K$'$(1.95-2.3
  \mumx).
}
\label{Tbl:obslist}
\end{table*}

\begin{table}\centering
\caption{Telescope/camera characteristics.}
\vspace{0.15in}
\begin{tabular}{c c  c  c  c c}
\hline\\[-0.1in]
           & Mirror &           &  Pixel & Diff. limit & \\
Telescope  & Diam. &  Camera &    size &    @ wavelength \\
\hline\\[-0.1in]
Gemini-N    &  8 m     &   NIRI    &   0.02138$''$    &  0.05$''$ @ 1.6 \mum  \\
Keck II   &  10 m    &   NIRC2-NA &  0.00994$''$     &  0.04$''$ @ 1.6 \mum  \\
\hline\\[-0.1in]
\end{tabular}
\vspace{0.1in}
\parbox{3.25in}{NOTES: Both groundbased telescopes have adaptive
  optics capability, but only Keck II can use Uranus itself as the
  wave front reference. Our Gemini observations had to use a satellite
  of Uranus for the wave front reference.}
\label{Tbl:cameras}
\end{table}

\begin{figure*}[!h]\centering
\includegraphics[width=5.5in]{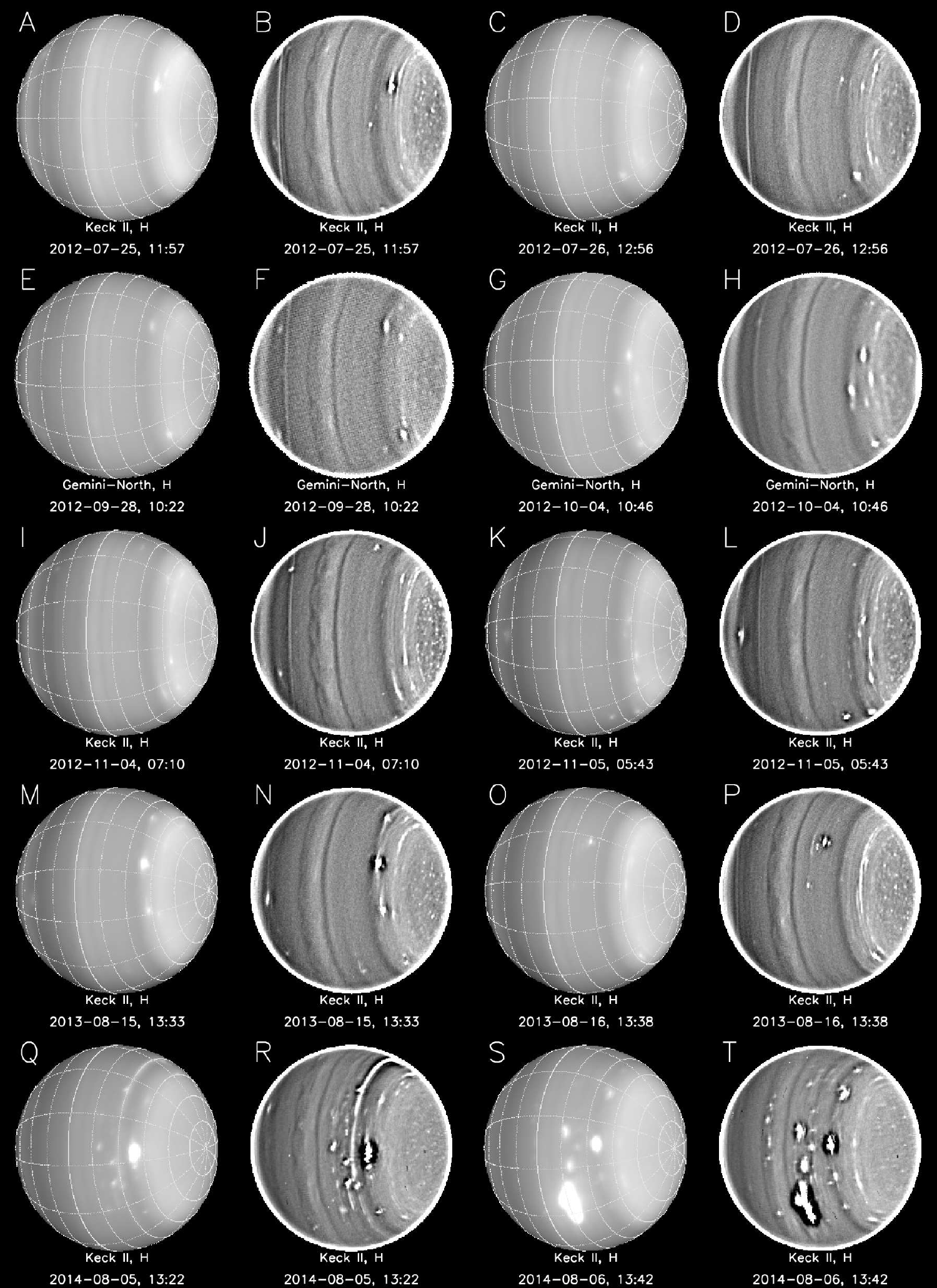}
\caption{Sample high-S/N images from the major data sets listed in Table\ \ref{Tbl:obslist}. Each
image is followed by a high-frequency amplified version, in which the difference between a smoothed
version and the original image is added back to the original image after the difference is
multiplied by a factor of 30. The box-car smoothing length was 0.13 arcseconds. The grids
shown are at intervals of 15\deg in planetographic latitude and 30\deg in longitude. Note the increased
activity in 2014.}
\label{Fig:imsamples}
\end{figure*}

Almost all our observations were designed for high S/N imaging and
were successful to varying degrees, mainly dependent on seeing
conditions.  A sampling of the mosaicked and averaged H-band images
from the Keck and Gemini data sets is provided in
Fig.\ \ref{Fig:imsamples}. Most of these images were averaged in
groups of eight 2-minute exposures.  The exceptions are for the Gemini
images in which all images during the night were averaged together
(this required removal of approximate zonal wind displacements as well as planet
rotation).  In Fig.\ \ref{Fig:imsamples} each image is followed by a
version with enhanced small-scale feature contrast, obtained by
replacing the original image I by I+30$\times$(I-smooth(I)), using a
smoothing length of 0.13$''$ (13 pixels in Keck images).  In most
cases the Keck AO images provide a level of detail far superior to
that obtained from Gemini or HST. In one case, however, a fortuitous
natural seeing of 0.37$''$ resulted in an excellent Gemini image (G
and H in Fig.\ \ref{Fig:imsamples}), providing slightly better S/N
though at a slightly lower resolution (see
Fig.\ \ref{Fig:imsamples}G).

Many of these images reveal a low-latitude waveform just south of the
equator, the best examples of which are seen in
Fig.\ \ref{Fig:imsamples}, panels B (25 July 2012) and J (4 November 2012). The
wave is also evident in several other images, including the Gemini
image from 4 October 2012.  When first observed in 2012, it was
suggested that this might be an unstable feature because of
morphological similarities to unstable waves in high-shear regions.
The persistence of the waveform, and the lack of evidence for high (or
any) zonal wind shear in the region of the wave, suggests that other
explanations might be needed.

These images also show a persistence of small cloud features at high
northern latitudes, although the contrast of these features in the 2014
images seems to be reduced, 
 perhaps by a developing haze in the north polar region
 \citep{DePater2015storm} (note the brighter near-pole region in 2014
 relative to that in 2012).  There is a general lack of discrete cloud
 features south of 30\degx N, with the striking exception of 2014
 images (Fig.\ \ref{Fig:imsamples}Q-T), which not only display
 numerous cloud features in this region, but also several unusually
 bright ones, including the brightest ever seen in K$'$ images
 \citep{DePater2015storm}, and the second brightest ever seen in
 H-band images. The brightest feature in H-band images
 \citep{Sro2007bright} was seen in 2005 at a planetographic latitude
 of 31\degx N.

\begin{figure}[!htb]\centering
\includegraphics[width=3.25in]{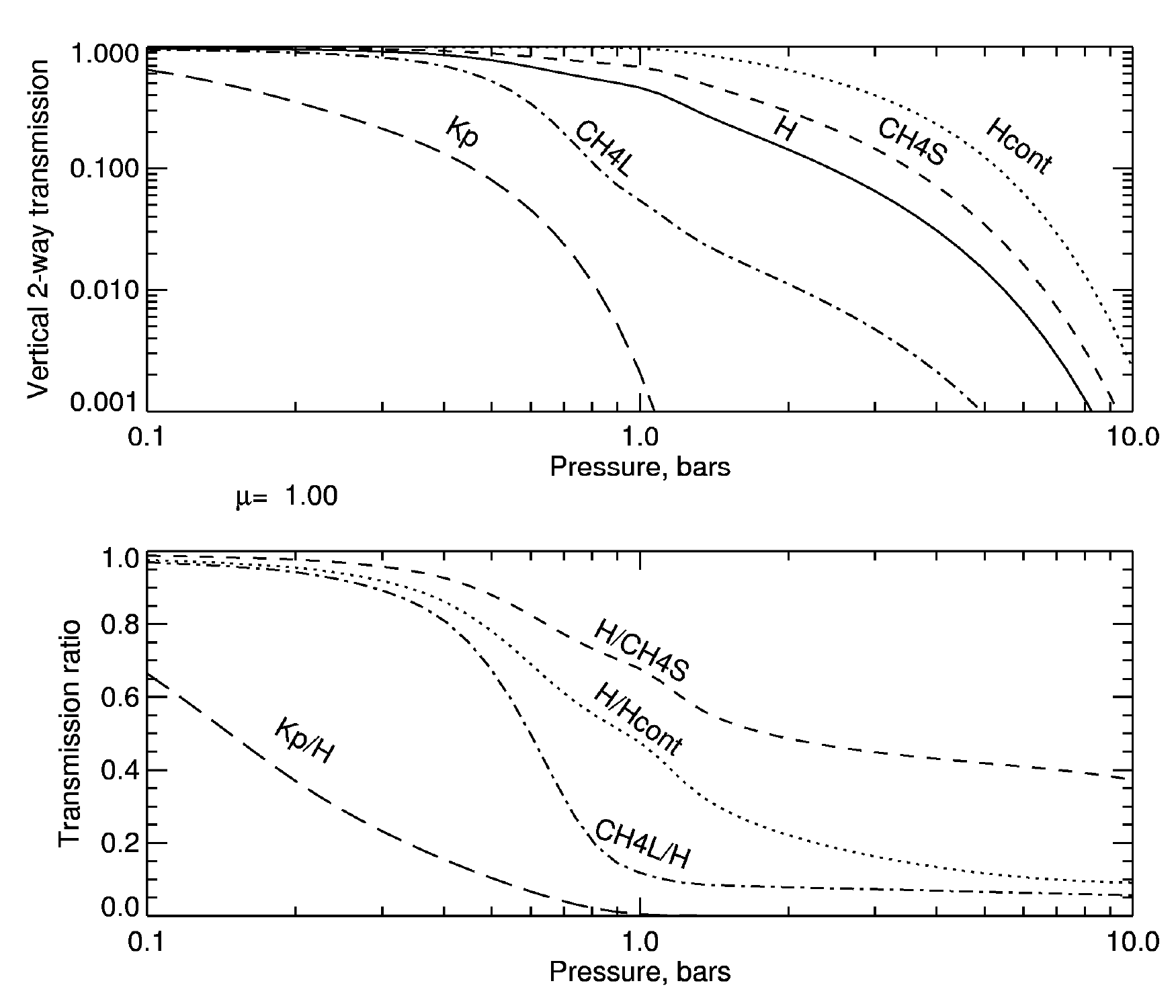}
\caption{Vertical 2-way transmission of a clear uranian atmosphere as a
  function of pressure for filter bands used in groundbased
  observations (top) and transmission ratios (bottom).}
\label{Fig:penprof}
\end{figure}

We also acquired Hcont images in groups of 12 (for Keck observations).
Images with the Hcont filter (not shown) differ from the H-filter
images in having relatively brighter mid latitudes, more noise (due to
less throughput in the narrow-band filter), generally lower contrast,
and fewer visible cloud features.  The average penetration depths of
light within these filter bands into a clear Uranus atmosphere are
shown in Fig.\ \ref{Fig:penprof}.  The different vertical weighting of
Hcont images allows the combination of H and Hcont images to constrain
the vertical location of cloud features. Limited imaging with CH4S and
K$'$ filters provided additional constraints, which are all discussed
in Section\ \ref{Sec:vertical}.

\section{Cloud tracking}

\subsection{Methodology}

\begin{figure*}[!htb]\centering
\includegraphics[width=6.in]{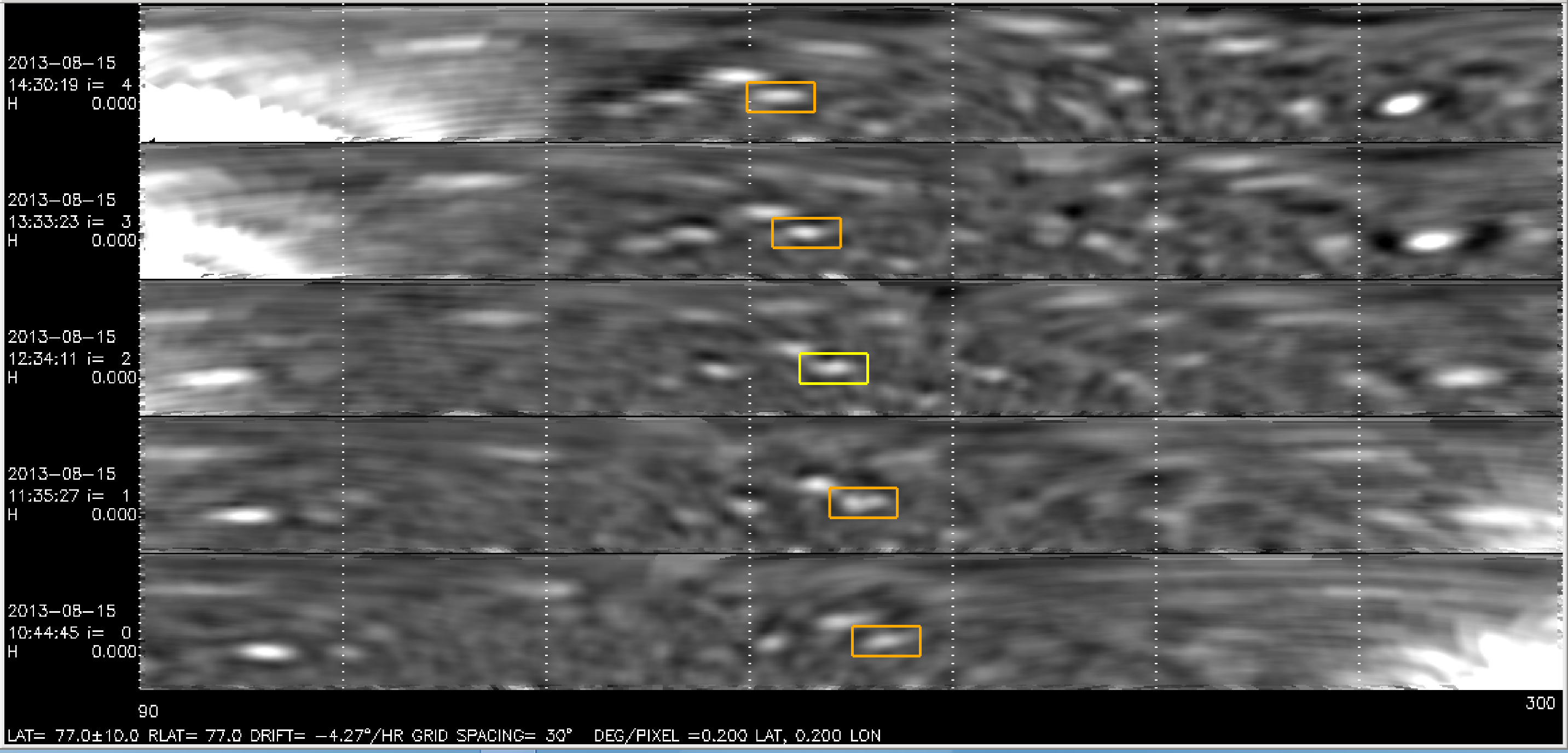}
\includegraphics[width=5.5in]{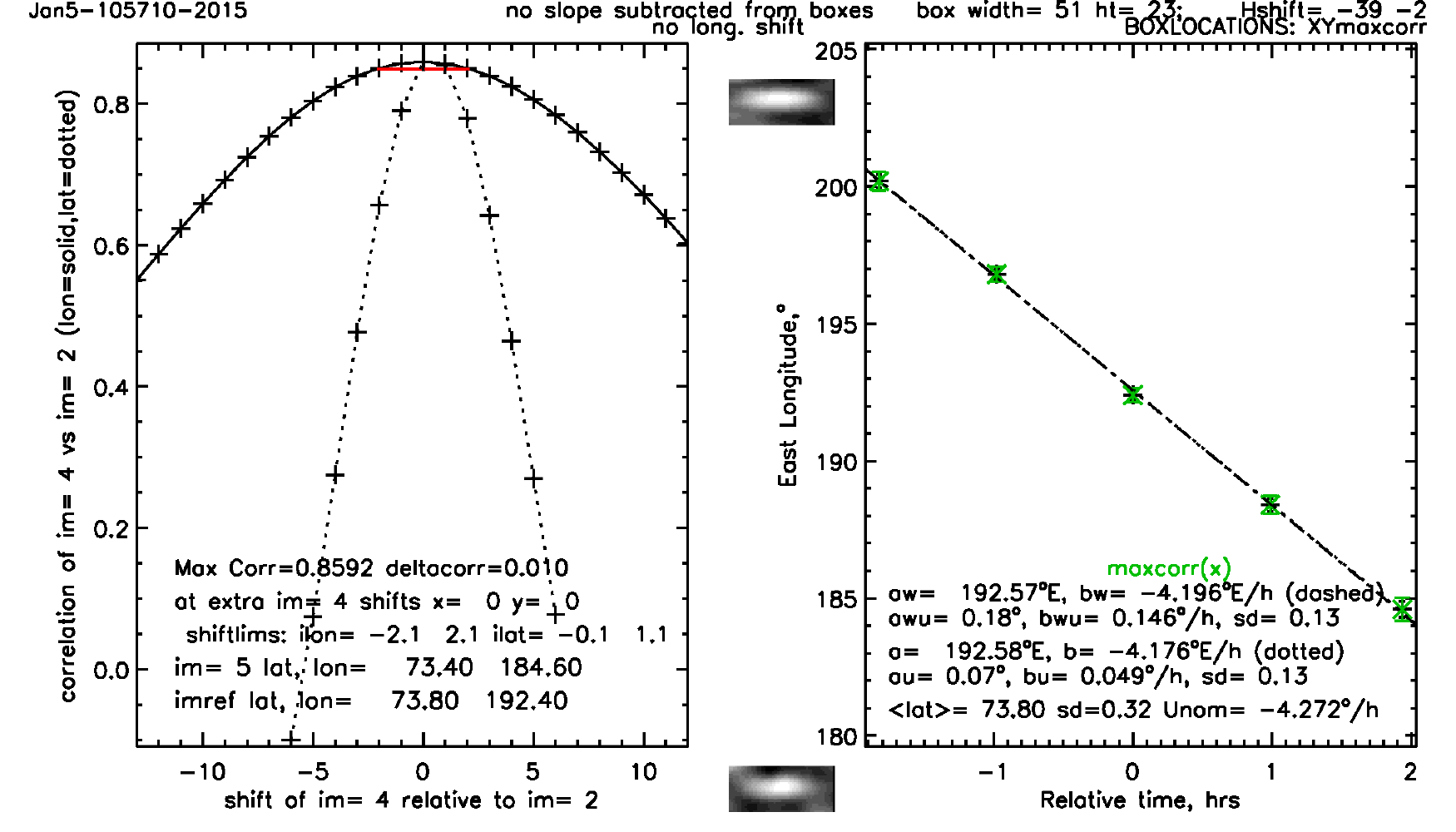}
\caption{Illustration of our manually guided correlated tracking
  method. The top panel displays five rectangular projections centered
  at 77\degx N, spanning 240\deg in longitude and 20\deg in latitude,
  vertically stacked with the earliest image at the bottom. At this
  point the box outlines surrounding the selected target feature have
  been moved to produce the maximum cross correlations with the
  feature in the reference box at the center of the screen.  Box
  longitudes vs. time are plotted at the bottom right along with
  best-fit linear drift rate coefficients for the model $\phi(t) = $a
  + b $(t-t_0)$) for both unweighted and weighted fits, with coefficients
(a,b) and  (aw,bw) respectively), with (au, bu) and (awu, bwu) denoting uncertainties.
  The initial guessed drift rate value (Unom) is also shown in the
  legend. Sample plots of correlation coefficient versus latitude and
  longitude displacements of the image-4 sampling box relative to
  image-2 sampling box are shown in the lower left panel.}
\label{Fig:scrnshot1}
\end{figure*}

We used maximum correlation tracking, as described by
\cite{Sro2012polar}, to maximize the number of cloud targets that could
be usefully tracked and to reduce errors in position measurements.  We displayed an image
sequence as a stacked series of narrow horizontal strips, each
containing a rectangular projection covering a specified range of
longitudes and a narrow range of latitudes, using
0.2\degx$\times$0.2\deg image pixels, as illustrated in
Fig.\ \ref{Fig:scrnshot1}.
For each cloud target visible in the selected
latitude range, a reference image is selected and a target box is
adjusted in size and position in the reference image so that the box
contains the cloud feature and a small region outside of it.  Target
boxes in other images are initially positioned using {\it a prior} wind
profile; then the guess is manually adjusted as needed to insure that all target
boxes contain the target feature.  The positions of the target boxes
in all but the reference images are then automatically
precision-adjusted to maximize the cross correlation between the
reference target box signal variations and those contained in each of
the other boxes. The correlation coefficient $r$ was computed using \begin{eqnarray}
 r = \frac{\Sigma (x_i-\bar{x})(y_i-\bar{y})}{\sqrt{\Sigma (x_i-\bar{x})^2}\sqrt{\Sigma (y_i-\bar{y})^2}}
\end{eqnarray}
where $x_i$ and $y_i$ are brightness values at the $i$th pixel
location in images $X$ and $Y$ respectively, $\bar{x}$ and $\bar{y}$
their average values, and the summation is over all corresponding pixels in the
target boxes.
 
To reduce the impact of large-scale variations such as produced by
latitude bands, we use high-pass filters to display local contrast and
remove large-scale image slopes before correlation.  Measurements of
longitude and latitude vs. time were fit to straight lines using both
unweighted and weighted fits, and assigned errors that were the larger
of the two estimates obtained from linear regression.  The results are
produced in real time as each target is tracked and displayed as
illustrated in Fig.\ \ref{Fig:scrnshot1}.  Errors in latitude and
longitude are estimated in two ways.  The first is an a priori
computation of the scaled root sum of squared errors produced by displacements
of one image pixel in both dimensions. These errors, which provide the
basis for the weighted fits, vary appropriately with view angle and
position on the disk and are important when the number of samples is
small.  But, when the number of samples is large, the second and more
accurate error estimate is computed from the RMS deviation of the
measurements from a straight line fit, as it includes both navigation
errors as well as target identification and tracking errors.  This
second estimate provides the basis for the unweighted fits.

\subsection{Cloud tracking statistics for 2012-2014 data sets}

We made over 850 measurements of cloud feature drift rates in the
2012-2014 images, mostly tracking small discrete cloud features, with
each drift rate based on from 3 to 11 individual position
measurements.  For comparison, the Voyager mission returned only 8
wind measurements \citep{Smith1986} and the recent extensive
reanalysis of Voyager images by \cite{Kark2015vgr} yielded 27
trackable discrete features.  The Keck data from 2014 provided an exceptional
number of measurements, which is a result of the increased activity of
Uranus in 2014, not due to exceptional image quality. In fact, image
quality during the 5-6 August observing run was a little below par for
our typical Keck II observations.

\begin{figure*}[!htb]\centering
\includegraphics[width=4.5in]{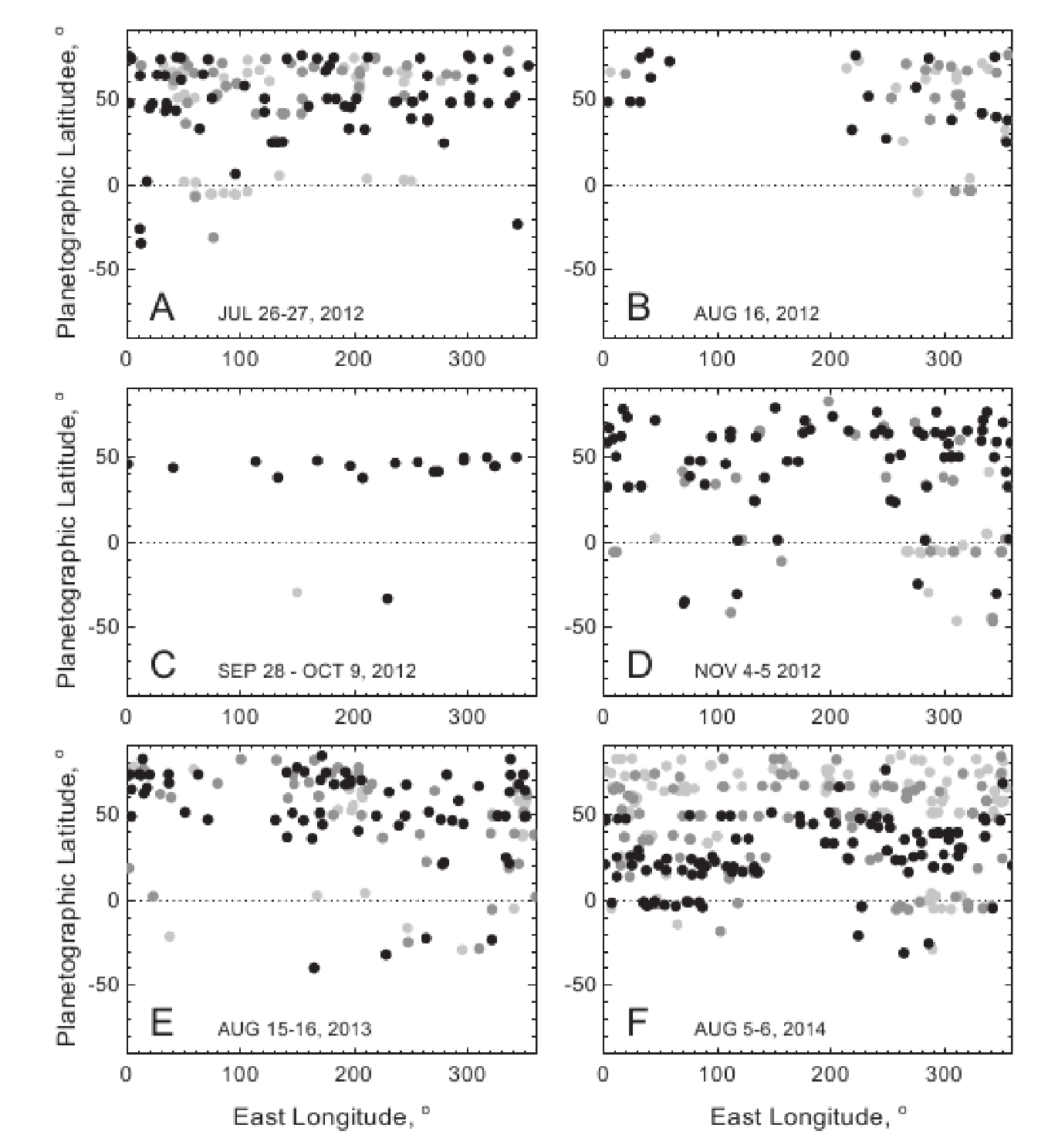}
\vspace{0.15in}
\caption{Spatial distribution of measured drift rates by observing
  run, except for C in which several runs are include. Filled circles
  indicate median correlation for the measurement plotted, using black
  for 0.9 or greater, dark gray for 0.8 to 0.9, and light gray for
  less than 0.8. The lower correlations seen just south of the equator
  in all but panel C, are based on wave feature tracking. The limited
  longitudinal coverage in B is typical of just one night of
  observing. The large number of measurements in 2014 (F) is due to
  increased activity of Uranus, not improved image quality, which is
  actually worse than most of the prior Keck observations. The broad
  longitude distribution in each data set show that results at most
  latitudes are not dominated by just a few discrete features. High correlations
indicate accurate tracking.}
\label{Fig:distribution}
\end{figure*}

\begin{figure*}[!htb]\centering
\includegraphics[width=5.6in]{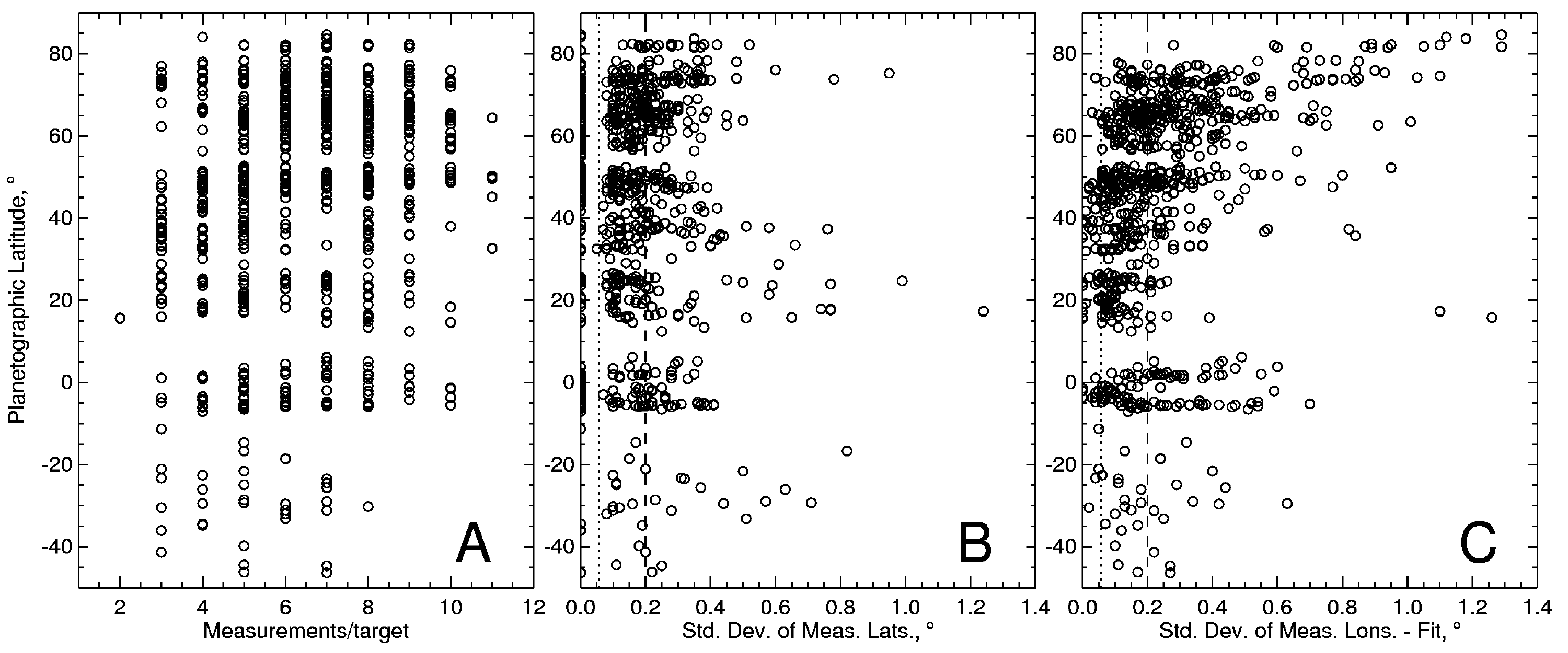}
\includegraphics[width=5.6in]{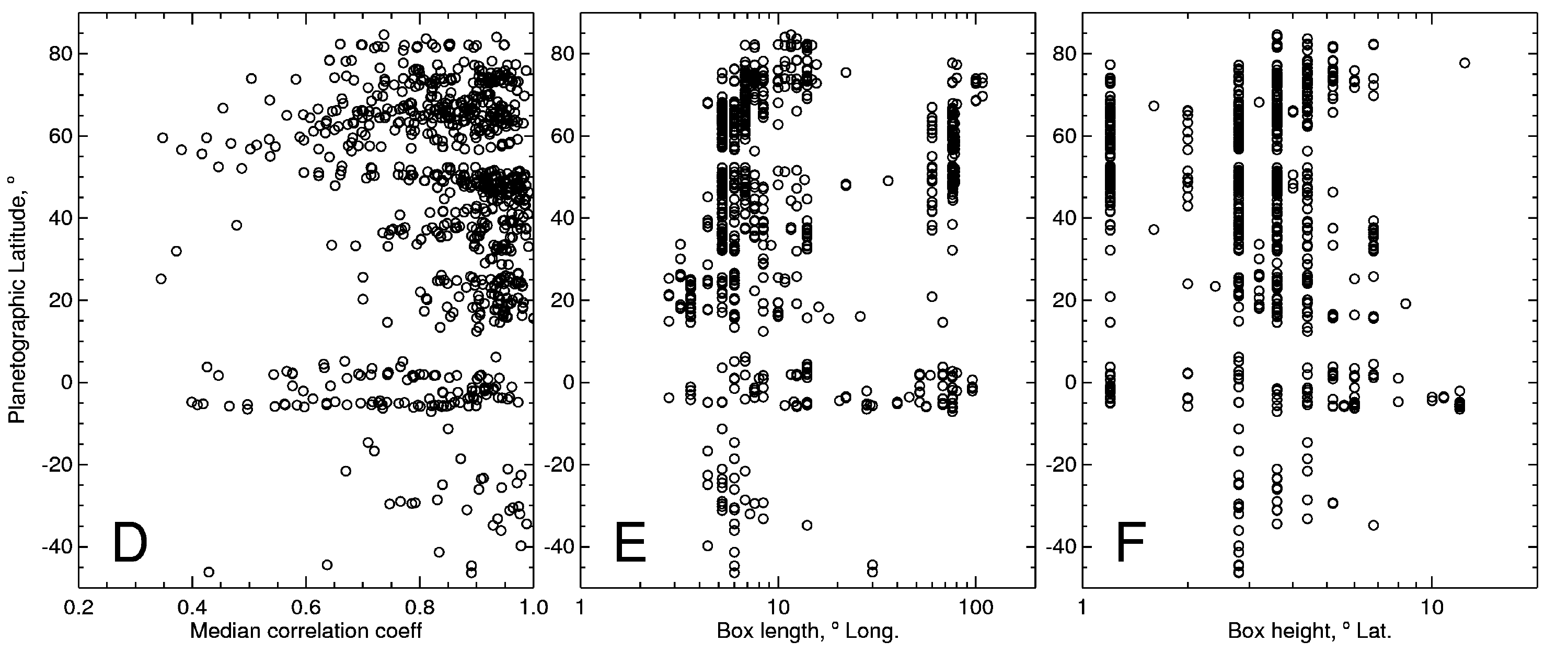}
\includegraphics[width=5.6in]{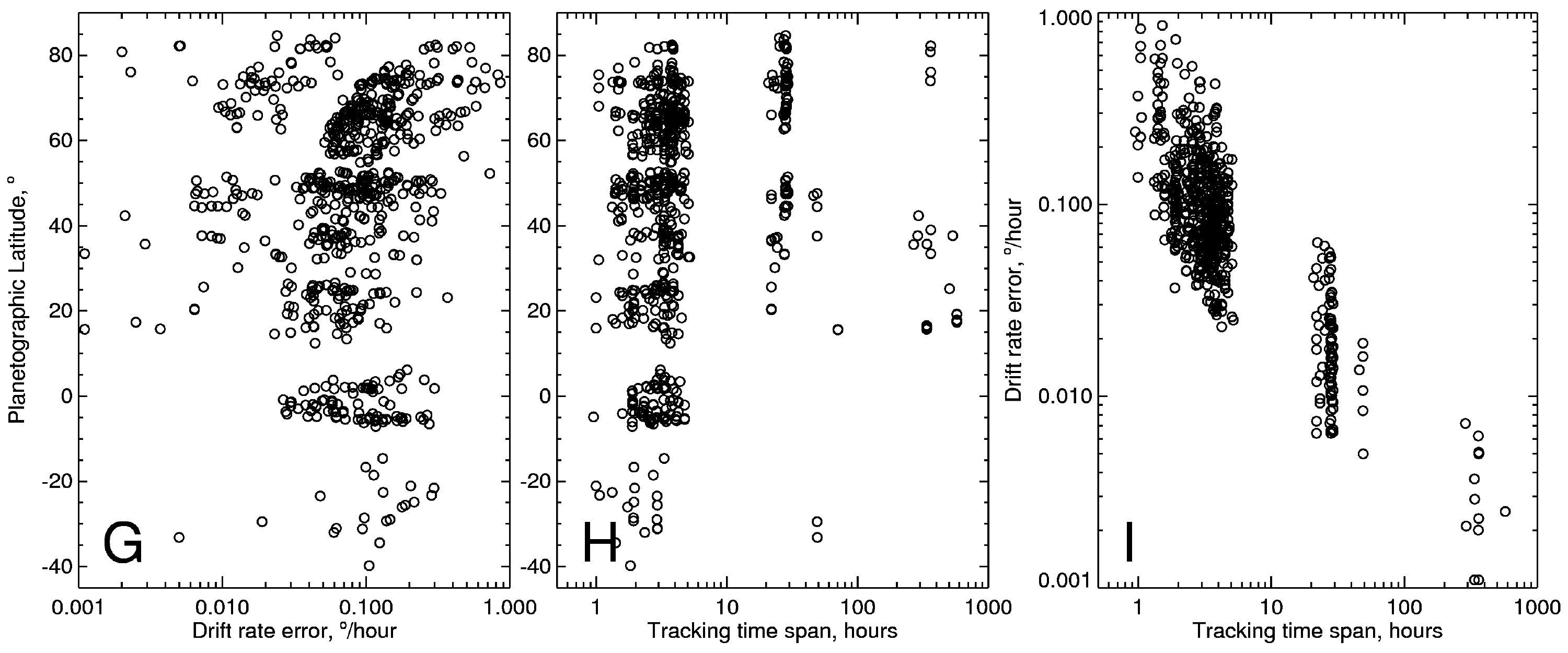}
\caption{Statistical properties of cloud tracked wind measurements vs. latitude: measurements
per target (A), standard deviation of measured latitudes from mean (B), standard deviation of measured
longitudes relative to the linear fit of longitude vs. time (C), median correlation coefficient between
reference cloud and other images of the same cloud (D), target box length (E), target box height (F),
drift rate error (G), and time over which a cloud target was tracked (H). Also shown is a plot of
drift rate error vs. tracking time span (I).}
\label{Fig:mstats}
\end{figure*}

The distribution of measurements in longitude and latitude segregated
by data set is provided in Fig.\ \ref{Fig:distribution}.  As can be
seen from this figure and the plot of statistical properties in
Fig.\ \ref{Fig:mstats}, the measurements are heavily weighted to the
northern hemisphere,  where discrete cloud targets are most
numerous.  Furthermore, because the sub-earth and sub-solar points are
in the northern hemisphere, the cloud features there are visible over
a longer time span during a given observing night than is possible at
southern latitudes. Consequently, even were the distribution of
targets uniform, there would still be more measurements and more
accurate measurements in the northern hemisphere. Especially numerous
were cloud features north of 45\deg N.  Not surprisingly, these data
sets provided no observations below about 40\deg S. 

The distribution of wind measurements over a wide range of longitudes
for most data sets achieves a major objective of our high S/N
observing program, i.e. to avoid having wind results dominated by a
few large features and features in their immediate surroundings, which
might be generated by the large feature and travel along with it, even
at a different latitude.  Fig.\ \ref{Fig:distribution} also classifies
each drift rate measurement according to median correlation of the set
of position measurements used to determine the drift rate.  The darker
the plotted point, the higher the median correlation, and the more
likely the cloud did not evolve much during the tracked interval, and
was elevated above the noise level, improving the accuracy of the
position tracking. 
Most of the measurements obtained median correlation peaks that were
between 0.8 and 1 (Fig.\ \ref{Fig:mstats}E).  Near-equatorial features
that appear to be associated with waves had some of the lowest median
correlation peaks (0.4-0.6).

Discrete cloud features were tracked in sequences of up to eleven
images, always in at least three, and most in 5-9 image sequences.
The standard deviation of latitude measurements for a given cloud
target (Fig.\ \ref{Fig:mstats}B) was usually within one or two map
pixels (0.2\degx/pixel), as were longitude measurements, which were
all determined by correlation.  Most features were tracked using
target boxes between 3\deg and 10\deg in longitude and 2-5\deg in
latitude (Fig.\ \ref{Fig:mstats}E-F).  For compact
features  both latitude and
longitude positions of the cloud boxes were determined by correlation,
which was done to compensate for small navigation errors between
images.  But navigation errors are responsible for only part of the
dispersion in latitude measurements.  The remainder is due to
evolution in cloud target morphology over time, and to some degree
it may be due to latitudinal drifts and/or oscillations, though
it is only tracking over long time intervals that revealed
any statistically significant latitudinal motion.

Figure\ \ref{Fig:mstats} also shows that a subset of our measurements
used correlation boxes that were wide (in longitude) and narrow (in
latitude), which were designed to group together a number of nearby
features at the same latitude to allow a more precise determination of
their group drift rate within a single transit. These were usually
1.2\deg in latitude and 50-100\deg in longitude.  For these boxes we
did correlation tracking of only the longitudinal displacement to
avoid latitude shifts due to the brightest cloud feature included.

A key factor in the resulting accuracy of the wind measurement was the
length of time that a target could be tracked
(Fig.\ \ref{Fig:mstats}H).  The shortest time that we considered even
marginally useful was one hour.  Typically the best we could obtain
within a single night of observing was a little over 4 hours.  When a
feature could be identified on successive nights, which was possible
mostly only for higher-latitude features, wind accuracy improved
dramatically.  For some targets tracked in Gemini data sets we were
able to track targets over a span of several hundred hours (with
significant coverage gaps of course).  Even longer tracking times were
achieved for six persistent discrete features identified mainly in
Keck imagery and discussed in Section\ \ref{Sec:long}.

\subsection{Wind measurements obtained from 2012-2014 data sets}

\begin{figure*}[!htb]\centering
\includegraphics[width=6.4in]{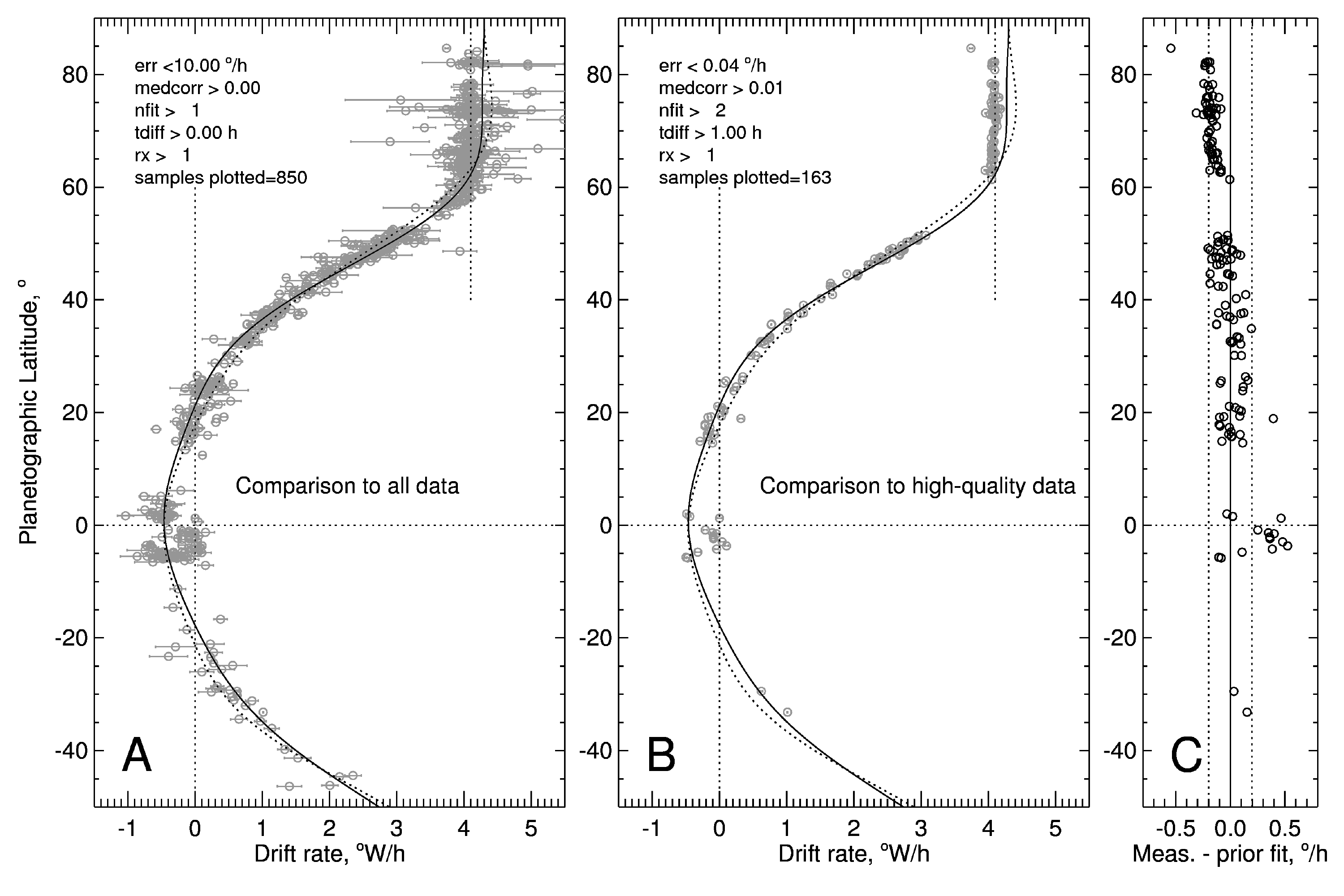}
\caption{Longitudinal drift rates for correlation tracking of cloud targets from 2012-2014 data sets.
The full data set is in A, the subset of measurements with estimated error less than 0.04\degx/h in B, both
in comparison with Model S13A (solid curve) of \cite{Sro2012polar}, and
the deviation of high-accuracy observations in B from the prior wind profile is shown in C.
The dotted curves in A and B trace the S13A  fit profile reflected about the equator to illustrate
its hemispheric asymmetry. Note that the dotted curve
north of 50\degx N comes from the solid curve south of 50\degx S latitude,
which is not well constrained by observations. Sromovsky et al. (2012c) constrained the profile in this
region using a single Voyager data point near 70\degx S (71\degx S planetographic), an assumed
value of 4.3\degx/h westward at the south pole, and the smoothing effects of
a finite number of terms fitting mainly data at other latitudes.}
\label{Fig:driftplot1}
\end{figure*}

The measurements of the westward drift rate relative to the planet's
longitude system are plotted in Fig.\ \ref{Fig:driftplot1}, where the
full data set of 852 measurements is shown in panel A and a subset
including only the 163 most accurate measurements in panel B.  These
two samples are both compared to the Model S13A, shown as a solid
curve, and its reflection about the equator shown as the dotted curve.
Points falling on the solid curve would tend to confirm the asymmetry
previously inferred.  Points between the solid and dotted curves would
be more consistent with hemispheric symmetry. Since many of our points
fall between these two curves, they suggest somewhat less asymmetry
than the prior fit.  A more detailed analysis of asymmetry is discussed
in Section\ \ref{Sec:asymm}.

The largest discrepancies between our new observations and the prior
Legendre polynomial fit are near the equator and at high latitudes
(Fig.\ \ref{Fig:driftplot1}).  In the north polar region, we find a
very robust determination of solid body rotation between 62\degx N and
83\degx N, with a westward rotation rate of 4.1\degx/h relative to the
planet's interior rotation rate of 360\degx/17.24h (=20.88\degx/h).
That corresponds to a rotational period of 14.41 hours.  The prior
Legendre polynomial fit has a relative solid body rotational period of
4.3\degx/h, which is less accurate because of fewer
measurements, shorter time tracks, and poorer views of high
latitudes.

The second discrepancy with the prior wind profile occurs within 6\deg
of the equator.  Measurements with the highest accuracy, based on small
discrete clouds of high contrast in 2014 images, yield a very small
longitudinal drift of 0.1\degx/h eastward between 1\deg N and 5\deg S,
while the Legendre fit suggests a drift rate of about 0.4\degx/h to
the east.  That latter drift rate is more compatible with our lower
accuracy measurements in this region, which are based on much more
diffuse low-contrast features that have relatively low median
correlations, or on wave patterns, such as the nearly sinusoidal dark
ribbon visible in several of the images in Fig.\ \ref{Fig:imsamples}.
It is clear that the wave features drift eastward at close to four
times the rate observed for the small discrete features tracked in
2014.  Drift rates similar to those measured for wave features in our
2012-2014 observations were also observed by \cite{Hammel2005winds},
who found, in their 2003 images of Uranus, ``fuzzy patches'' within
2\deg of the equator moving with an average eastward velocity of 47
m/s, or 0.38\degx/h eastward.  Similar features measured by
\cite{Sro2012polar} 2-3\deg north of the equator, were found to move
eastward at a rate of 0.51\degx/h.  Interestingly, \cite{Sro2012polar}
found the features to have roughly a 40\deg spacing in longitude,
while \cite{Hammel2005winds} found a 30\deg spacing, but with gaps.
The features in our 2012-2014 observations seem to be near 36\degx,
and not quite twice the period of the ribbon wave. (This is presented
in Section\ \ref{Sec:waves}.) These high S/N images indicate that not all the
equatorial features match the average spacing of 36\degx; some have
considerably different spacings.  There are also gaps and regions
where features are spread out and difficult to locate. It is not clear
that the wave producing these features extends entirely around the
planet. The variability of these features may explain the
discrepancies between prior measurements of their spacings.

\subsection{Binned wind results}

\begin{figure*}[!htb]\centering
\includegraphics[width=6.5in]{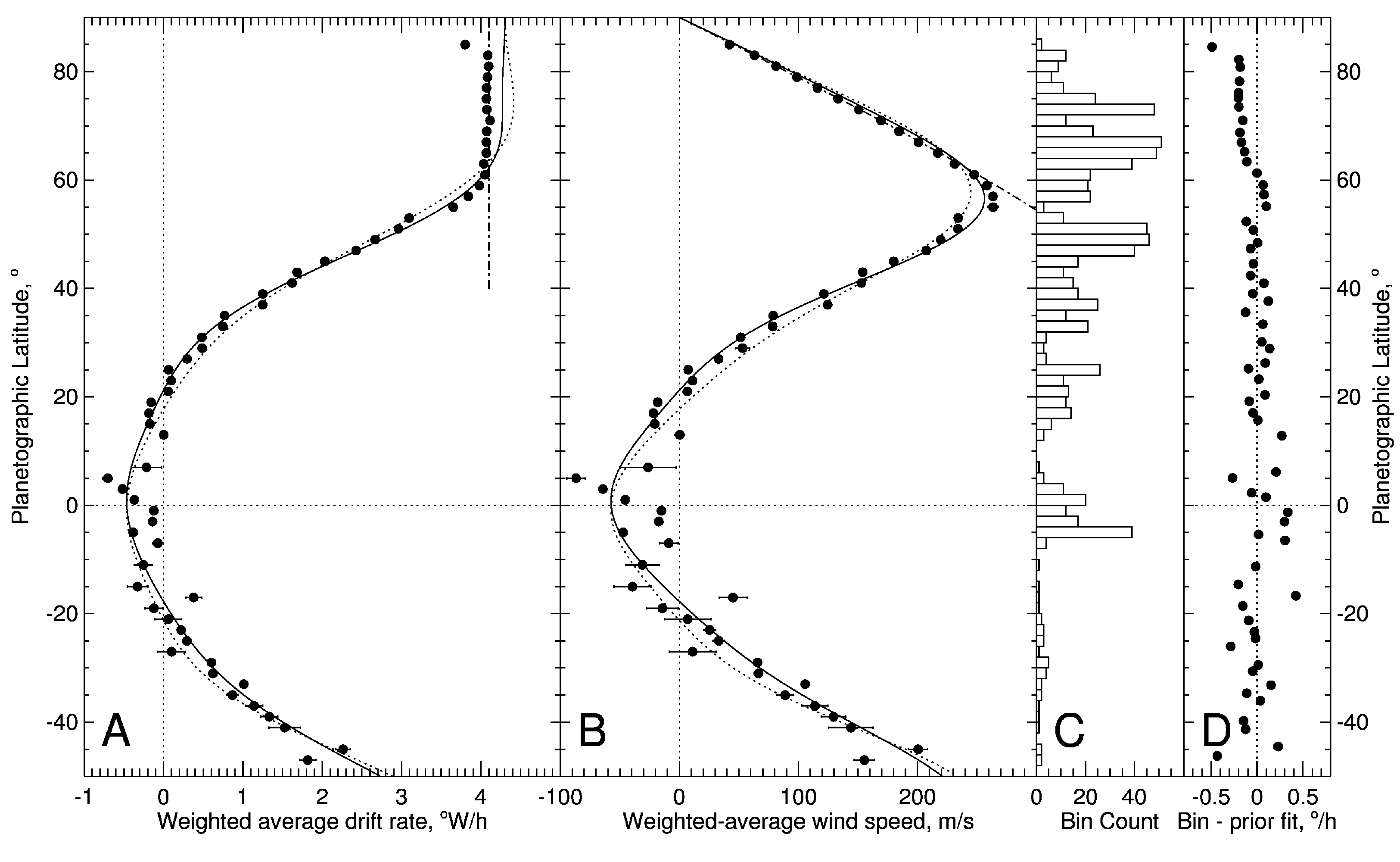}
\caption{Weighted-average longitudinal drift rates for correlation
  tracking of cloud targets from 2012-2014 data sets, averaged in
  2\deg bins (A), corresponding wind speeds (B), the number of
  measurements/bin (C), and differences (D) from Model S13A of
  \cite{Sro2012polar}.  The dotted curves in A and B are the prior
  profile reflected about the equator to illustrate its hemispheric
  asymmetry. The dot-dash lines in C and D are lines of constant
  angular drift rate at 4.1\degx/h. Note that both binned wind values
  and differences are computed at weighted average latitudes for
  features within the bins, not at the central latitude of the bin.}
\label{Fig:bin1}
\end{figure*}

Selecting only the high-accuracy observations leaves significant gaps
in latitude coverage.  To fill in those gaps we average observations
within latitude bins, using the estimated variance of each observation
as its weight in computing a weighted average.  The results for 2\deg
bins are shown in Fig.\ \ref{Fig:bin1} and tabulated in
Tables\ \ref{Tbl:northbin} and \ref{Tbl:southbin}. This procedure does
indeed fill in many of the missing latitudes in the high-accuracy
subset.  The estimated error of the weighted average measurements is
so small in most bins that it is within the plotted symbol.  The
region of solid body rotation is seen to be very uniform in these
binned results. Between 65\deg N and 83\deg N, we obtain an average of
the binned measurements of 4.079\degx/h with a standard deviation of
0.015\degx/h, compared to a range of weighted average expected errors
of .002-0.022\degx/h.

Near the equator, binned results display significant scatter.  Just
north of the equator, the binned results are in rough agreement with
the old (S13A) wind profile.  But just south of the equator, the
high accuracy winds from small high-contrast features has dominated
the winds inferred from larger scale low contrast and wave features.

An odd feature of the binned results is that the wind profile in the
region from about 18\degx N to 45\degx N seems to have a sequence of
stair steps of local solid body rotation extending for a few degrees
with sharp, high shear transitions between them.  It could very well
be the case that the zonal wind profile of Uranus is simply not a very
smooth function of latitude.  We also tried binning at 1\deg
intervals, and found the same stair-step structure between 10\degx N
and 30\degx N, but hints of that structure at other latitudes were no
longer evident, perhaps due to increased variability. The north polar
region of solid-body rotation was largely unchanged by reducing the
bin size, except for small increases in noise due to reduced numbers
of samples per bin.  No fundamentally new structures (that could be
distinguished from noise) were found in the wind profile by using
smaller bin sizes of 1\deg or 0.5\degx.

\begin{table*}\centering
\caption{Binned 2012-2014 wind observations for the northern hemisphere.}
\vspace{0.15in}
\begin{tabular}{c c c c c c c}
\hline\\[-0.1in] 
\multicolumn{3}{c}{Planetographic bin latitude}    & Drift rate& Std. Dev. & Wind Speed & Bin \\
Center & Avg. & Std. Dev. & \degx E/h & \degx /h  & m/s & Count \\
\hline\\[-0.1in] 
  85 & 84.58 &  0.41 & -3.801$\pm$0.022 &   0.32 &    42.0$\pm$0.2 &  2 \\
  83 & 82.24 &  0.44 & -4.085$\pm$0.003 &   0.13 &    63.1$\pm$0.1 & 12 \\
  81 & 80.84 &  0.32 & -4.096$\pm$0.002 &   0.39 &    81.2$\pm$0.0 &  9 \\
  79 & 78.23 &  0.15 & -4.081$\pm$0.015 &   1.00 &    98.7$\pm$0.4 &  6 \\
  77 & 76.11 &  0.58 & -4.070$\pm$0.002 &   0.36 &   116.0$\pm$0.1 & 11 \\
  75 & 75.16 &  0.69 & -4.068$\pm$0.007 &   0.31 &   133.3$\pm$0.2 & 24 \\
  73 & 73.51 &  0.50 & -4.074$\pm$0.004 &   0.26 &   150.7$\pm$0.1 & 48 \\
  71 & 71.00 &  0.60 & -4.114$\pm$0.008 &   0.52 &   169.4$\pm$0.3 & 12 \\
  69 & 68.74 &  0.66 & -4.072$\pm$0.006 &   0.28 &   184.5$\pm$0.3 & 23 \\
  67 & 66.93 &  0.64 & -4.068$\pm$0.004 &   0.18 &   200.8$\pm$0.2 & 51 \\
  65 & 65.23 &  0.59 & -4.067$\pm$0.005 &   0.24 &   217.0$\pm$0.3 & 49 \\
  63 & 63.41 &  0.51 & -4.035$\pm$0.004 &   0.22 &   231.2$\pm$0.2 & 39 \\
  61 & 61.30 &  0.49 & -4.050$\pm$0.009 &   0.18 &   247.7$\pm$0.6 & 22 \\
  59 & 59.11 &  0.56 & -3.979$\pm$0.017 &   0.14 &   258.3$\pm$1.1 & 21 \\
  57 & 57.33 &  0.43 & -3.841$\pm$0.017 &   0.15 &   263.4$\pm$1.1 & 22 \\
  55 & 55.16 &  0.41 & -3.651$\pm$0.057 &   0.07 &   263.5$\pm$4.1 &  3 \\
  53 & 52.34 &  0.18 & -3.094$\pm$0.023 &   0.15 &   234.2$\pm$1.7 & 11 \\
  51 & 50.77 &  0.48 & -2.959$\pm$0.005 &   0.20 &   234.0$\pm$0.4 & 45 \\
  49 & 48.42 &  0.50 & -2.665$\pm$0.005 &   0.23 &   219.5$\pm$0.4 & 46 \\
  47 & 47.34 &  0.54 & -2.426$\pm$0.003 &   0.22 &   207.6$\pm$0.2 & 40 \\
  45 & 44.54 &  0.53 & -2.031$\pm$0.004 &   0.16 &   180.0$\pm$0.3 & 17 \\
  43 & 42.38 &  0.57 & -1.681$\pm$0.002 &   0.22 &   154.0$\pm$0.2 & 11 \\
  41 & 40.95 &  0.41 & -1.620$\pm$0.012 &   0.15 &   153.0$\pm$1.1 & 15 \\
  39 & 39.05 &  0.46 & -1.250$\pm$0.000 &   0.11 &   121.4$\pm$0.0 & 17 \\
  37 & 37.67 &  0.53 & -1.248$\pm$0.000 &   0.19 &   124.5$\pm$0.0 & 25 \\
  35 & 35.60 &  0.34 & -0.771$\pm$0.001 &   0.13 &    78.8$\pm$0.1 & 12 \\
  33 & 33.43 &  0.53 & -0.748$\pm$0.001 &   0.14 &    78.3$\pm$0.1 & 21 \\
  31 & 30.14 &  1.07 & -0.482$\pm$0.012 &   0.13 &    51.5$\pm$1.3 &  4 \\
  29 & 28.89 &  0.21 & -0.487$\pm$0.054 &   0.15 &    53.1$\pm$5.9 &  3 \\
  27 & 26.26 &  0.26 & -0.297$\pm$0.020 &   0.15 &    32.9$\pm$2.3 &  4 \\
  25 & 25.21 &  0.53 & -0.065$\pm$0.000 &   0.17 &     7.4$\pm$0.0 & 26 \\
  23 & 23.27 &  0.62 & -0.097$\pm$0.017 &   0.33 &    11.1$\pm$2.0 & 11 \\
  21 & 20.38 &  0.53 & -0.058$\pm$0.004 &   0.17 &     6.7$\pm$0.5 & 13 \\
  19 & 19.19 &  0.51 &  0.155$\pm$0.000 &   0.22 &   -18.2$\pm$0.0 & 12 \\
  17 & 17.05 &  0.64 &  0.184$\pm$0.000 &   0.15 &   -21.8$\pm$0.0 & 14 \\
  15 & 15.70 &  0.62 &  0.173$\pm$0.001 &   0.16 &   -20.7$\pm$0.1 &  6 \\
  13 & 12.85 &  0.56 & -0.003$\pm$0.034 &   0.09 &     0.4$\pm$4.1 &  3 \\
   7 &  6.17 &  0.00 &  0.214$\pm$0.193 &   0.00 &   -26.3$\pm$23.7 &  1 \\
   5 &  5.04 &  0.39 &  0.703$\pm$0.063 &   0.12 &   -86.8$\pm$7.8 &  3 \\
   3 &  2.31 &  0.74 &  0.519$\pm$0.017 &   0.12 &   -64.3$\pm$2.2 & 11 \\
   1 &  1.51 &  0.41 &  0.367$\pm$0.014 &   0.23 &   -45.5$\pm$1.7 & 20 \\
\hline
\end{tabular}\label{Tbl:northbin}
\end{table*}

\begin{table*}\centering
\caption{Binned 2012-2014 wind observations for the southern hemisphere.}
\vspace{0.15in}
\begin{tabular}{c c c c c c c}
\hline\\[-0.1in] 
\multicolumn{3}{c}{Planetographic bin latitude}    & Drift rate& Std. Dev. & Wind Speed & Bin \\
Center & Avg. & Std. Dev. & \degx E/h & \degx /h  & m/s & Count \\
\hline\\[-0.1in] 
  -3 & -3.02 &  0.74 &  0.139$\pm$0.011 &   0.30 &   -17.2$\pm$ 1.4 & 17 \\
  -5 & -5.38 &  0.48 &  0.383$\pm$0.009 &   0.23 &   -47.3$\pm$ 1.1 & 39 \\
  -7 & -6.49 &  0.45 &  0.072$\pm$0.063 &   0.36 &    -8.8$\pm$ 7.7 &  4 \\
 -11 &-11.30 &  0.00 &  0.255$\pm$0.116 &   0.00 &   -31.1$\pm$14.1 &  1 \\
 -15 &-14.60 &  0.00 &  0.328$\pm$0.130 &   0.00 &   -39.4$\pm$15.6 &  1 \\
 -17 &-16.68 &  0.00 & -0.379$\pm$0.099 &   0.00 &    45.0$\pm$11.8 &  1 \\
 -19 &-18.56 &  0.00 &  0.121$\pm$0.113 &   0.00 &   -14.2$\pm$13.3 &  1 \\
 -21 &-21.24 &  0.35 & -0.060$\pm$0.168 &   0.36 &     6.9$\pm$19.5 &  2 \\
 -23 &-23.36 &  0.47 & -0.221$\pm$0.044 &   0.37 &    25.3$\pm$5.1 &  3 \\
 -25 &-24.55 &  0.56 & -0.293$\pm$0.042 &   0.25 &    33.0$\pm$4.7 &  3 \\
 -27 &-26.03 &  0.00 & -0.101$\pm$0.178 &   0.00 &    11.2$\pm$19.7 &  1 \\
 -29 &-29.43 &  0.41 & -0.603$\pm$0.018 &   0.19 &    65.7$\pm$2.0 &  5 \\
 -31 &-30.61 &  0.45 & -0.622$\pm$0.033 &   0.11 &    66.4$\pm$3.5 &  4 \\
 -33 &-33.17 &  0.82 & -1.010$\pm$0.005 &   0.19 &   105.7$\pm$0.5 &  2 \\
 -35 &-34.67 &  0.25 & -0.869$\pm$0.071 &   0.20 &    88.8$\pm$7.2 &  2 \\
 -37 &-36.06 &  0.00 & -1.140$\pm$0.110 &   0.00 &   113.8$\pm$11.0 &  1 \\
 -39 &-39.78 &  0.00 & -1.333$\pm$0.105 &   0.00 &   129.6$\pm$10.2 &  1 \\
 -41 &-41.31 &  0.00 & -1.526$\pm$0.198 &   0.00 &   144.2$\pm$18.7 &  1 \\
 -45 &-44.50 &  0.15 & -2.262$\pm$0.093 &   0.17 &   200.5$\pm$8.2 &  2 \\
 -47 &-46.22 &  0.11 & -1.816$\pm$0.101 &   0.24 &   155.4$\pm$8.7 &  2 \\
\hline
\end{tabular}\label{Tbl:southbin}
\end{table*}
\subsection{The problem of measuring near-equatorial wind speeds}

Very few compact discrete cloud features have ever been observed in the
near-equatorial region (within about 10\deg of the equator).  Most of
the zonal drift rates in this region have been estimated by tracking
wave features, either the ribbon wave that can be seen near 5\degx S
in the most of the high S/N images, or diffuse bright
features that also seem to be associated with waves.  An unusual
appearance of compact discrete cloud features in the 2014 images in
the 0-5\degx S region provided a distinctly different drift rate
estimate that is close to zero.  Which of these is the best estimate
of the atmospheric mass flow is not certain, although waves are often
seen traveling relative to the mass flow.  Thus we tend to favor the small discrete clouds as
more representative of the mass flow.  On the other hand, the only
independent measure of the mass flow, inferred from radio occultation
measurements \citep{Lindal1987} suggests that the mass flow is more
eastward than either the wave or discrete feature results, although the
measurement has such a large error bar that it does not provide a very
firm constraint.

\section{Legendre polynomial fits and symmetry properties}

\subsection{Legendre fitting methodology}

Our aim in carrying out polynomial fits to the wind
observations was to provide a smooth
profile for atmospheric modelers and other researchers.
We modeled longitudinal drift rates and
wind speeds  using the following Legendre expansion and conversion
equations:\begin{eqnarray} d\phi/dt = \sum_{i=0}^{n} C_i \times
  P_i(\sin(\theta))\\ U = -4.8481\times10^{-3} R(\theta) \times
  d\phi/dt \\ R(\theta) = R_E/\sqrt{1+(R_P/R_E)^2 \tan(\theta)^2}
\end{eqnarray}
where $C_i$ are the coefficients given in Table\ \ref{Tbl:legfits},
$P_i(\sin(\theta))$ is the $i$th Legendre polynomial evaluated at the
sine of planetographic latitude $\theta$, $d\phi/dt$ is the
longitudinal drift rate in \degx /h (using planetographic longitude,
which is east longitude for a retrograde rotator like Uranus), $U$ is
the westward wind speed in m/s, $R$ is the radius of rotation in km at
latitude $\theta$, which is the distance from a point on the 1-bar
surface to the planet's rotational axis, and $R_E$ and $R_P$ are the
equatorial and polar radii of Uranus, for which we use 25559 km and
24973 km \citep{Archinal2011}. We fit the longitudinal drift rates
rather than wind speeds, because the former do not change as dramatically
at high latitudes as the latter, and thus are much easier to fit
without generating large deviations in data-sparse latitude regions.
To limit such large deviations, we also need to limit the order of the
polynomials. For the symmetric fits, the summation over $i$ is only
over the even polynomials.  The model coefficients are found by
minimizing $\chi^2$, but with error estimates for the observations
modified as described by \cite{Sro2012polar} and paraphrased in the
following paragraph.  Note that while we tabulate drift rates in east
longitude, the plots show westward drift rates to be consistent with
the prior practice of making prograde (in this case westward) winds
positive
\citep{Allison1991uranbook,Hammel2001Icar,Hammel2005winds,Sro2005dyn,Sro2009eqdyn}.

Because very accurate measurements of drift rates at nearly the same
latitude often differ by many times the value expected from those
uncertainties, it appears that one or more of the following must be
true: (1) the circulation is not entirely steady, (2) the features
that we measure do not all represent the same atmospheric level, (3)
the feature tracked has evolved or been misidentified, or (4) the
cloud features we track are not always at the same latitude as the
cloud-generating circulation feature that is moving with the zonal
flow. Examples of the latter possibility are the companion clouds to
Neptune's Great Dark Spot \citep{Sro1993Icar}.  In Section
\ref{Sec:AB} we provide examples from our current data set. If we
weighted highly accurate measurements by their expected inverse
variance, they would dominate the fit, leading to wild variations in
regions where there are less accurate measurements.  Since these high
accuracy measurements clearly do not all follow the mean flow, we must
add an additional uncertainty to characterize their deviations from
it.  We do this by root sum squaring the estimated error of
measurement with an additional error of representation denoted by
$\sigma_{R}$, which is adjusted so that the $\chi^2$ value of the
complete fit is approximately equal to the number of measurements.
The need for $\sigma_{R}$ might be partly due to slight discrepancies
in latitude that arise from a cloud feature being generated by a
circulation feature that is latitudinally offset from the bright cloud
feature that is tracked.  

Another factor in evaluating wind measurement accuracy is the rate at
which latitudinal accuracy and drift rate accuracy improve with
extended tracking time spans.  Drift rate accuracy improves roughly
linearly with time span, and can improve dramatically from just a few
widely separated measurements, while latitude accuracy improves as the
square root of the number of measurements, and is not much improved by
adding just a few measurements or by extending the time span of the
measurements. If we consider a shear of 2\degx/h per 10\deg of
latitude (about the maximum observed shear), a half-degree error in
latitude is equivalent to a 0.1\degx /h error in wind measurement,
which is about the value of $\sigma_{R}$ that we need to make the
\chisq value for the fit equal to the number of degrees of freedom (or
\chisqx/N$_F$ $\approx$ 1).  The apparent local stair steps suggested
by the binned wind profile would also cause local errors relative to a
smooth profile.  We found $\sigma_R\approx$ 0.1\degx/h for both the
2009-2011 and 2007-2011 data sets analyzed by \cite{Sro2012polar}. A
similar value was needed in our current analysis.

\subsection{Symmetric fit results}

\begin{figure*}[!htb]\centering
\includegraphics[width=6.5in]{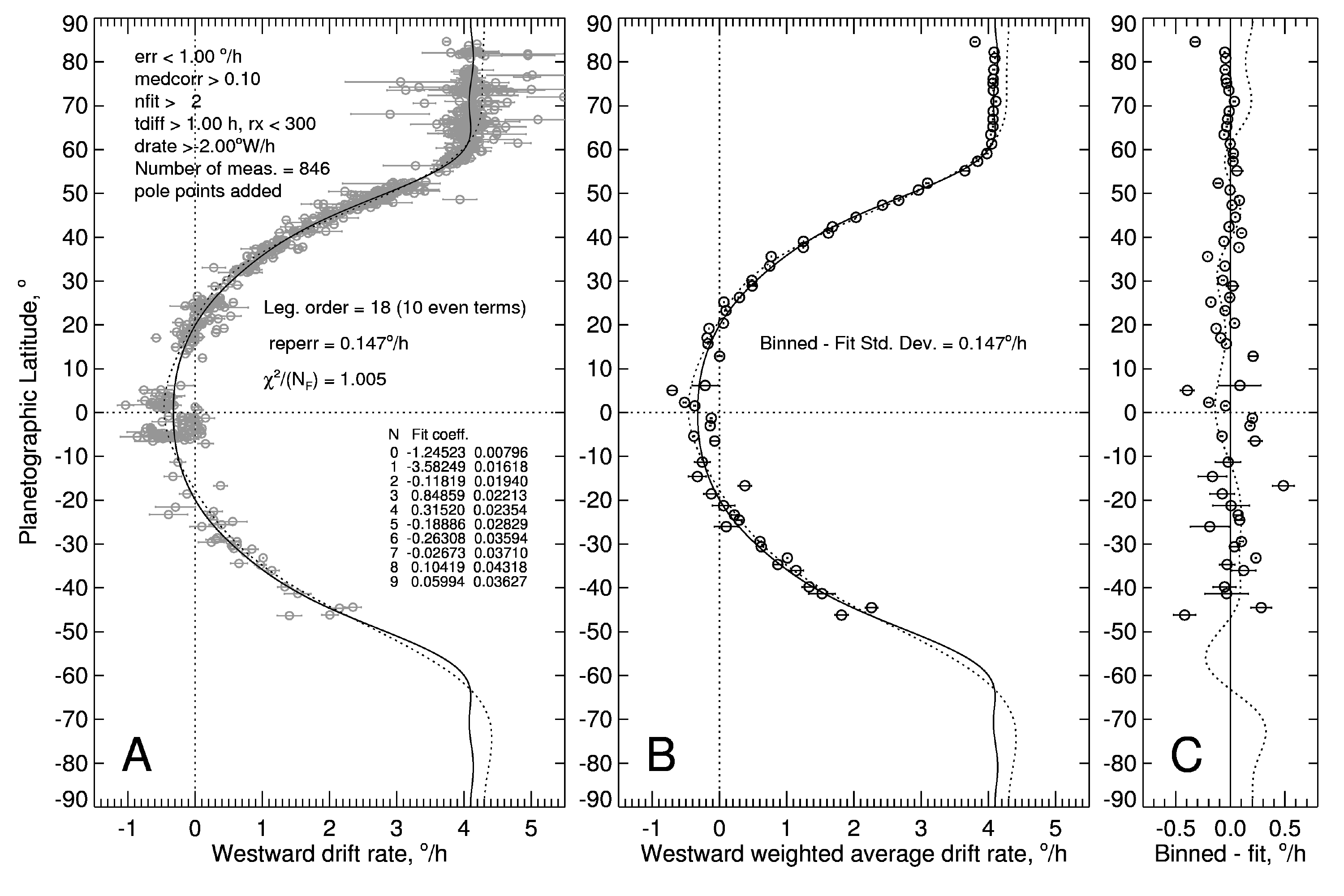}
\caption{Ten-term order-18 Legendre polynomial fit to nearly our
  entire data set using a $\sigma_R$ (reperr) value of
  0.147\degx/h, which is shown as a solid curve, in comparison with our
  large data set in (A) and with our binned data set in (B).  The
  difference between binned results and the fit are shown in C.  The
  dotted curves in A and B show Model S13A of \cite{Sro2012polar}.}
\label{Fig:symfitall}
\end{figure*}

Because our 2012-2014 wind data set contains such a sparse sampling of
the southern hemisphere, with no samples poleward of 47\degx S, it is
difficult to reliably constrain any north-south asymmetry in the wind
profile.  Accordingly, we first considered fits using a series of even
Legendre polynomials, which guarantees hemispheric symmetry and
provides a useful reference for detecting asymmetries. To fit the
relatively sharp kinks near 55\degx N and 62\deg N, we needed to use a
fairly large number of polynomials.  After finding fit problems with
orders of 10-14, we settled on a ten-term series of even polynomials
up to order 18.  The results are shown in Fig.\ \ref{Fig:symfitall}
for the case in which we fit essentially the entire data set,
including 846 measurements and eliminating only 6. This is shown as a
solid line, while the most recent fit to prior observations, Model
S13A, is shown as a dotted line.  Here we also added synthetic points
at both north and south poles to help straighten out our drift rate
fit at high northern polar latitudes. We used a $\sigma_R$ value of
0.147\degx/h. The Legendre coefficients and their uncertainties are
given in Table \ref{Tbl:legfits}. There we also provide coefficients
for alternate fits discussed later in this section.

In comparison with our new symmetric fit, the prior asymmetric fit
provides a degree of asymmetry in the 20\deg - 40\deg range that seems
to provide slightly better agreement with the new observations in that
region.  But in the region from 55\degx N to 85\degx N, the old fit
has a solid body rotation rate that is 0.2\degx/h too fast.  This
might represent a slight decrease over time, but the small number of
2011 measurements and their error bars of 0.3\degx/h or more, make
this a difference of dubious significance.  Near the equator, our
new fit has a lower eastward drift rate, which is a result of the
large number of high quality 2014 measurements between -6\deg and
1\deg that pull the profile closer to zero.  Our new fit is not quite
straight enough to fit the constant rotation rate observed at high
northern latitudes, but is within about 0.1\degx/h of the correct
drift rate.  Near the equator, this fit splits the difference between
small discrete results and the large pattern results.  This leads to a
nearly solid body rotation within about 7.5\deg of the equator, at a
rate of about 0.12\deg/h eastward.

\begin{figure*}[!htb]\centering
\includegraphics[width=6in]{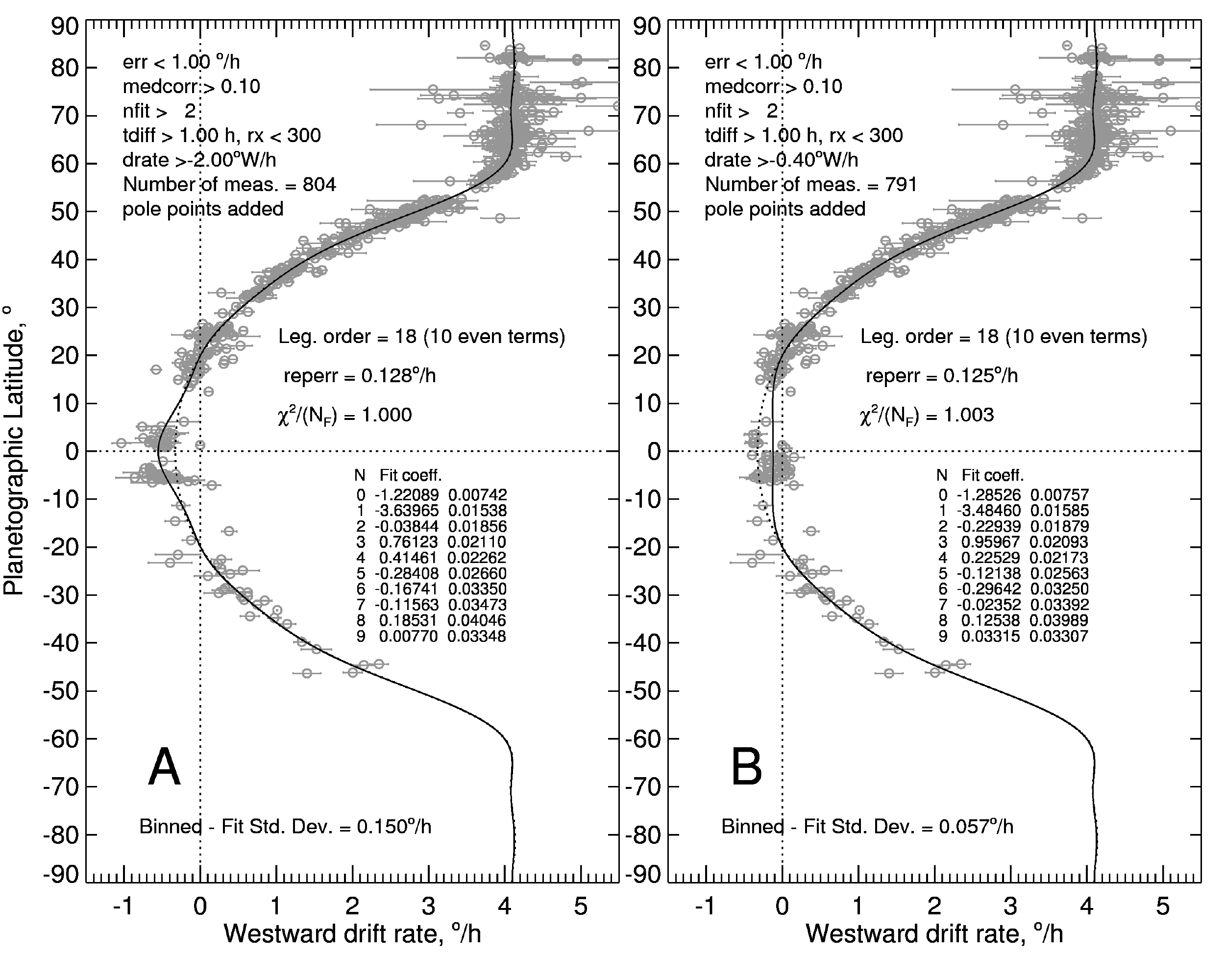}
\caption{As in Fig.\ \ref{Fig:symfitall}, except that (A) 2014
  measurements of discrete features between 7\degx S and 2\degx N are
  excluded from the fit, and a $\sigma_R$ (reperr) of 0.15\degx/h was
  used, and for (B) observations of near equatorial wave motions were
  excluded from the fit, and a $\sigma_R$ = 0.125\degx/h was used. The
  fits are shown by solid lines, while the dotted line here shows the
  fit that includes all the data.}
\label{Fig:dualsymfits}
\end{figure*}

An alternative fit that eliminates the high-correlation low-latitude
2014 observations (those associated with small discrete features), but
leaving the drift rates derived from wave features, is shown in
Fig.\ \ref{Fig:dualsymfits}A.  For this fit we reduced $\sigma_R$ to
0.128\degx /h. This fit displays a clear retrograde (eastward)
equatorial peak.  Realizing that the observations near 6\deg S and
near 2\deg N, with drift rates less than -0.3\degx/h are measurements
of wave motions, while the 2014 measurements in this region are almost
entirely of small discrete features that are more likely indicators of
mass flow, we also carried out a fit in which the putative wave
tracking results are eliminated.  That fit, displayed in
Fig.\ \ref{Fig:dualsymfits}B, contains a region within 15\deg of the
equator that appears to be in solid body rotation at a rate of
0.1\degx/h eastward.

\begin{table*}\centering
\caption{Fit coefficients for symmetric Legendre polynomial fits to 2012-2014 wind observations.}
\vspace{0.15in}
\begin{tabular}{c c c c c c c c}
\hline\\[-0.1in] 
 & Legendre & \multicolumn{2}{c}{Fit 1 (846 pts)} & \multicolumn{2}{c}{Fit 2 (Eq. spots excl.)}  & \multicolumn{2}{c}{Fit 3 (Eq. waves excl.)}\\
Term & Order & Coeff. & Unc. & Coeff. & Unc.  & Coeff. & Unc.\\
\hline\\[-0.1in] 
    0  &  0& -1.245225&    0.0079& -1.220885&    0.0074& -1.285259&    0.0076\\[0.05in]
    1  &  2& -3.582487&    0.0162& -3.639650&    0.0154& -3.484602&    0.0158\\[0.05in]
    2  &  4& -0.118185&    0.0194& -0.038439&    0.0186& -0.229385&    0.0188\\[0.05in]
    3  &  6&  0.848593&    0.0221&  0.761227&    0.0211&  0.959671&    0.0209\\[0.05in]
    4  &  8&  0.315199&    0.0235&  0.414609&    0.0226&  0.225287&    0.0217\\[0.05in]
    5  & 10& -0.188857&    0.0283& -0.284082&    0.0266& -0.121381&    0.0256\\[0.05in]
    6  & 12& -0.263077&    0.0359& -0.167412&    0.0335& -0.296423&    0.0325\\[0.05in]
    7  & 14& -0.026728&    0.0371& -0.115626&    0.0347& -0.023522&    0.0339\\[0.05in]
    8  & 16&  0.104192&    0.0432&  0.185315&    0.0405&  0.125384&    0.0399\\[0.05in]
    9  & 18&  0.059944&    0.0362&  0.007696&    0.0335&  0.033149&    0.0331\\[0.05in]
\hline
\end{tabular}\label{Tbl:legfits}
\end{table*}

\subsection{Evidence for mid-latitude asymmetry}\label{Sec:asymm}

\begin{figure*}[!htb]\centering
\includegraphics[width=6.in]{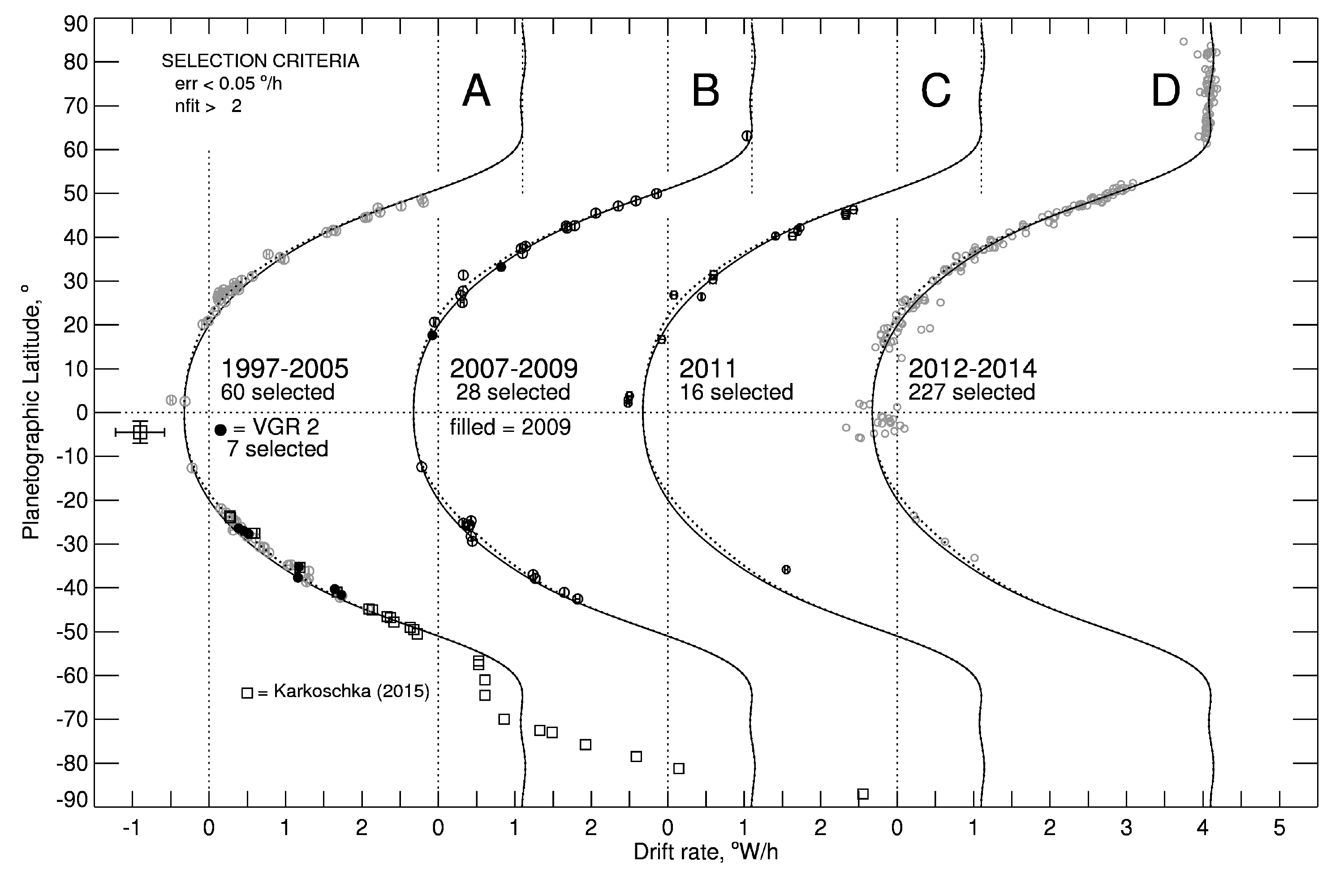}
\caption{Higher accuracy wind measurements from 1997-2005 (A),
  2007-2009 (B), 2011 (C), and 2012-2014 (D), compared to our new even
  polynomial fit to the entire 2012-2014 data set (solid curves).  In
  (A) we also show Voyager 2 observations \cite{SmithBA1986} as filled
  black circles and one radio occultation result (large square) from
  \cite{Lindal1987}, as well as the discrete tracking results (small
  squares) from the reanalysis of Voyager 2 images by
  \cite{Kark2015vgr}.  See text for further references. The dotted
  curves are the sum of the even (symmetric fit) and the asymmetric
  fit to the differences, as shown in the left panel of
  Fig.\ \ref{Fig:winddiff}. }
\label{Fig:priorwinds}
\end{figure*}

\begin{figure*}[!htb]\centering
\includegraphics[width=2.36in]{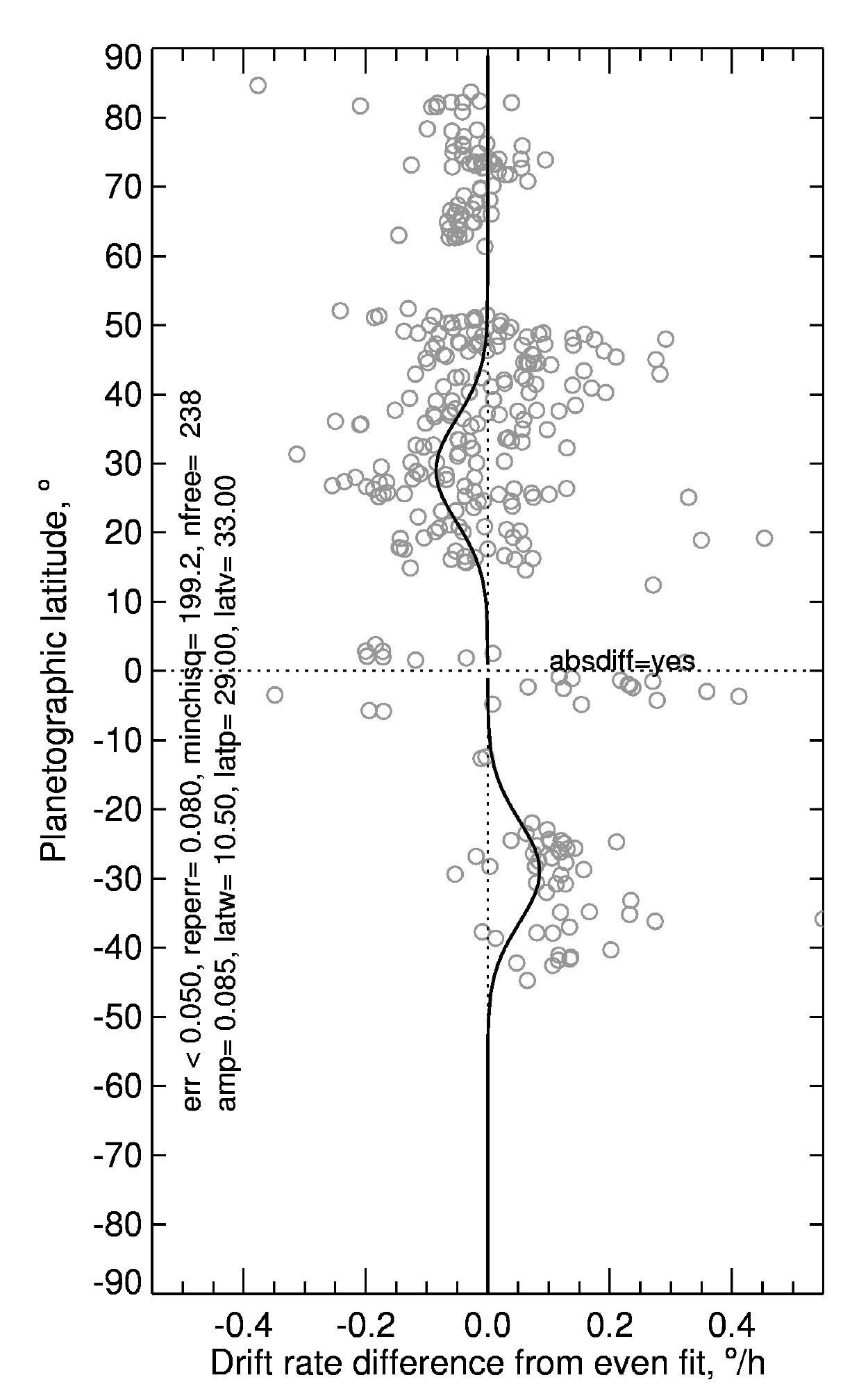}
\includegraphics[width=2.36in]{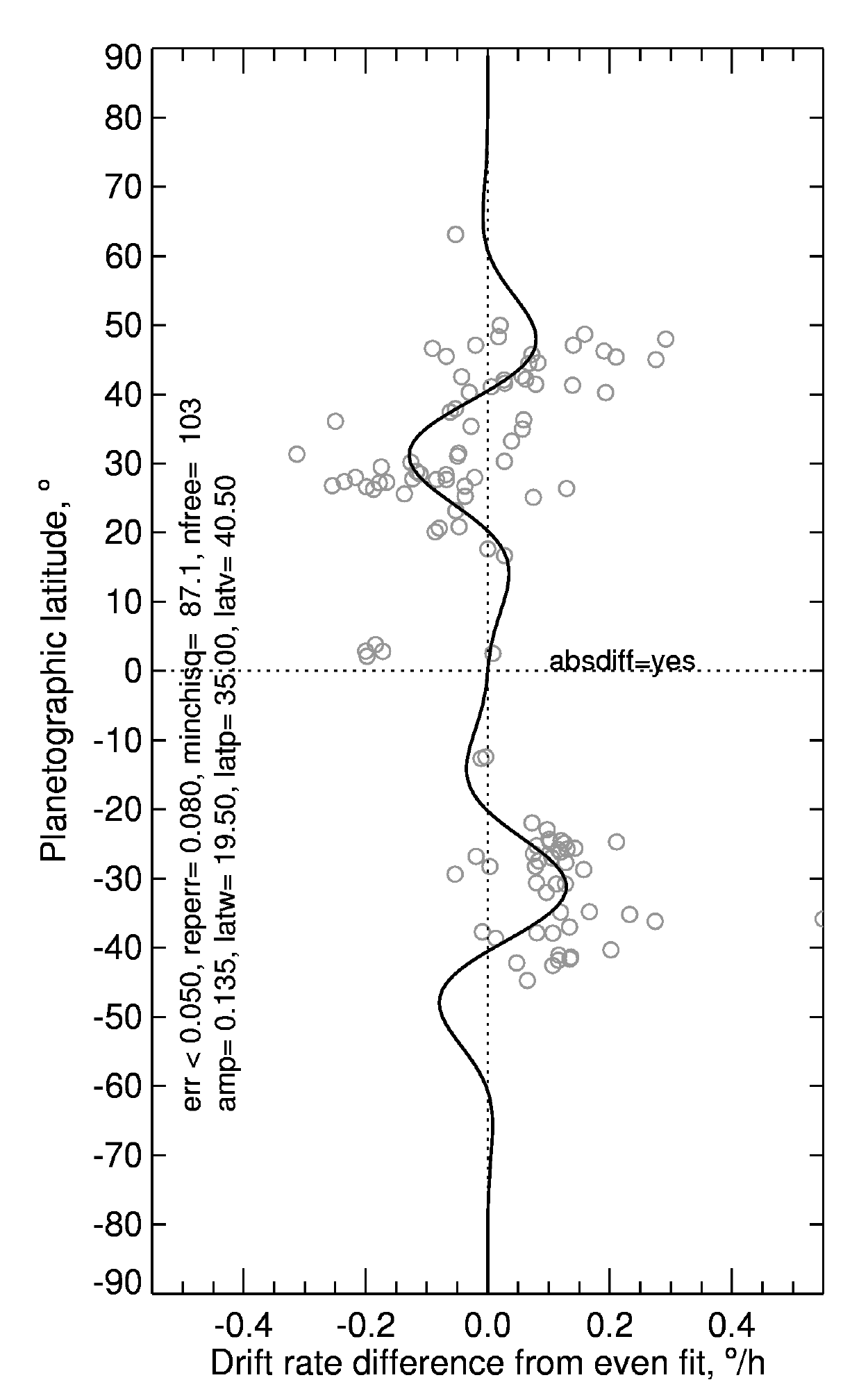}
\caption{Wind difference between our symmetric model and all
  measurements shown in Fig.\ \ref{Fig:priorwinds} (Left) and
  differences excluding 2012-2014 observations (Right).  Simple
  best-fit models are shown by solid lines. The left fit uses an
  asymmetric Gaussian and the right fit uses a damped sine function
  (both defined in the text). To reduce the impact of outliers, both
  fits minimized the sum of absolute differences scaled by expected
  errors. The earlier observations seem to have somewhat more
  pronounced asymmetry.}
\label{Fig:winddiff}
\end{figure*}

The very small asymmetry we see in the 2012-2014 observations is
interesting to compare with prior observations of Uranus.  As shown in
Fig.\ \ref{Fig:priorwinds}, the strongest indication of mid-latitude asymmetry
comes from observations made between 1997 and 2005, which are based on
HST imaging \citep{Kark1998Sci,Hammel2001Icar} and Keck imaging
\citep{Hammel2001Icar,Sro2005dyn,Sro2007bright,Hammel2009Icar,Sro2009eqdyn}.
The dotted curve in this figure is the sum of our symmetric fit and
the all-data asymmetric fit to the differences from the symmetric fit,
described later in the section and shown in the left panel of
Fig.\ \ref{Fig:winddiff}.  Observations from 2007 Keck imaging
\citep{Sro2009eqdyn} and 2009 HST imaging \citep{Fry2012} are in close
agreement with 2012-2014 results.  The 2011 results
\citep{Sro2012polar} from Keck and Gemini imaging, provide
insufficient mid-latitude sampling to constrain the asymmetry
properties of the circulation on their own, but they do contribute to
the impression that there is indeed an asymmetry.  This is made more
apparent by the plots of measurements relative to the symmetric model
in Fig.\ \ref{Fig:winddiff}.  There we also plot a simple model that
provides a crude fit to the residuals.  We considered two empirical
models of the asymmetry.  To fit the asymmetry average for all the
high-quality observations from Voyager through 2014, we used the
following model:
\begin{eqnarray}
  d\phi/dt(\theta)= A \times(-|(\theta)|/\theta))\exp(-((\theta-b)/c)^2) \label{Eq:asymm1}
\end{eqnarray}
where we found $A$=0.085\degx/h to be the best-fit amplitude, $b$=29\deg to be the best-fit
peak location in latitude, and $c$=10.5\deg to be the best-fit latitudinal width parameter. 
The second factor in Eq.\ \ref{Eq:asymm1} provides the sign reversal between hemispheres and is replaced by
zero at the equator.  A comparison of our fit to
the observations can be found in Fig.\ \ref{Fig:winddiff} (left panel).  Measurements prior to 2012-2014
suggest a larger and more complex asymmetry structure, which we fit using the following model:
\begin{eqnarray}
  d\phi/dt(\theta)= A  \sin[2\pi(\theta/a)]\exp(-((\theta-b)/c)^2) \label{Eq:asymm2}
\end{eqnarray}
where in this case $A$=0.135\degx/h provides the best-fit asymmetry amplitude, $a$=40.5\degx provides
the best latitudinal period of variation, $b$= 35\deg the best peak of the exponential factor, and
$c$=19.5\deg the best damping width of the exponential.  These parameters are only
crudely constrained by the observations, as can be seen from a comparison of this model with the
difference plots in Fig.\ \ref{Fig:winddiff} (right panel). 

\subsection{Comparison with new southern winds from Voyager}

\begin{figure*}[!htb]\centering
\includegraphics[width=6in]{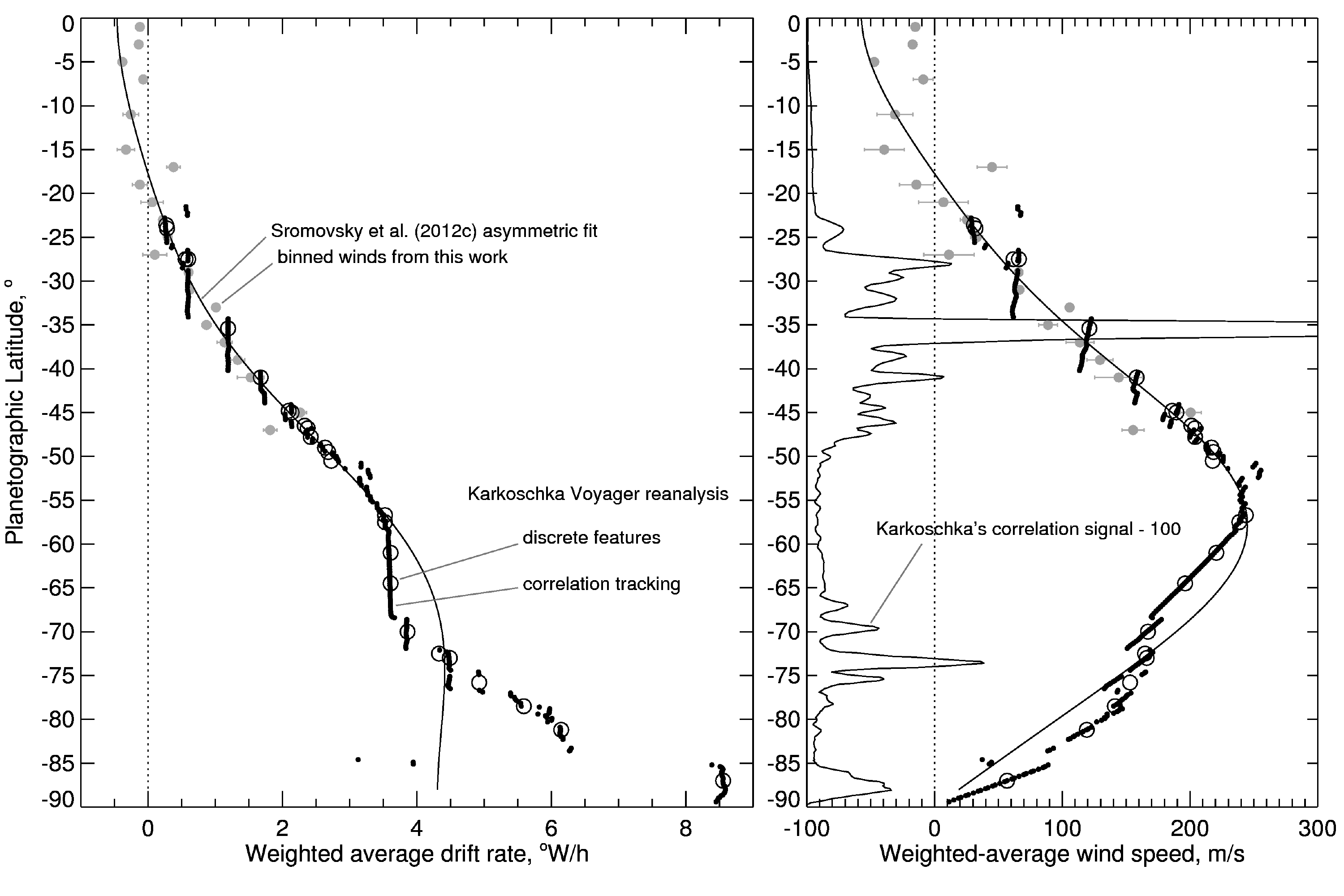}
\caption{Southern hemisphere longitudinal drift rates (left) and
  corresponding wind speeds (right) obtained from reanalysis of 1986
  Voyager images by \cite{Kark2015vgr}, compared to the asymmetric fit
  (solid curve) of \cite{Sro2012polar} and our measurements from
  2012-2014 observations (gray dots). Two Karkoschka results are
  shown: tracking of discrete cloud features (open circles) and
  correlation tracking of patterns in narrow latitude bands (small
  black dots). Also plotted at the right is the Karkoschka correlation
  signal-100 versus latitude, where the signal unit is 10$^{-5}$ in
  I/F. Only points above correlation signals of 6 are shown, due to
  scatter obtained for lower signal levels.}
\label{Fig:southcomp}
\end{figure*}

Voyager approached Uranus in 1986, when the southern hemisphere
was facing the sun, but very few winds were obtained because very
few cloud features were detected. Recently,
\cite{Kark2015vgr} carried out an extensive reprocessing of Voyager
imagery to remove artifacts, improve non-linearity corrections, and
carry out shift-and-add averaging, similar to the high S/N approach
we used in our analysis, except that many more images were averaged and much higher
S/N ratios were achieved, allowing the detection of very low contrast cloud patterns and
discrete cloud features over most of the southern hemisphere.  Twenty-seven
discrete cloud features were tracked between 87\degx S and 23.6\degx S.
These are shown in Fig.\ \ref{Fig:southcomp} in comparison to the asymmetric profile of \cite{Sro2012polar}
and to southern hemisphere measurements from our 2012-2014 data set.
Also shown is the result of Karkoschka's correlation tracking at 395 latitudes between
the equator and the pole.  These new results are in generally good agreement
with current and prior results north of 55\degx S, where the non-Voyager
results exist. But there are many substantial deviations and remarkable
asymmetries implied by the new results, as discussed in the following
paragraphs.

\begin{figure}[!htb]\centering
\includegraphics[width=3.5in]{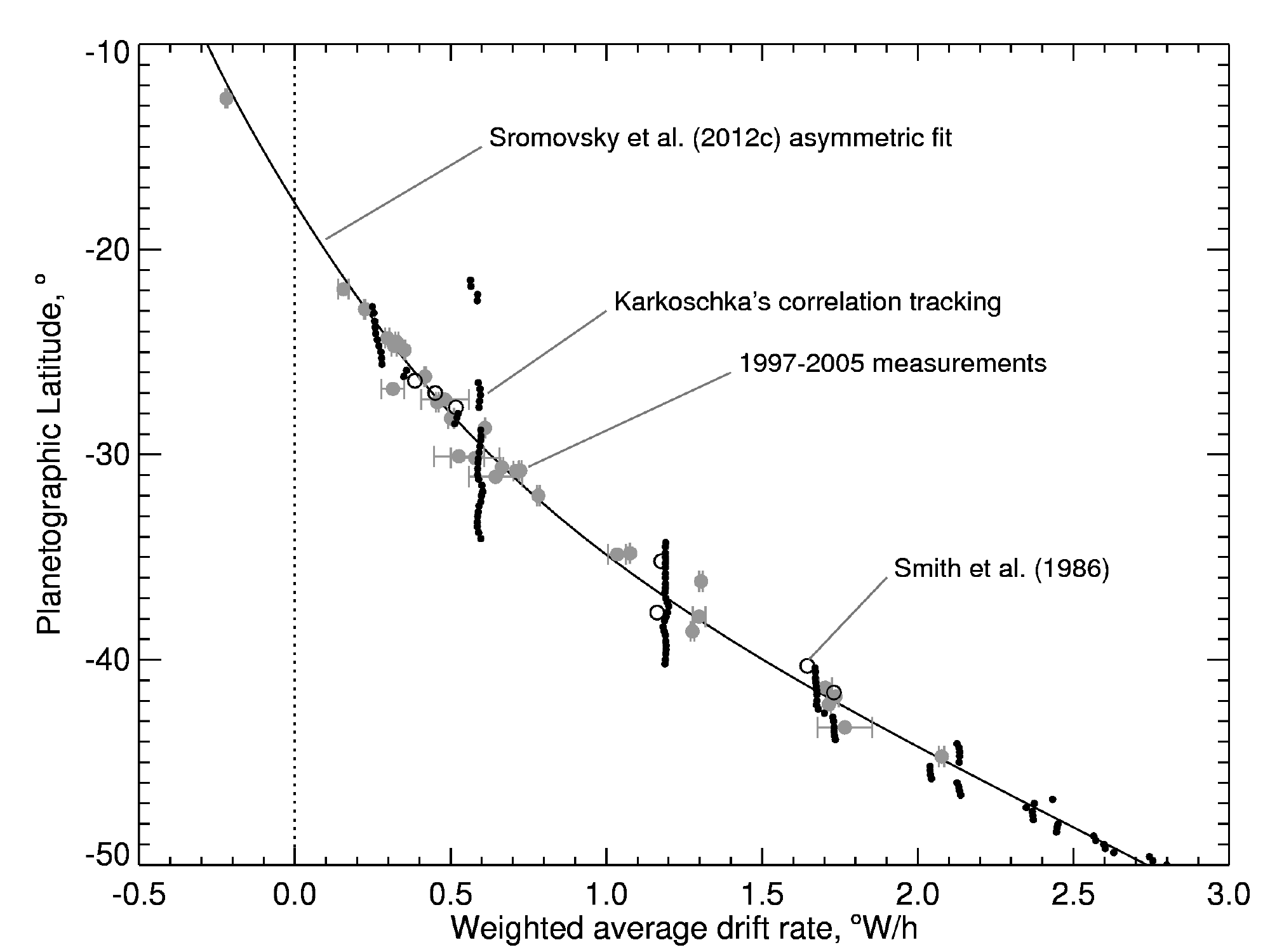}
\caption{The \cite{Kark2015vgr} correlation tracking
  results (small black dots), compared to the asymmetric fit (solid
  curve) of \cite{Sro2012polar}, 1997-2005 measurements (larger gray
  dots), and prior Voyager results (open circles) of
  \cite{Smith1986}. The uncertainty in the latter results is comparable
  to the size of the plotted symbols, while error bars are shown for
  1997-2005 measurements.}
\label{Fig:corrcomp}
\end{figure}

First, we consider the correlation tracking results (small black dots
in Fig.\ \ref{Fig:southcomp}).  These show regions of constant
longitudinal drift rate, or solid-body rotation (26\degx S - 36\degx
S, 36\degx S - 42\degx S, and 58\degx S - 68\degx S, for example). The
regions north of 50\degx S are not in agreement with prior
measurements, as shown in Fig.\ \ref{Fig:corrcomp}, most of which
follow a smooth variation with latitude. Exceptions are probably due
to the fact that larger vortex features can generate cloud features
over a range of latitudes that travel along with the generating
feature rather than following the zonal wind profile.  Major
examples of this effect are provided by the Great Dark Spot and other
dark spots on Neptune \citep{Sro2002spots}. We also identify a pair of
uranian features with this characteristic in Section
\ref{Sec:AB}. Thus, it is conceivable that, at least where we have
contradicting observations, the regions where Karkoschka's Voyager
correlation results show solid-body rotation are due to large
circulation features that influence an extended latitude region,
generating clouds that travel with the circulation features rather
than following the zonal flow.  This is less plausible as an
explanation for the 58\degx S to 68\degx S region because of its size.
While we find small regions of apparent solid body rotation in our
data set in the northern hemisphere, these are not as extensive as the
mid-latitude regions in Karkoschka's correlation profile.  The major
region of solid body rotation that we find, from 62\degx N to 83\degx
N, is clearly not an artifact, as it is based on numerous well-defined
discrete cloud features.  There is an amazing amount of shear in
Karkoschka's correlation profile, just where one solid body region
transitions to another.  When plotted as wind speeds, there regions of
enormous jumps in wind speed over a very short distance, zig-zagging
to higher values as latitude increases toward the pole, then
decreasing for a while, then jumping again to a new high. This
characteristic also suggests that these features might not represent
the zonal wind structure.


The most extraordinary result from \cite{Kark2015vgr} is the huge
north-south asymmetry it implies at high latitudes. Between 70\degx S
and 87\degx S, his inferred westward drift rates rises from 3.8\degx/h
to 8.6\degx/h, which is more than double what we measured at high
northern latitudes, where drift rates are invariant with latitude over
a comparable latitude range. Given that the middle latitude winds have
changed by barely measurable amounts from 1986 through 2014, a 28-year
period, it is hard to understand how such enormous seasonal changes
might occur. If the north-south difference is seasonal, and the
Voyager analysis represents the winds at the southern summer solstice
(October 6, 1985), then we would expect that the same wind profile
would appear in northern latitudes by the time of the northern summer
solstice in 2030 (on April 29 according to \cite{Meeus1997}). That is
less than 16 years over which this enormous change should occur, while
almost nothing has changed at middle latitudes in the last 28 years.
Evidence for change at high latitudes is only beginning to be
accumulated.  There may have been a small decrease in the solid body
rotation rate from 4.3\degx /h to 4.1\degx /h between 2011 and 2014,
but that is probably within the error of the 2011 measurements, and
provides little evidence for the rapid change needed to match in 2030
what seems to have existed in 1986 in the southern hemisphere,
according to \cite{Kark2015vgr}.  On the other hand, perhaps this
north-south difference is a permanent feature of the wind profile,
just as it appears that Jupiter and Saturn have asymmetries that
survive over a complete change of seasons.  It will certainly be
worthwhile to monitor the winds of Uranus over the next two decades to
determine whether this asymmetry is really a seasonal effect.

\subsection{A complete zonal profile for Uranus}

\begin{figure*}[!htb]\centering
\includegraphics[width=6.in]{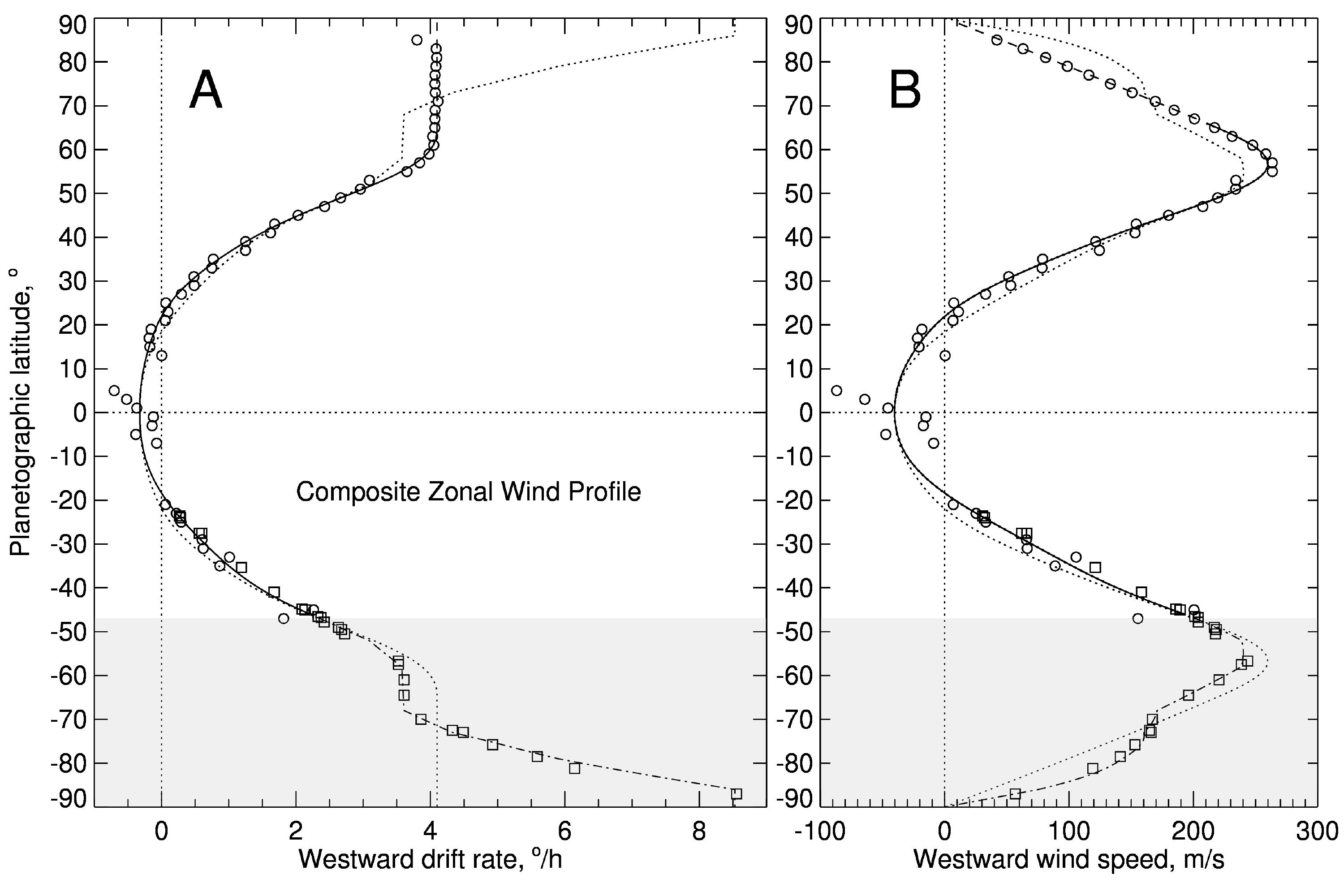}
\caption{Composite zonal profiles of longitudinal drift rate (A) and
  wind speed (B) versus latitude.  The grayed region from 47\degx S to
  90\degx S is where we used the adopted profile of
  \cite{Kark2015vgr}, which is based on his reanalysis of 1986 Voyager
  images (dot-dash curve). The adopted profile for higher latitudes
  (solid curve) is based on a symmetric fit to the 2012-2014
  observations defined in Fig.\ \ref{Fig:symfitall} and the asymmetry
  fit to all selected observations defined in
  Fig.\ \ref{Fig:winddiff}A.  One exception is that we replaced the
  nearly constant angular velocity region north of 67\degx N with an
  exactly constant profile (dashed line). In each panel the dotted
  curve gives the adopted profile reflected about the equator, which
  shows the degree of north-south asymmetry that is present. Also
  shown are binned results from the current work (open circles) and
  discrete cloud tracking results from \cite{Kark2015vgr}. Our adopted
profile is also given in Tables \ref{Tbl:adoptdrate} and \ref{Tbl:adoptwind}.}
\label{Fig:compositewinds}
\end{figure*}

Here we create a complete pole-to-pole zonal wind profile for Uranus
by combining results from groundbased observations at latitudes
spanning 47\degx S to 83\degx N with the recent \cite{Kark2015vgr}
results for the southern hemisphere, obtained from 1986 Voyager 2
imaging.  In Fig.\ \ref{Fig:compositewinds} we show our composite
profile in comparison with binned observations and the discrete cloud
tracking results of \cite{Kark2015vgr}. The composite profile uses our
symmetric fit to our 2012-2014 observations
(Fig.\ \ref{Fig:symfitall}) summed with the asymmetric fit to
differences from that profile, as displayed in
Fig.\ \ref{Fig:winddiff}A, to cover the latitude range from 46\degx S
to 67\degx N. From 67\degx N to 90\degx N we used a solid-body
rotation profile of 4.1\degx/h westward.  From 47\deg S to 90\degx S
we used the adopted profile of \cite{Kark2015vgr}.  The latter profile
does not pass exactly through the discrete measurements in this
region, probably because the profile also took into account
correlation measurements.  The composite profile is also provided in
Tables \ref{Tbl:adoptdrate} and \ref{Tbl:adoptwind}.  This represents
our best estimate of the zonal wind profile of Uranus under the
assumption that the indicated asymmetry is a permanent feature of the
atmosphere.  This may be an incorrect assumption.  Since the southern
winds from Voyager are based on 1986 observations, while the remaining
winds (from 47\degx S to 90\degx N) are based on observations heavily
weighted towards the 2012-2014 time period, this may not represent the
current profile.

\begin{table*}\centering
\caption{Adopted westward drift rate (D) profile in \degx/hour vs. planetographic latitude (L).}
\vspace{0.15in}
\begin{tabular}{r r r r r r r r r r r}
L, \deg& D(L)&D(L-1)&D(L-2)&D(L-3)&D(L-4)&D(L-5)&D(L-6)&D(L-7)&D(L-8)&D(L-9)\\
\hline\\[-0.15in] 
 90&   4.100&   4.100&   4.100&   4.100&   4.100&   4.100&   4.100&   4.100&   4.100&   4.100\\
 80&   4.100&   4.100&   4.100&   4.100&   4.100&   4.100&   4.100&   4.100&   4.100&   4.100\\
 70&   4.100&   4.100&   4.100&   4.100&   4.097&   4.100&   4.099&   4.090&   4.072&   4.043\\
 60&   4.002&   3.946&   3.875&   3.789&   3.687&   3.571&   3.443&   3.302&   3.153&   2.997\\
 50&   2.836&   2.673&   2.510&   2.349&   2.191&   2.039&   1.893&   1.753&   1.620&   1.493\\
 40&   1.373&   1.258&   1.149&   1.045&   0.946&   0.850&   0.759&   0.672&   0.588&   0.509\\
 30&   0.434&   0.363&   0.298&   0.237&   0.181&   0.129&   0.083&   0.040&   0.002&  -0.032\\
 20&  -0.064&  -0.092&  -0.118&  -0.142&  -0.164&  -0.185&  -0.204&  -0.222&  -0.238&  -0.253\\
 10&  -0.266&  -0.278&  -0.289&  -0.298&  -0.305&  -0.312&  -0.317&  -0.321&  -0.323&  -0.325\\
  0&  -0.325&  -0.324&  -0.322&  -0.319&  -0.314&  -0.308&  -0.300&  -0.290&  -0.278&  -0.264\\
-10&  -0.247&  -0.228&  -0.206&  -0.181&  -0.154&  -0.124&  -0.090&  -0.054&  -0.015&   0.027\\
-20&   0.072&   0.120&   0.170&   0.223&   0.278&   0.335&   0.394&   0.455&   0.518&   0.582\\
-30&   0.647&   0.714&   0.783&   0.854&   0.927&   1.002&   1.081&   1.164&   1.253&   1.346\\
-40&   1.447&   1.554&   1.669&   1.793&   1.924&   2.064&   2.210&   2.359&   2.501&   2.645\\
-50&   2.790&   2.937&   3.086&   3.166&   3.247&   3.328&   3.410&   3.492&   3.575&   3.578\\
-60&   3.581&   3.585&   3.588&   3.591&   3.595&   3.598&   3.601&   3.605&   3.608&   3.749\\
-70&   3.891&   4.035&   4.181&   4.328&   4.596&   4.869&   5.120&   5.377&   5.638&   5.904\\
-80&   6.250&   6.605&   6.970&   7.344&   7.729&   8.124&   8.530&   8.530&   8.530&   8.530\\
-90&   8.530&   &   &   &   &   &  &   &   &   \\
\hline
\end{tabular}\label{Tbl:adoptdrate}
\end{table*}

\begin{table*}\centering
\caption{Adopted westward zonal wind (U) profile in m/s vs. planetographic latitude (L).}
\vspace{0.15in}
\begin{tabular}{r r r r r r r r r r r}
L, \deg & U(L)&U(L-1)&U(L-2)&U(L-3)&U(L-4)&U(L-5)&U(L-6)&U(L-7)&U(L-8)&U(L-9)\\
\hline\\[-0.15in] 
90&    0.00&    9.07&   18.15&   27.21&   36.27&   45.31&   54.34&   63.35&   72.33&   81.29\\
 80&   90.23&   99.13&  108.00&  116.83&  125.62&  134.36&  143.06&  151.72&  160.32&  168.86\\
 70&  177.35&  185.77&  194.14&  202.43&  210.53&  218.84&  226.82&  234.33&  241.20&  247.23\\
 60&  252.26&  256.13&  258.70&  259.87&  259.57&  257.78&  254.54&  249.90&  243.99&  236.94\\
 50&  228.93&  220.13&  210.74&  200.94&  190.88&  180.72&  170.58&  160.54&  150.67&  140.99\\
 40&  131.53&  122.27&  113.20&  104.30&   95.56&   86.97&   78.54&   70.28&   62.22&   54.39\\
 30&   46.83&   39.60&   32.73&   26.26&   20.20&   14.58&    9.40&    4.63&    0.26&   -3.75\\
 20&   -7.43&  -10.82&  -13.96&  -16.88&  -19.60&  -22.16&  -24.55&  -26.78&  -28.85&  -30.75\\
 10&  -32.49&  -34.05&  -35.43&  -36.63&  -37.64&  -38.49&  -39.16&  -39.67&  -40.02&  -40.22\\
  0&  -40.24&  -40.11&  -39.85&  -39.42&  -38.81&  -37.99&  -36.95&  -35.66&  -34.12&  -32.29\\
-10&  -30.17&  -27.74&  -24.99&  -21.92&  -18.53&  -14.81&  -10.77&   -6.41&   -1.76&    3.19\\
-20&    8.41&   13.87&   19.57&   25.47&   31.55&   37.76&   44.08&   50.48&   56.92&   63.39\\
-30&   69.87&   76.35&   82.83&   89.33&   95.88&  102.50&  109.26&  116.20&  123.37&  130.83\\
-40&  138.62&  146.78&  155.32&  164.22&  173.44&  182.92&  192.53&  201.83&  210.02&  217.83\\
-50&  225.25&  232.26&  238.84&  239.62&  240.08&  240.23&  240.04&  239.52&  238.66&  232.26\\
-60&  225.77&  219.19&  212.53&  205.78&  198.95&  192.04&  185.05&  177.98&  170.85&  169.86\\
-70&  168.32&  166.19&  163.48&  160.17&  160.37&  159.58&  156.88&  153.20&  148.50&  142.75\\
-80&  137.54&  130.97&  122.96&  113.47&  102.43&   89.78&   75.45&   56.61&   37.75&   18.88\\
-90&   0.000&   &   &   &   &   &  &   &   &   \\
\hline
\end{tabular}\label{Tbl:adoptwind}
\end{table*}

\section{Persistent cloud patterns and discrete features}\label{Sec:patterns}

\begin{figure*}[!htb]\centering
\includegraphics[width=5.85in]{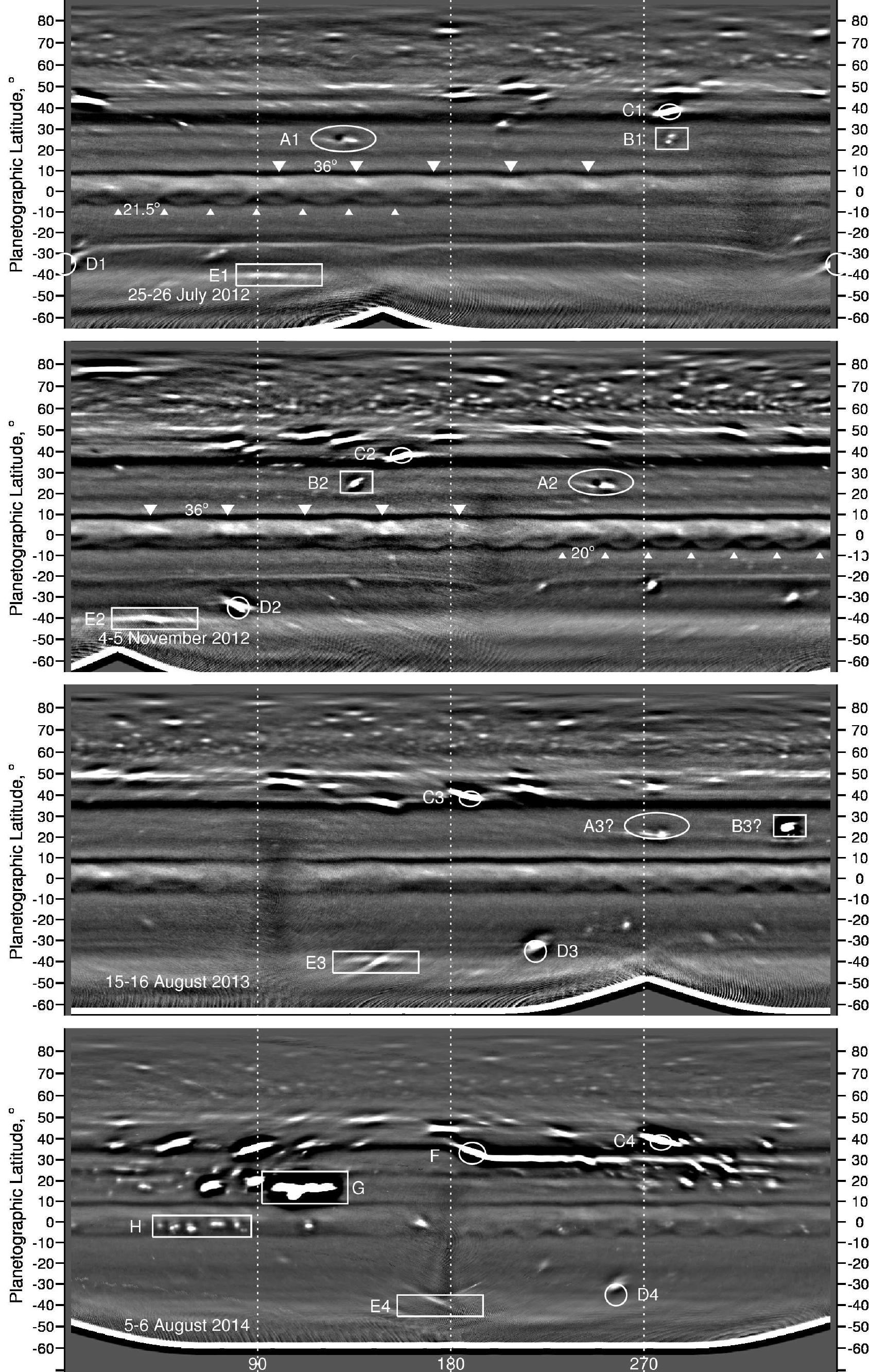}\\[0.01in]
\caption{Rectangular projections of 2-night mosaics from July 2012
  (top), November 2012, August 2013, and August 2014 (bottom). Each is
  high-pass filtered so that low-contrast features could be enhanced
  without being overwhelmed by brightness variations as a function of
  latitude (see text).}
\label{Fig:rectpro}
\end{figure*}

A number of discrete features and cloud patterns were found to persist
for long time periods.  To facilitate their identification we created
a mosaic of high S/N H-filter images for each high-resolution data set
for which successive observing nights allowed us to take multiple
images of both sides of the planet, totaling between 144 and 179
images per data set. This made it possible to combine all longitudes into one
rectangular map.  Overlapping images were blended with a weighting of
$\mu^2T/T_\circ$, where $\mu$ is the observer zenith angle cosine and
$T/T_\circ$ is the image exposure ratio.
Also, pixels were mosaicked only if $\mu$ and $\mu_\circ$ (solar
incidence angle cosine) were greater than or equal to 0.025.  These
two cosines are so close together that a Minnaert model of brightness
variation, i.e. $I\mu = I_\circ (\mu\mu_\circ)^k$, which is fit
well with $k=0.6$ for H-filter images, collapses to $I=I_\circ \mu^{0.2}$.
Because this is a such a weak dependence on view angle we did not
attempt to correct for it. 
High-pass filtered versions of these mosaics are shown as rectangular
projections in Fig.\ \ref{Fig:rectpro} at a scale
0.25\degx/pixel. These were obtained by subtracting from the mosaicked
images a version smoothed with a 6.25\deg (25-pixel) boxcar average.
To display the resulting low-contrast variations, we enhanced these
images so that black and white respectively correspond to I/F
deviations of -0.4$\times 10^{-4}$ and +0.8$\times 10^{-4}$ from the
smoothed versions, which have a central disk I/F of about 0.01.  The
combination of images taken at different times requires that each
longitudinal image line from each component image be shifted by the
zonal wind drift rate times the time difference between the image time
and the reference time of the map (chosen to be the midpoint of the
observations for the two nights).  This does not distort features at
low and high latitudes, because wind shear is very low at low
latitudes and zero above 55\degx N. However, at middle latitudes
(25-55\degx), where wind shear is significant, round features become
stretched into slanted ellipses.  This distortion is a reasonable
price to pay for obtaining a global view of where features are located
and how many features are present in a given band of latitude.  For
key features that are identified, we also provide in the following
section undistorted views that more accurately display morphological
characteristics.

\subsection{Discrete feature identification based on morphology}

\begin{figure*}[!htb]\centering
\begin{minipage}[b]{1.6in}{
\includegraphics[width=1.6in]{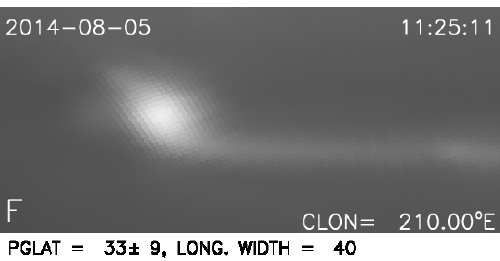}
\includegraphics[width=1.6in]{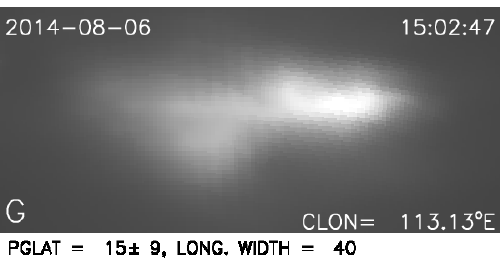}}
\includegraphics[width=1.6in]{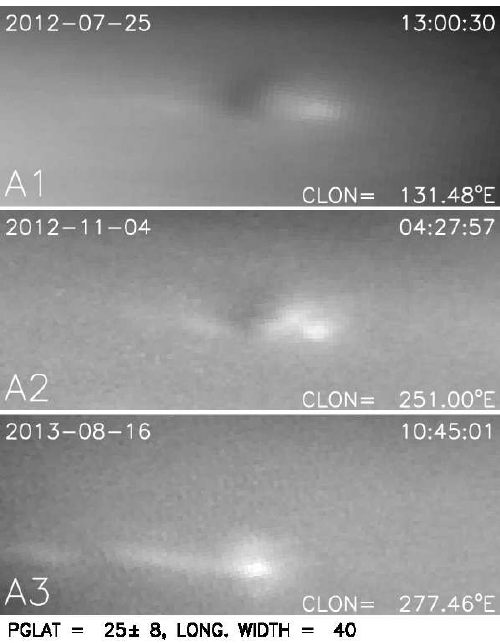}
\end{minipage}
\includegraphics[width=1.2in]{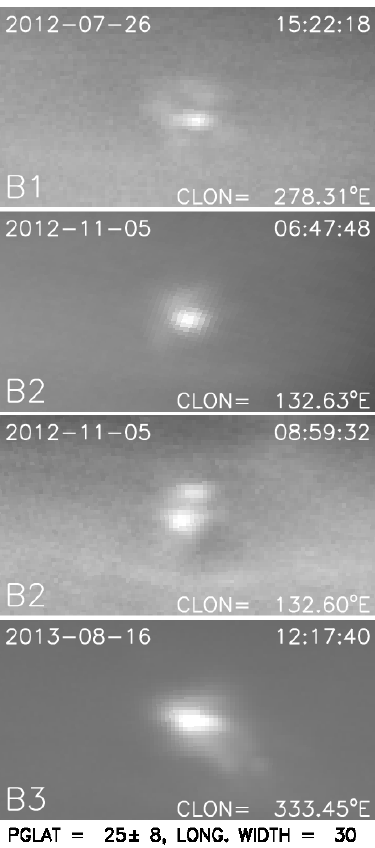}
\includegraphics[width=1.2in]{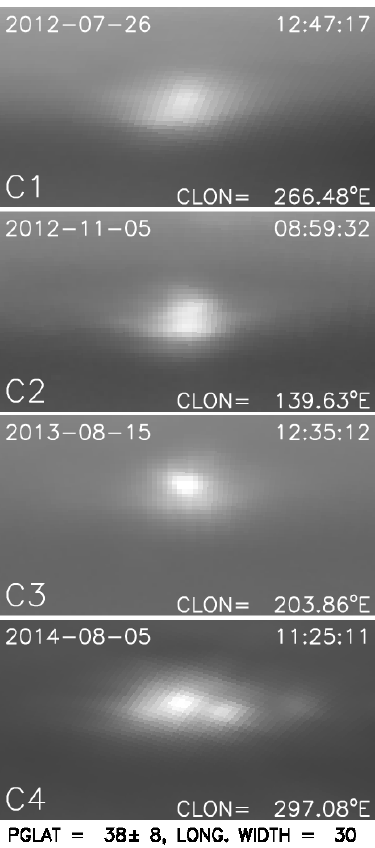}
\includegraphics[width=1.2in]{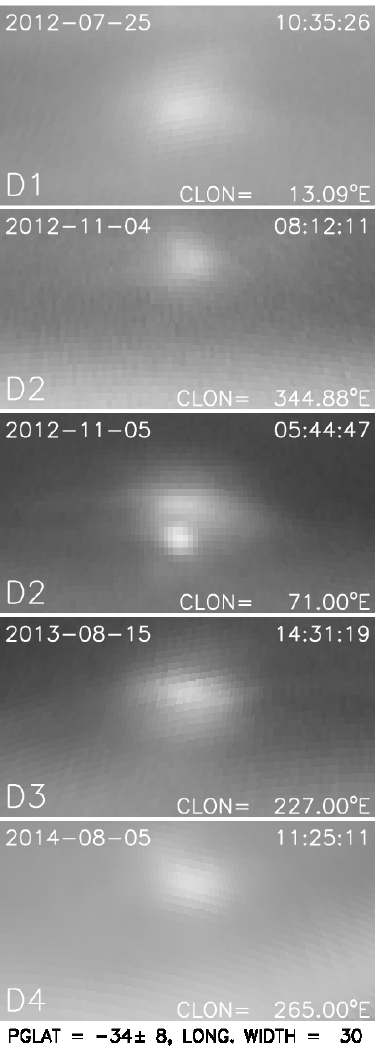}
\caption{Rectangular projections of discrete features A-D, F, and G. These
are stretched but not high-pass filtered.}
\label{Fig:featuresABCDFG}
\end{figure*}

Inspecting the four rectangular maps in Fig.\ \ref{Fig:rectpro}, we
see that there are too many features of similar appearance in the
north polar region to identify which, if any, might have lasted for
long time periods.  The band from 45\degx N to 50\degx N also has too
many features to make any unambiguous matches between data sets.  In
other, less populated regions, we were able to identify six discrete
long-lived features (A-F), which are also shown enlarged and without
high-pass filtering in Fig.\ \ref{Fig:featuresABCDFG}.

Between 20\degx N and 30\degx N, only two significant features are seen in July 2012 (A1 and B1),
November 2012 (A2 and B2), and August 2013 (A3 and B3).  A1 and A2
have nearly identical morphologies and latitudes ($\approx$25\degx N),
making identification rather easy.  In the high-pass filtered images
they appear as small dark spots with what look like companion clouds
traveling with them, consistent with the idea that the bright features are orographic
clouds generated by vertical deflections associated with flow around
an oval vortex.  In this case vertical deflections cause adiabatic
cooling and condensation of cloud particles.  The dark feature so
clearly visible in the high-pass filtered version is also apparent
without such filtering, as shown in
Fig.\ \ref{Fig:featuresABCDFG}. Feature A3 is likely associated with
the same vortex, but the morphological match is not as good, with no
dark spot showing.  

Feature B is at essentially the same latitude as feature A, but its
bright features can be found both north and south of the bright
features associated with A.  From a comparison of A2 and B2 subimages
in Fig.\ \ref{Fig:featuresABCDFG}, we see that the bright features of
B are indeed north of the bright features of A, but there seems to be
a dark spot associated with B that is at a lower latitude than the
dark spot associated with A.  It is likely that the latitude of the dark
spot is what really matters because  the bright features are likely to
follow the motion of the dark spot, rather than the zonal wind
profile, a behavior already well established for dark spots on
Neptune.

\begin{figure*}[!htb]\centering
\begin{minipage}[b]{1.65in}{
\includegraphics[width=1.65in]{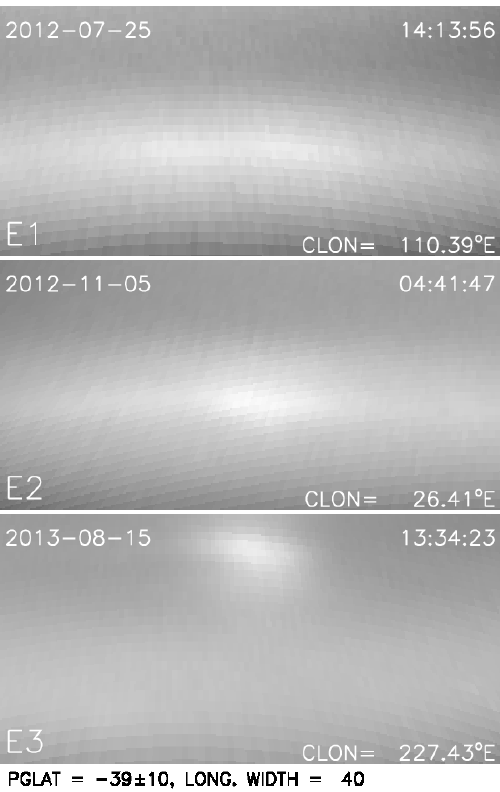}
\includegraphics[width=1.65in]{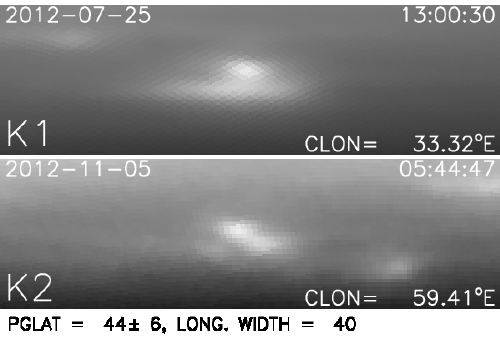}}
\end{minipage}
\includegraphics[width=3.6in]{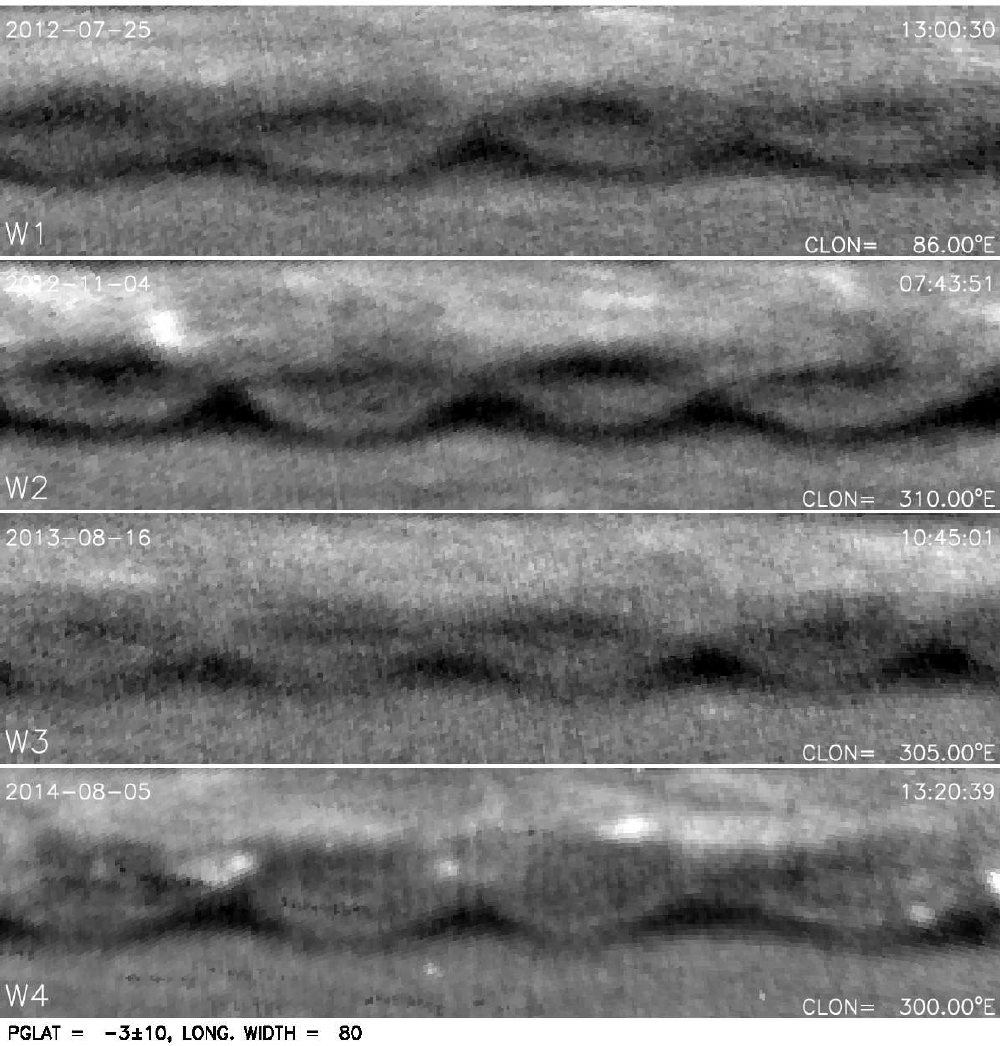}
\caption{Rectangular projections of discrete features E (40\deg in
  longitude $\times$ 20\deg in latitude), K (40\deg $\times$ 12\degx),
  and near equatorial waves (W1-W4)(80\deg $\times$ 20\degx). Only
  W1-W4 are high-pass filtered, which was needed to remove a strong
  latitudinal gradient in brightness.}
\label{Fig:featuresEKW}
\end{figure*}

One feature that seems to have survived at the same latitude all the way from July 2012 to
August 2014, is the feature labeled C1-C4, which appears at latitude 38\deg N.

\subsection{Long-lived discrete feature tracking}\label{Sec:long}

To solidify the identification of discrete features, we tried to add
longitudinal continuity to the morphological character and zonal
uniqueness evident at several discrete times.  If the features we have
identified as unique are indeed the same at each time period, then
their longitudes should be a continuous function of time that is
roughly consistent with the zonal winds within the latitude region
where they are found.  During 2012 we acquired observations over a
4-month period, with a similar range covered in 2014, providing strong
constraints on longitudinal and latitudinal motions, but only one
sample was obtained between these times, during 15-16 August 2013.

We first tried to fit observations to the simplest model in which
longitude is a linear function of time, implying that the drift rate
is constant and that the latitude of the feature (or that of the 
circulation feature generating the visible cloud feature) is
constant.  None of the long-lived features followed this simple model
for the entire time period.  Two other models were then considered:
(1) a sinusoidal variation in drift rate (and latitude), which leads
to a sinusoidal variation in longitude relative to a linearly
increasing longitude, and (2) a linear variation in drift rate (and a
linearly increasing latitude), which implies that the longitude should
vary as the square of the time difference (at least over small ranges
of latitude for which a constant wind shear is a plausible
assumption). The linear latitudinal drift model can be expressed as
follows: \begin{eqnarray} \phi(t) = \phi_\circ + a \times (t-t_0) + b
  \times (t-t_0)^2\label{Eq:longquad}\\ \omega (t) = d\phi(t)/dt = a +
  2b\times (t-t_0)\label{Eq:driftquad}\\ \theta(t) = \theta_\circ +
  2b\times (t-t_0)/(d\omega/d\theta)\label{Eq:latquad}
\end{eqnarray}
where $\phi(t)$ is longitude at time $t$, $\omega$ is rate of change
of longitude, $\theta$ is latitude, and $d\omega/d\theta$ is the
latitudinal shear in the zonal wind profile.  The corresponding
equations for the sinusoidal variation model are as follows:
\begin{eqnarray}
      \phi(t) &=& \phi_\circ + a \times (t-t_0) + c \times \sin(2\pi(t-t_0)/P)\label{Eq:longsine}\\
      \omega (t) &=& d\phi(t)/dt = a + (2c/\pi)\times  \nonumber \\
      & & \cos(2\pi(t-t_0)/P)\label{Eq:driftsine}\\
     \theta(t) &=& \theta_\circ + (2c/\pi)\times \cos(2\pi(t-t_0)/P)/(\frac{d\omega}{d\theta})\label{Eq:latsine}
\end{eqnarray}
where $P$ is the period of variation. When the period becomes significantly longer than the
span of the observations, it becomes difficult to distinguish the two models. 
For one feature (D), we needed to include both types of variations:
\begin{eqnarray}
      \phi(t) &=& \phi_\circ + a \times (t-t_0) + b \times (t-t_0)^2 +  \nonumber \\
   & &c \times \sin(2\pi(t-t_1)/P)\label{Eq:longcomb}\\
     \omega (t) &=& d\phi(t)/dt = a + 2b\times (t-t_0) +   \nonumber \\ 
   & &(2c/\pi)\times \cos(2\pi(t-t_1)/P)\label{Eq:driftcomb}\\
     \theta(t)& =& \theta_\circ + 2b\times (t-t_0)/(d\omega/d\theta) + \nonumber \\ 
   & & (2c/\pi)\times 
      \cos(2\pi(t-t_1)/P)/(d\omega/d\theta)\label{Eq:latcomb}
\end{eqnarray}               
where we also introduced a new time offset $t_1$.  As a result of the
additional time offset, this model has $\phi(t_\circ) = \phi_\circ + c
\times \sin(2\pi(t_\circ-t_1)/P)$ and $\theta(t_\circ) = \theta_\circ
+ (2c/\pi)\times \cos(2\pi(t_\circ-t_1)/P)/(d\omega/d\theta)$, while
for all the other models, $\phi(t_\circ) = \phi_\circ$ and
$\theta(t_\circ) = \theta_\circ$. In all cases $\theta_\circ$ is
determined by solving for $a=\omega(\theta_\circ)$, where $\omega
(\theta)$ is the measured zonal wind profile.  All the adjustable constants
are constrained by fits to the longitude measurements. Our best-fit models are
summarized in Table \ref{Tbl:longtrack} and discussed in the following
paragraphs.

\begin{table*}\centering
\caption{Models of long-lived cloud feature motions.}
\vspace{0.15in}
\begin{tabular}{c c c c c c c c c }
\hline\\[-0.1in] 

   & $t_\circ$  & $\theta_\circ$  & $\phi_\circ$ &         $a$ &                    $b$     & $c$ & $P$ & $\sigma$\\
ID & days  & \deg N & \deg E &    \degx/d (east)  &    0.001\degx/d$^2$       & \deg & days & \deg\\
\hline\\[-0.1in] 
 A      & 2.3$\pm$0.1  & 27.6   & 120.0 & -6.48$\pm$0.014 & 5.74$\pm$0.04 & & & 1.7 \\
 B      & 433.5$\pm$1.7 & 22.4  & 5.0 & -0.422$\pm$0.007 &  & 118.4$\pm$0.1 & 686$\pm$10 & 1.4\\
 C      & 148.5$\pm$0 & 38.6   & 271.0 & -29.177$\pm$0.002 &  & 27.5$\pm$0.9 & 258.5$\pm$1.4 & 2.9\\
 D      & 18.43$\pm$0.1& -33.2  & -6.8$\pm$0.2 & -20.87$\pm$0.01 & 2.54$\pm$0.01 & 35.0 & 750.0 & 1.9\\
 E      & 84.4$\pm$4.6& -41.7     & 20.0 & -39.24$\pm$0.05 & 5.04$\pm$0.01 &  &  & 6.5\\
 F      & 24.5$\pm$0.1&  34.1    & 125.0& -18.49$\pm$0.014 & & 16.0$\pm$1.2 & 111.0$\pm$2.9 & 1.6\\
\hline\\[-0.1in]
\end{tabular}
\vspace{0.2in}
\parbox{5.0in}{NOTES: Model D uses an additional time offset $t_1$=25.0 days after noon on 25 July 2012;
  $t_\circ$ is given in days after noon on 25 July 2012 for
  features A-E, and after noon on 5 August 2014 for feature F.  }
\label{Tbl:longtrack}
\end{table*}

\subsubsection{Features A and B}\label{Sec:AB}

These low-latitude features were seen during 2012 and 2013, but not in
the unusual 2014 images, where an abundance of low latitude features
developed, none of which seen to have any relationship to A and B.
Both A and B seem to involve a dark spot, as suggested in
Fig.\ \ref{Fig:featuresABCDFG}, and both moved toward the equator
between 2012 and 2013, though it is not clear if that trend continued
after August 2013. Their rate of equatorward drift was about
2\degx/year.  An equatorial drift (accelerating from 2\degx /y to more
than 10\degx /y) was also observed for the large southern uranian
feature named the Berg \citep{DePater2011}, and for the Great Dark
Spot on Neptune, which averaged 15\degx /y \citep{Sro1993Icar}.  The
drift of these new uranian features suggests that they might also be
produced by vortex circulations.

\begin{figure*}[!htb]\centering
\includegraphics[width=6.5in]{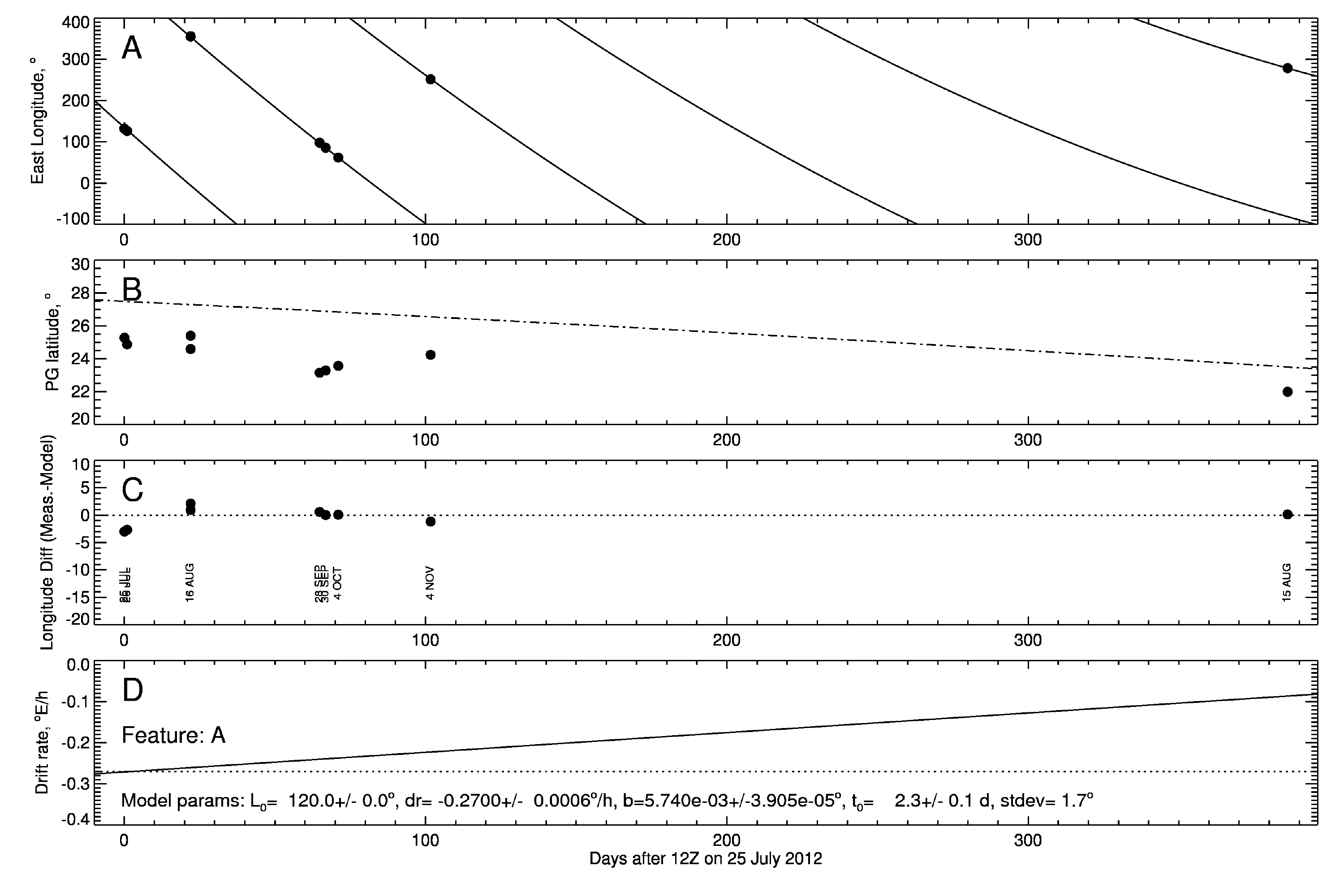}
\caption{(A): Longitude measurements for feature A (points) compared
  to a constant latitudinal drift model (solid line) (fit to Eq.
  \ref{Eq:longquad}).  (B:) latitude measurements (points) compared to
  the latitudes (dot-dash line) computed by plugging fit coefficients
  $a$ and $b$ into Eq. \ref{Eq:latquad}, where $d\omega/d\theta$ was
  obtained from the S13A Model of drift rate versus latitude. This
  model is compatible with the observed latitudinal rate of change in
  (B), but not with the absolute value of the latitude, which the
  model exceeds by about 2\degx. (C) measured longitudes (points)
  minus model longitudes. (D) Drift rate versus time, inferred from
  fitting Eq. \ref{Eq:longquad} to the longitude observations and then
  plugging the coefficients $a$ and $b$ into
  Eq. \ref{Eq:driftquad}. The x coordinate in each panel is the time
  in days relative to 25 July 2012 at 12:00 UT. }
\label{Fig:Atrack}
\end{figure*}

It is possible to fit Feature A longitude observations extremely well
with the sinusoidal model using a period of 408.2 days, but this model
does not reproduce the decreasing latitude from 2012 through 2013.
The alternative increasing drift rate model (Fig.\ \ref{Fig:Atrack})
provides an excellent fit to longitude measurements and also is
consistent with the rate of equatorward motion of the feature from
2012 to 2013.  However, the model latitude that is inferred from the
feature's longitudinal drift rate, is about 2\deg N of the observed
latitude of the bright features, suggesting that the underlying
circulation feature is to the north of the bright clouds that it
generates, which is also suggested by the images for A1 and A2 in
Fig.\ \ref{Fig:featuresABCDFG}.

\begin{figure*}[!htb]\centering
\includegraphics[width=6.5in]{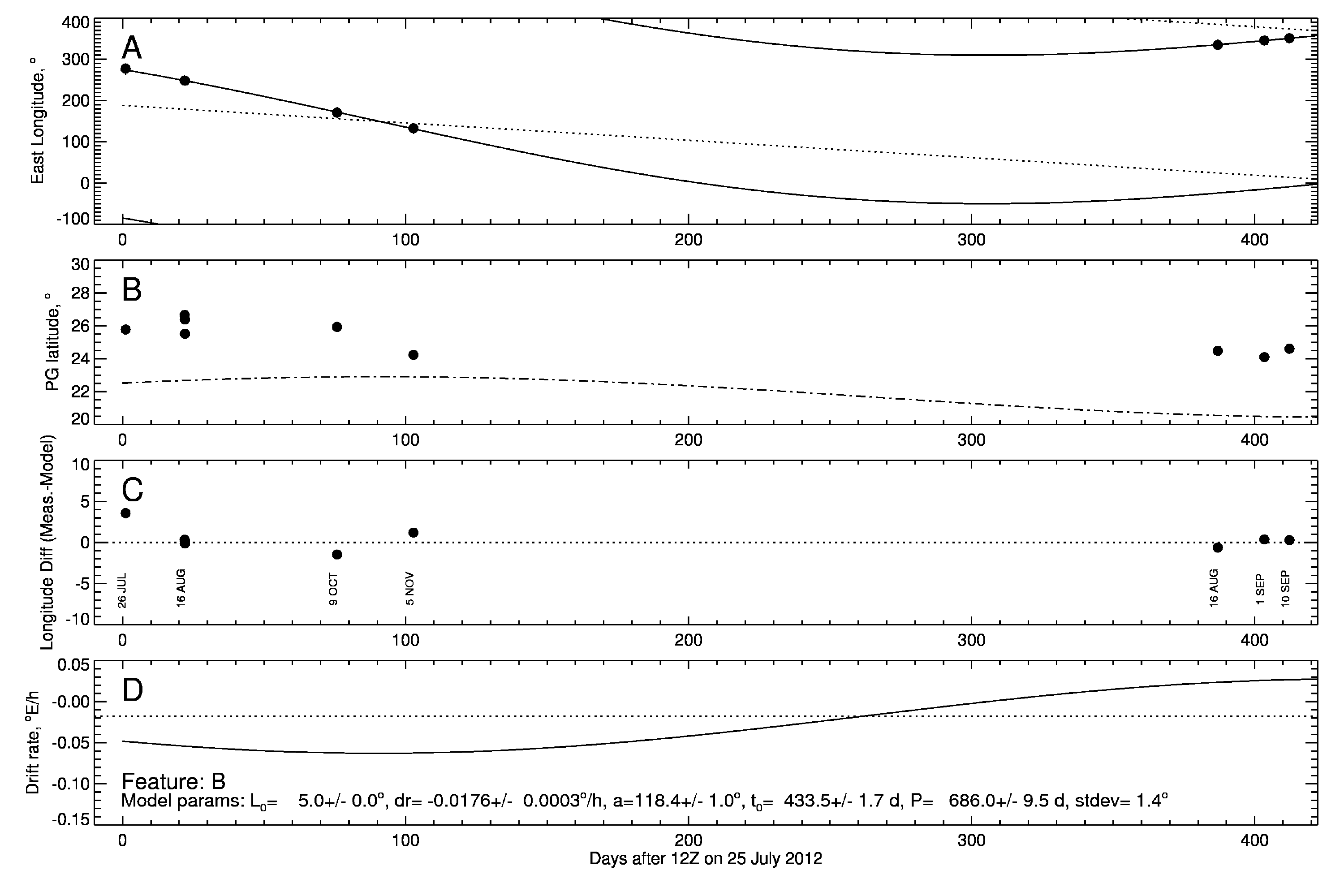}
\caption{As in Fig.\ \ref{Fig:Atrack}, except that longitudes here are
  for feature B and the model is defined by Eqs. \ref{Eq:longsine} -
  \ref{Eq:latsine}. This feature has a drift rate of -0.060\degx/h for
  the first 100 days (from July to November 2012) and 0.026\degx/h
  during August-September 2013.  This variation is compatible with a
  sinusoidal variation in longitude, as shown in panel C, a sinusoidal
  variation in drift rate (D), and a sinusoidal variation in latitude,
  as shown by the dot-dash curve in B, where the latitudinal model is
  obtained from Eq.\ \ref{Eq:latsine}, which in this case shows a
  2\deg deficit relative to the observed latitudes (points).}
\label{Fig:Btrack}
\end{figure*}

Feature B had a nearly linear drift rate of -0.060\degx/h during 2012,
and a drift rate of +0.0259\degx/h during 2013.  This can be joined
and well fit with a sinusoidal model, using an average drift rate of
-0.0176\degx/h (-0.442\degx /day), an oscillation amplitude of 118\degx, a period of 686
days (shown in Fig.\ \ref{Fig:Btrack}).  This model is also consistent
with the observed latitudinal drift over that time period (see panel B),
but the model latitude is about 2\deg S of the observed latitude, an offset
opposite to that observed for A, but consistent with what is shown in image B2
of Fig.\ \ref{Fig:featuresABCDFG}.

A feature seen in 2006 and either the same or another feature seen in
2007 \citep{Sro2012polar} appeared at about 27\degx N, close to the
latitude observed for features A and B.  These earlier features also
seemed to be associated with dark spots, and the 2006 feature was seen
as a dark spot in an HST ACS image at 658 nm \citep{Hammel2009Icar}.
These might be related to either A or B in our current data set, based
on similarity in morphology as well as latitude.

\subsubsection{Feature C}

\begin{figure*}[!htb]\centering
\includegraphics[width=6.5in]{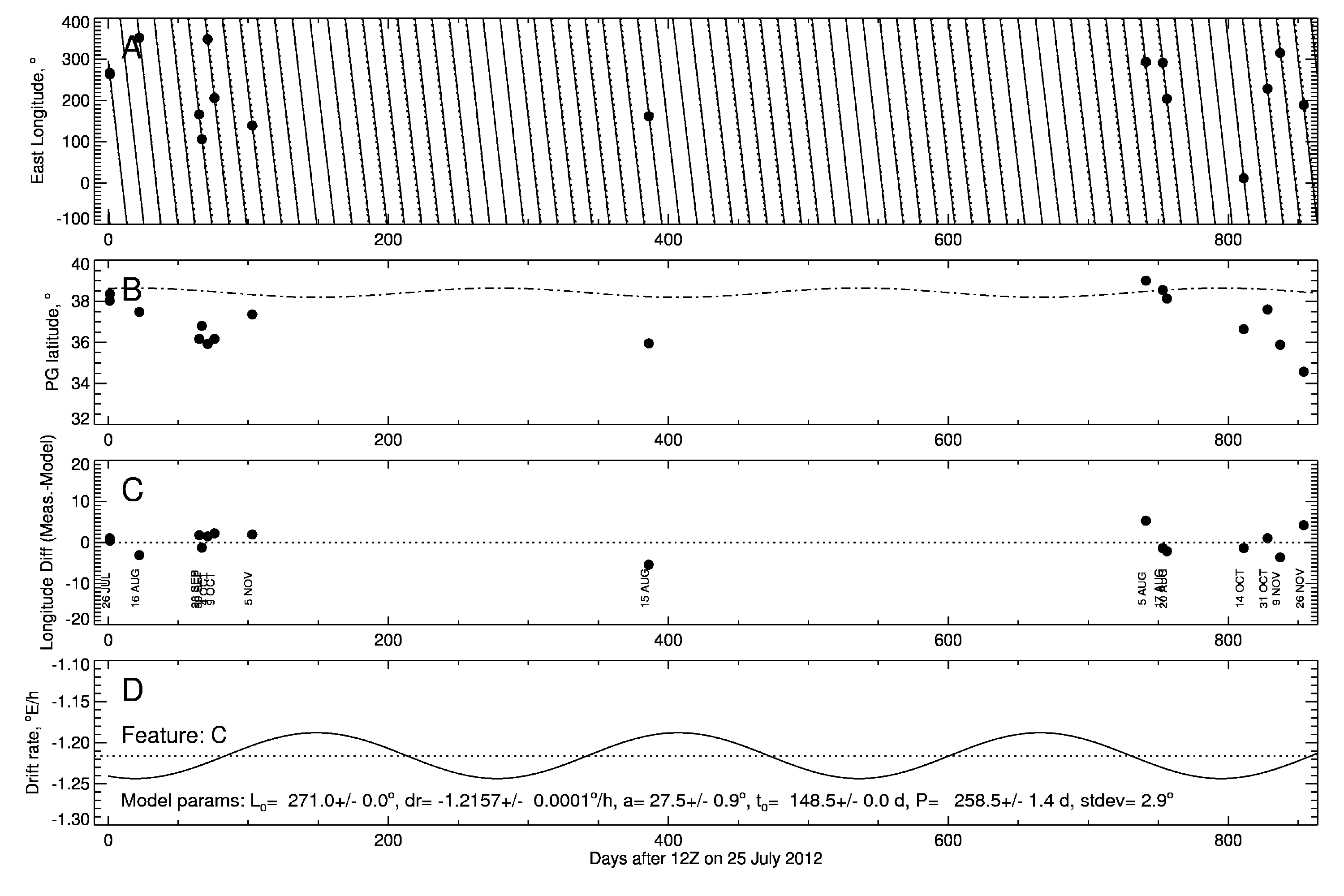}
\caption{As in Fig.\ \ref{Fig:Btrack}, except that longitudes here are for feature C. 
This feature appears to have a relatively short period sinusoidal variation in longitude (A), relative
to a baseline drift rate of -1.2157\degx/h, which implies a sinusoidal variation in drift rate (B),
and a small sinusoidal variation in latitude (dot-dash curve in B), where the latitudinal
model is obtained as in Fig.\ \ref{Fig:Btrack} and exceeds measured latitudes (points) by about 1\degx.}
\label{Fig:Ctrack}
\end{figure*}

Feature C, located at about 36.5\deg N, is the longest living feature
we observed, appearing in August and November 2012 data sets, as well
as in 2013 and 2014 data sets, including August as well as November
2014 images.  It is also a feature that does not seem to have drifted
in latitude over this time period, although its varying drift rate
suggests a small oscillation in latitude of less than 1\deg in
amplitude.  Feature C can also be roughly fit with model that includes
a sinusoidal variation about a mean drift rate, with a much greater
average drift rate of -1.2157\degx/h, compared to A or B, a much
shorter period of variation (258 days) compared to B, and a much
smaller amplitude of 27.5\deg (see Fig.\ \ref{Fig:Ctrack}), which
leads to a much smaller latitudinal variation (panel C) than feature
B.  However, the latitudinal measurements of C exhibit considerable
scatter, especially in the latter 2014 observations, although for most
of the range the observed and model latitudes are within 1-2 degrees,
with the model being north of the measurements.  A possible reason for
the poorer fit of Feature C in comparison with features A and B is
that C is near a latitude region containing many features that might
perturb its motions, as well as being in a region of relatively high
zonal wind shear, so that slight latitude shifts might yield
significant changes in drift rate.

\subsubsection{Features D and E}

The southern hemisphere features D and E both were observed to move towards the equator
during the two years spanned by our observations. And both appear to have existed for
the entire time period, but are harder to observe because at their latitudes we can only
see a small range of longitudes in any given image, a consequence of the sub-observer point being
in the northern hemisphere.  Feature D is relatively compact and is not difficult to measure
when it can be seen.  The longitude of feature E is particularly difficult to measure accurately because of the feature's long narrow morphology. 

\begin{figure*}[!htb]\centering
\includegraphics[width=6.5in]{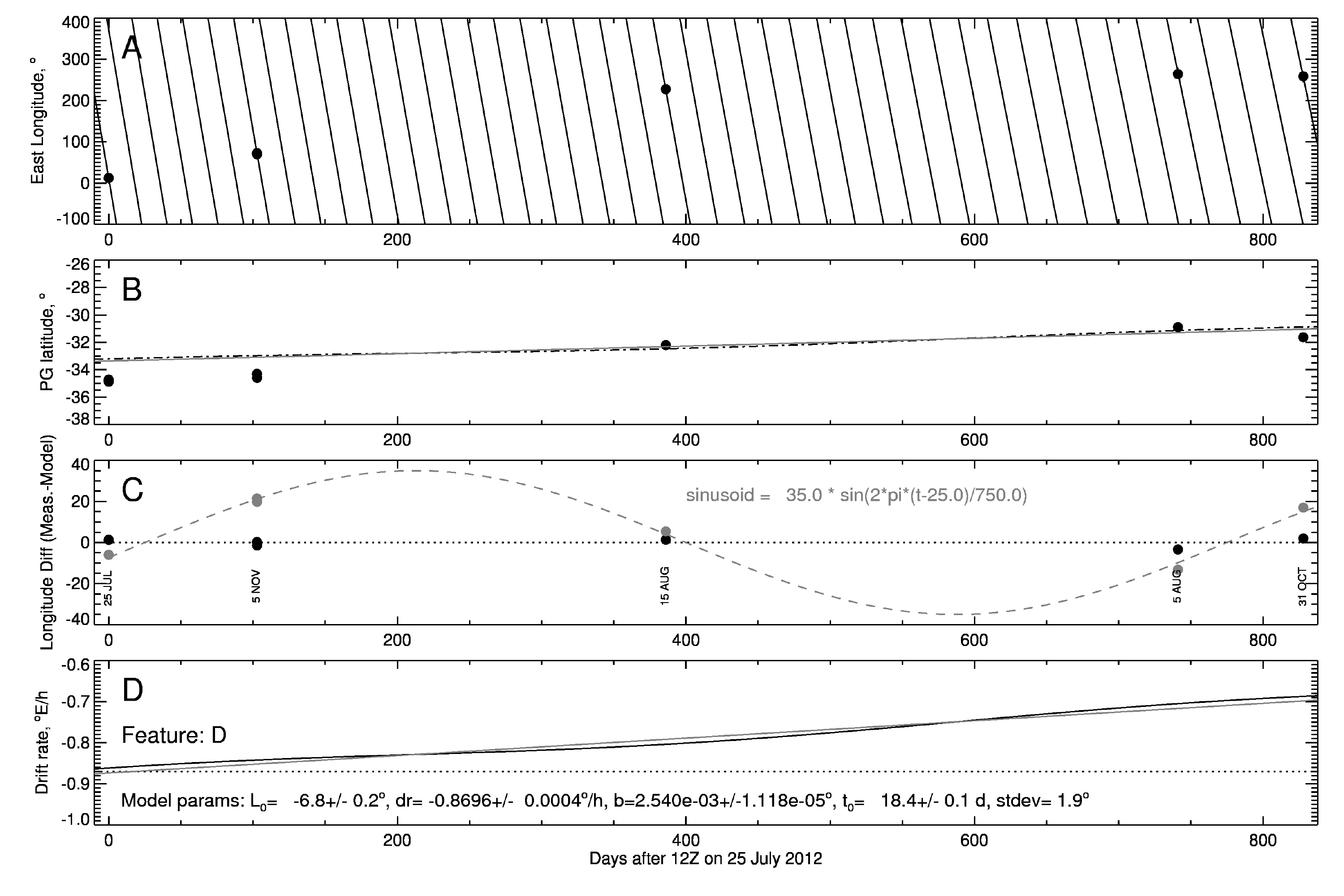}
\caption{As in Fig.\ \ref{Fig:Atrack}, except that longitudes here are
  for feature D and the model is given by Eqs. \ref{Eq:longcomb} -
  \ref{Eq:latcomb}.  This feature appears to be drifting to the north
  and decreasing the magnitude of its negative drift rate in accord
  with the zonal wind profile. A model in which longitude relative to
  a baseline drift model (at -0.8695\degx/h) is assumed to vary as the
  square of a time difference provides a rougher fit that is
  compatible with a linearly varying drift rate (gray line in panel D)
  and a linearly varying latitude (gray line in panel B), where the
  latitudinal model is obtained as in Fig.\ \ref{Fig:Atrack}, and in
  this case agrees as well with measured latitudes (points) as the
  more complete and accurate model (shown as solid black lines in A
  and D, dot-dash line in B, and dotted line in C). The additional
  sinusoidal variation contained in this model is shown by the dashed
  curve in panel C.}
\label{Fig:Dtrack}
\end{figure*}

Our model fit to D is displayed in Fig.\ \ref{Fig:Dtrack}.  As it
moves from about 35\deg S to about 32\deg S, its drift rate is seen to
increase from -0.87\degx/h to -0.7\degx/h.  The drift rate model, when
converted to a latitudinal variation using Model S13A to relate drift rate to
latitude, yields a latitudinal slope in good agreement with
observations, and also agrees well with the absolute value of the
latitude.  A more accurate fit to the measured longitudes can be
obtained with a baseline drift of -0.7145\degx/h, with drift rate
varying from -0.73\degx/h to -0.52\degx/h, which reduces the RMS
deviation from 14.1\deg to 9.0\degx, but the inferred latitude then
becomes about 3\deg greater than the observed latitude. This might be
a consequence of the cloud features being displaced in latitude from
the circulation feature that is actually following the zonal mass
flow.  Both fits can be substantially improved by adding another
component of variation, namely a sinusoidal deviation from the
quadratic model, as shown by the dashed line in
Fig.\ \ref{Fig:Dtrack}, which reduces the RMS deviation down to
1.9\deg for the model that agrees well with the observed latitudes.
That variation has a period of 750 days and an amplitude of 35\deg in
longitude.  An even longer period of nearly three years was observed
for a prior southern hemisphere long-lived feature named the Berg
\citep{Sro2009eqdyn}. That feature spent most of its time oscillating
$\pm$2\deg about a mean latitude of 35.2\deg S (planetographic), then
in 2005 began to drift towards the equator, reaching 27\degx S by 2007
and reaching 5\degx S and dissipating in late 2009
\citep{DePater2011}.


Feature E (model fit in Fig. 23) was first seen near 41\deg S and last
seen near 38\deg S.  Its latitude seems to follow a linear trend of
about 1\degx/year towards the equator, and the longitudinal model fit
is compatible with latitudinal variation obtained by interpolation of
Model S13A of \cite{Sro2012polar}. However, there are so few
observations of this feature that we cannot entirely rule out other
drift rate models.

All the features in the 2012-2014 data sets that were observed to have
substantial latitudinal motions moved towards the equator, as did the
previously mentioned Berg feature last seen in 2009.

\begin{figure*}[!htb]\centering
\includegraphics[width=6.5in]{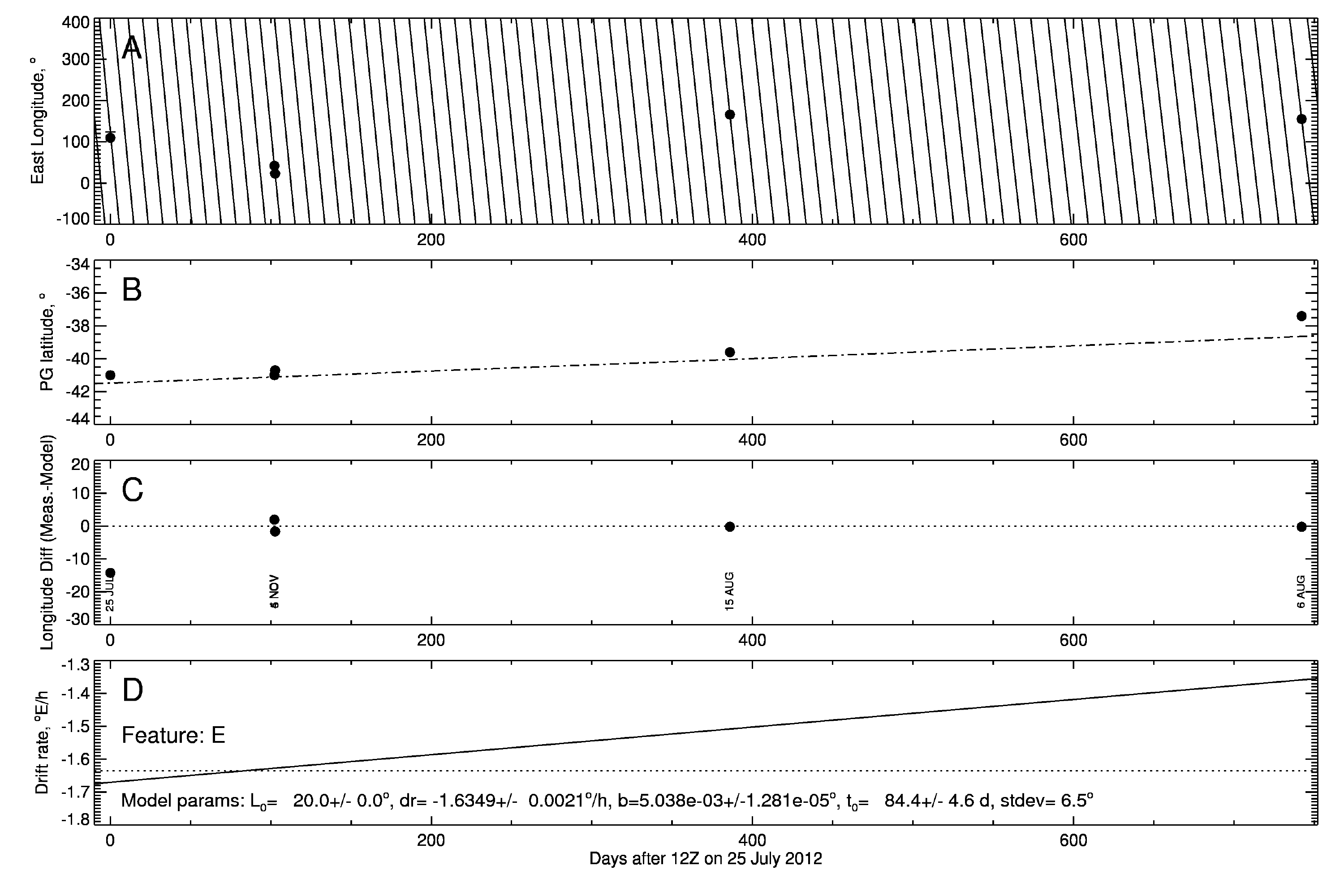}
\caption{As in Fig.\ \ref{Fig:Atrack}, except that longitudes here
  are for feature E.  This feature appears to be drifting in latitude
  and following the local wind profile. Observations of longitude (A)
  can be roughly fit with a constant drift rate, but the varying
  latitude (B) suggests a varying drift rate, and a fit to the
  difference from a constant drift rate using a quadratic function of
  time (C) provides a comparable fit that is also compatible with a
  linear latitudinal variation (B).}
\label{Fig:Etrack}
\end{figure*}

\subsubsection{Feature F}

Found near 34\degx N planetographic latitude (Figs.\ \ref{Fig:rectpro}
and \ref{Fig:featuresABCDFG}), feature F, which is feature 2 of
\cite{DePater2015storm}, is the first discrete cloud feature ever
clearly detected by amateurs using CCD detectors and small telescopes
\citep{DePater2015storm}. Compared to most other prominent
northern-hemisphere features seen in 2014, feature F was not
particularly bright at near-IR wavelengths and did not extend to as
high an altitude.  It was not seen in K$'$ images, placing it deeper
than the 1-bar level according to Fig.\ \ref{Fig:penprof}, while
several other 2014 features were spectacularly bright at that
wavelength, with G being the most notable \citep{DePater2015storm}.
Yet, it was F that was detected by amateurs, probably because it had a
substantially greater optical depth than surrounding clouds.  Feature
F turned out to have an exceptionally long life and was tracked by
amateurs as well as by HST observations using a ToO program 13712
(K. Sayanagi, PI) and OPAL program 13937 (A. Simon, PI), and by
observations from Gemini, Keck, and Palomar, which are all summarized
by \cite{Sayanagi2015too}. Our model fit to its motions is displayed
in Fig. 24. We first saw F in our 2014 Keck images, but were not able
to identify it in 2013 or 2012 images.  This is not because of
confusion about which feature is which.  When we project our model
backward we find no feature at all anywhere near the predicted
location.  Thus it appears that F developed sometime between August
2013 and August 2014.

F also has an unusual morphology. It appears to have a very short
plume extending to the west on its north side and a very extensive
plume extending eastward over more than 100\degx of longitude on its
south side (see Figs.\ \ref{Fig:rectpro} and
\ref{Fig:featuresABCDFG}).  The direction of the plume is consistent
with cloud particles spreading out from F mainly to the south, then
following the zonal flow and falling behind the faster (more westward)
motion of F. Given the F-to-plume latitude distance of 3-4\deg and the
local wind shear of 0.084\degx/h per degree of latitude (from
Table\ \ref{Tbl:adoptdrate}). The time to extend the plume 100\deg of
longitude would be 30 to 40 hours.  Of course the length of the plume
is also limited by particle fallout, so the time span of convective
activity and plume generation could of course have been much longer.
The prominence of the F plume is a marker of its unusual convective
strength relative to other features in similar shear regions that
generate no plumes and is consistent with a large optical
depth that would facilitate amateur detection of the feature.  The last
observation of this feature at the time of this writing was obtained
on 8 January 2015, using the Gemini NIRI imager.  At that time it did
not appear to have a prominent plume, suggesting that its convective
activity has substantially declined. It was found just 12\deg east
of the location predicted using the model in Table 8.

\begin{figure*}[!htb]\centering
\includegraphics[width=6.5in]{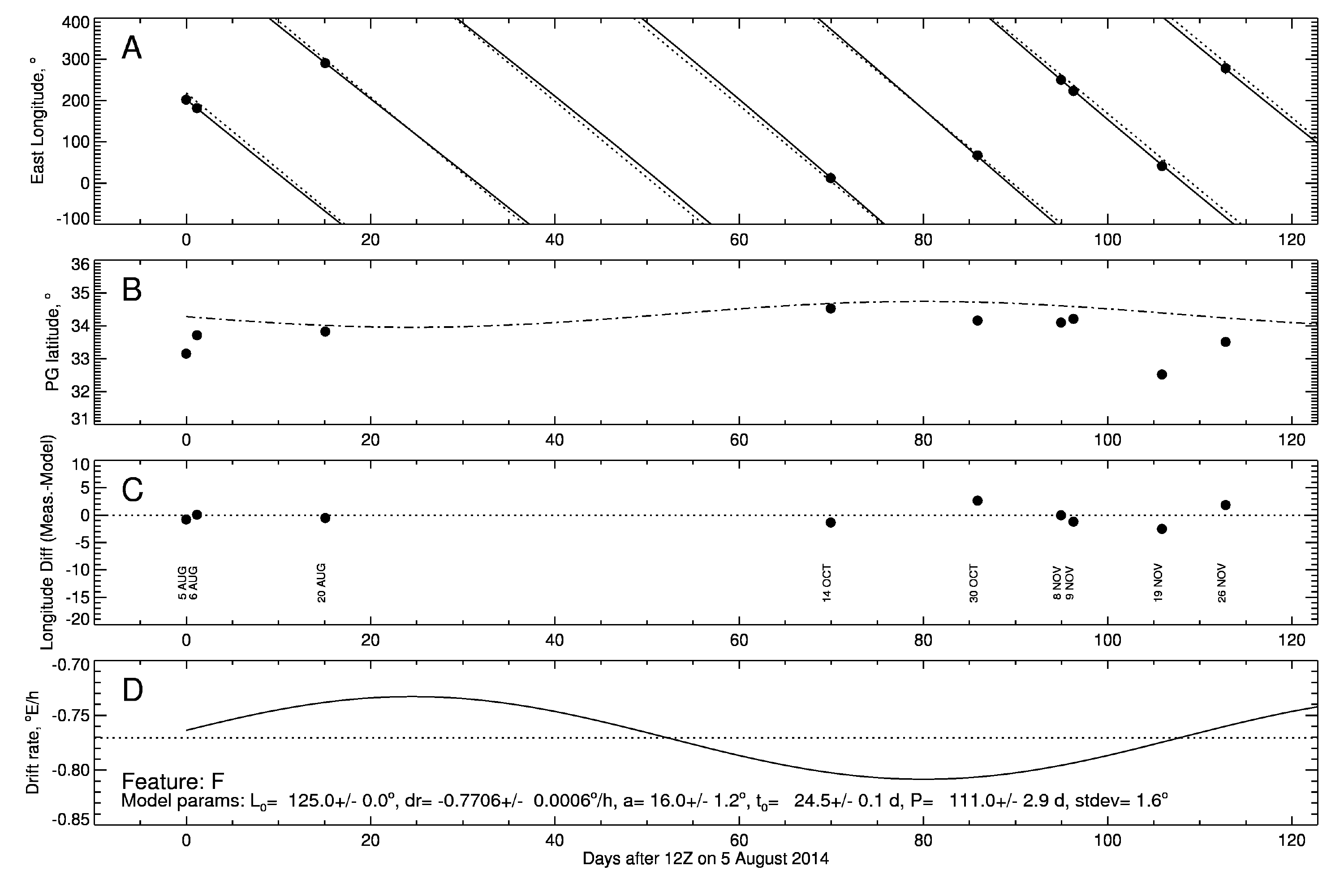}
\caption{As in Fig.\ \ref{Fig:Btrack}, except that longitudes here are
  for feature F.  This feature has a varying drift rate (solid lines
  in A and D), following a slow oscillation in longitude about a mean
  drift rate (dotted lines in A and D) of -0.7706$\pm$0.0006\degx/h,
  which is nearly consistent with the zonal flow at a mean latitude
  about 0.5\deg north of the observed latitudes (filled dots in
  B). The latitudinal model shown as the dot-dash line in B is
  obtained as in Fig.\ \ref{Fig:Atrack}.}
\label{Fig:Ftrack}
\end{figure*}

\subsection{Possible interactions of long-lived features}\label{Sec:interact}

Long-lived features at nearby latitudes move at different longitudinal
speeds and will eventually approach each other at close range,
raising the possibility of some sort of interaction.  An example of
this occurred on 25 December 2011, when two bright features separated
by 2\deg in latitude had a close approach that resulted in formation
of a small dark spot and companion clouds \citep{Sro2012bs}.  A close
approach observed on Saturn \citep{Sro1983JGR} resulted in a more
dramatic development of bright clouds during the interaction.  It is
also possible that features might merge or dissipate.  Examples of
dark spots merging on Saturn can be found in Fig.\ 3 of
\cite{Porco2005}.  If the features have underlying vortex circulations
of similar vorticities, we might expect to see latitudinal deflections
in opposite directions as they pass by each other, followed by a
return to their original latitudes.  It is less clear what changes to
expect in the structure of their companion clouds.  We found no
evidence of strong interactions for the two feature pairs described
below.

\subsubsection{Close approaches of A and B}

\begin{figure}[!htb]\centering
\includegraphics[width=3.5in]{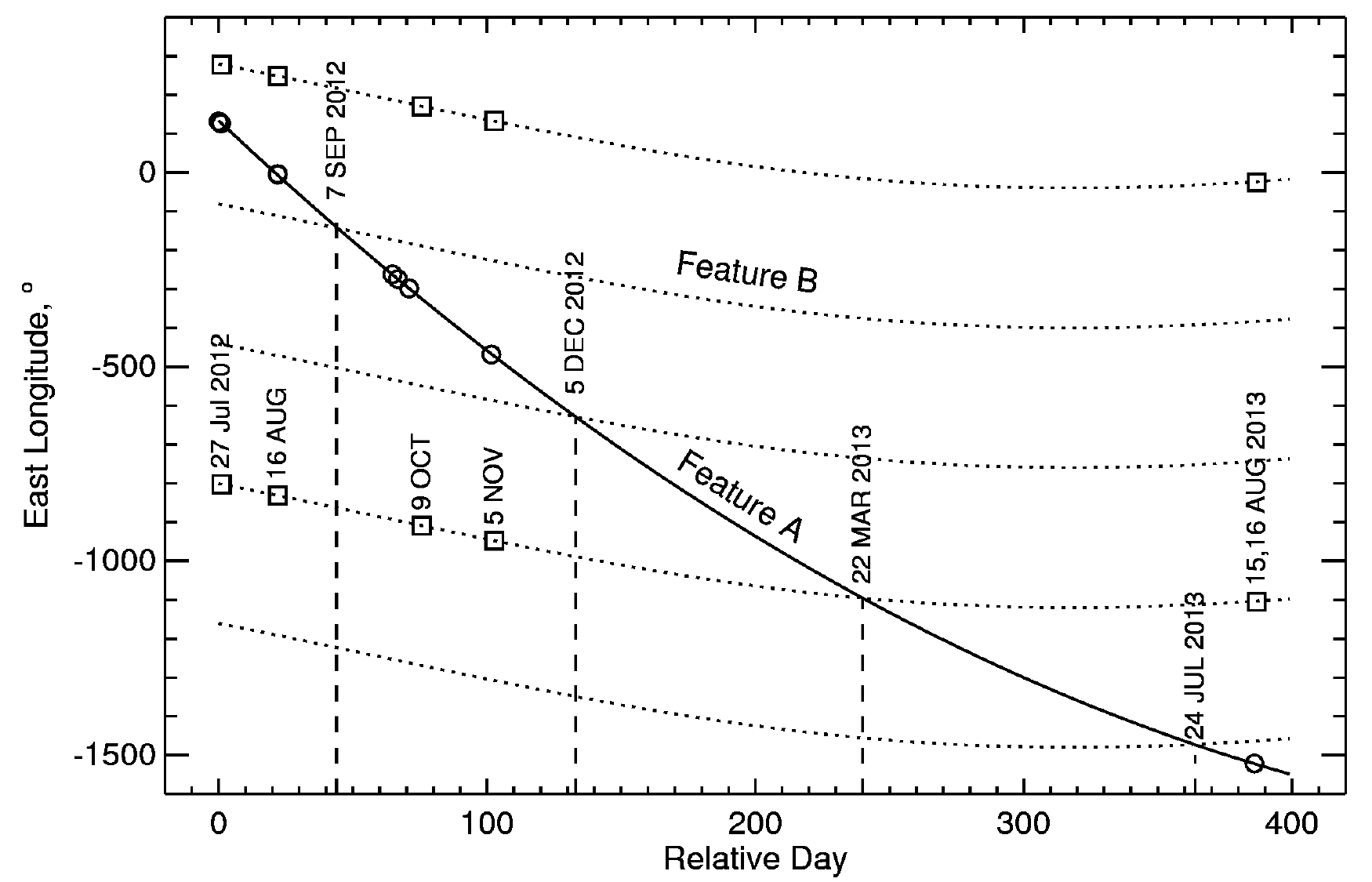}
\caption{Comparison of long term tracking results for features A and B. The x axis shows time
in days relative to 25 July 2012 at 13:32:39 UT. The models of longitude versus time are those
given in Figs.\ \ref{Fig:Atrack} and \ref{Fig:Btrack}, the latter plotted with multiple 360\deg
offsets to show crossing events more clearly (these are marked by vertical dashed lines and
annotated by date at which they reach the same longitude). }
\label{Fig:ABtrack}
\end{figure}

Long-term tracking of A and B features show that B has an average
drift rate of -0.0597$\pm$0.0002\degx/h, while A has an average drift
rate of -0.248$\pm$0.002\degx/h, with this average taken over a period
from late August 2012 to 4-5 November 2012.  Given that the winds
become more westward at higher latitudes, the drift rate comparison
suggests that the putative vortex generating the A features is
actually north of the vortex generating the B features.  This is
consistent with the November 4-5 appearance of the features in
Fig.\ \ref{Fig:featuresABCDFG}.  The long term tracking of these two
features, illustrated in Fig.\ \ref{Fig:ABtrack}, provides another
surprising result.  Since their sizes (judged by the extent
of the bright clouds associated with them) are somewhat greater than the
latitudinal difference between them, we would expect some sort of
interaction when they pass by each other.  However, even though they
reached the same longitude on September 7, 2012, their drift rates
before and after that close encounter were not perceptibly changed.

Features A and B had three more close encounters before our next
observing run on 15-16 August 2013, yet they remained distinguishable
features. However, there is considerable ambiguity as to what happened
during the nearly 10 months without any observations. Extending the
drift track of these features to the August 2013 date, we find close
approaches should have occurred on 5 December 2012 and 25 March 2013.
It is plausible that the multiple close approaches altered their
latitudes and drift rates slightly, and our assumed model of a linear
variation in latitude (and in drift rate) may not be correct.  A
non-uniformly varying drift rate has been observed for other long
lived features on Uranus, so this would not be surprising.  It is also
possible that one or both of the features disappeared during that
unobserved period and that at least one of them in our 2013 images is
a new feature.

Also noteworthy, is that the latitudes we infer from our drift rate
models interpolated to latitudes using the zonal profile of
\cite{Sro2012polar}, suggest that Feature A is following a circulation
feature that is about 2\deg N of the observed bright clouds and that
Feature B is following a circulation feature this is about 3\deg S of
the bright clouds associated with it. In this connection it is worth
noting that the rectangular projections of A2 and B2 in
Fig.\ \ref{Fig:featuresABCDFG} for November 2012 (2nd row from bottom)
provide evidence of a dark spot associated with A that is north of the
bright clouds accompanying it, and a dark spot associated with B that
is south of the bright clouds accompanying it. The inferred separation
of about 5\deg in latitude makes it somewhat more plausible that the
features seem to suffer no major interactions.

\subsubsection{Close approaches of C and F}

\begin{figure*}[!htb]\centering
\includegraphics[width=6in]{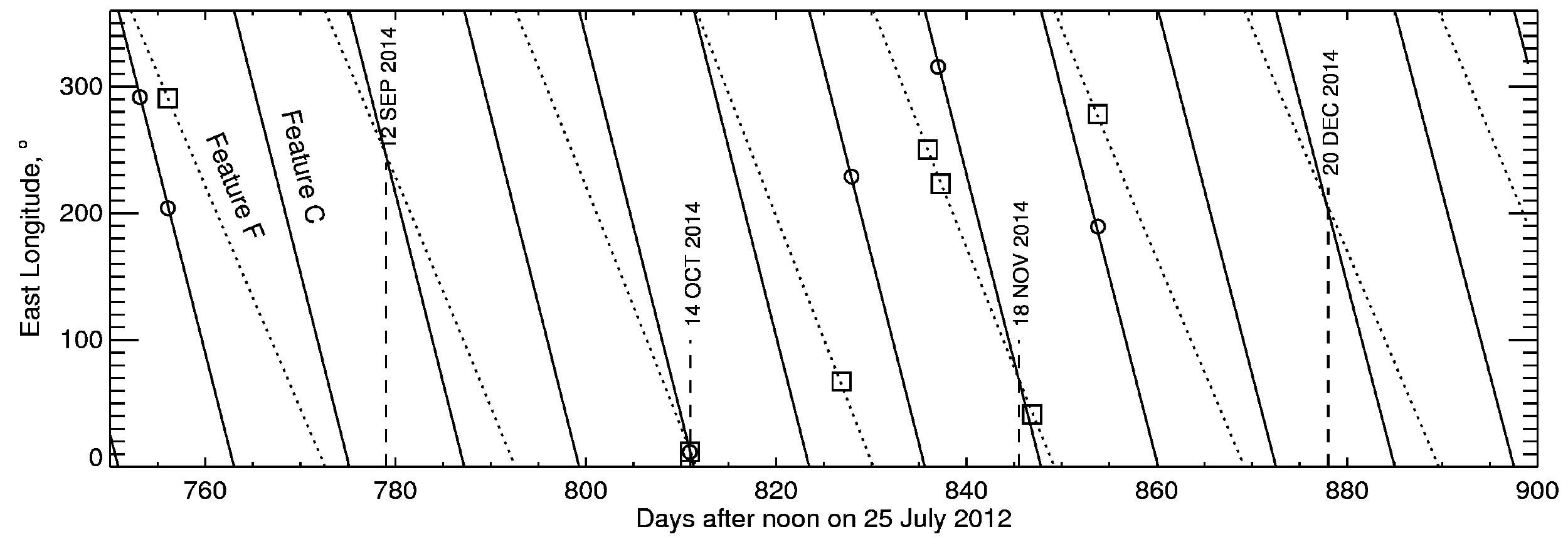}
\caption{Comparison of long term tracking results for features C and F. The x axis shows time
in days relative to 25 July 2012 at 12:00 UT. The models of longitude versus time are those
given in Figs.\ \ref{Fig:Atrack} and \ref{Fig:Btrack}, each plotted with multiple 360\deg
offsets to show crossing events (these are marked by vertical dashed lines and
annotated by date at which they reach the same longitude). Measured points are
plotted as circles (for feature C) and squares (for F). Note the overlap of measurements
on 14 October 2014.}
\label{Fig:CFtrack}
\end{figure*}

Close approaches of features C and F are shown by intersections of
longitude versus time plots in Fig.\ \ref{Fig:CFtrack}. The average
latitudes for Features C and F are 36\deg N and 34\deg N respectively.
Since each of these appears to extend over several degrees of latitude
(see Fig.\ \ref{Fig:featuresABCDFG}), it would be surprising if they
did not display evidence of some interaction.  In fact, these were
observed in such close proximity in HST images acquired on 14 October
2014 (taken as part of a Target of Opportunity program, with Kunio
Sayanagi as PI), that we initially thought we were observing a single
feature.  We also observed them in close proximity in a Gemini NIRI
image acquired on 19 November, about 1 day after close approach.  Both
of these close approaches are consistent with the longitude versus
time plots shown in Fig. \ \ref{Fig:CFtrack}.  Besides the 14 October
and 18 November approaches, we also found unobserved close approaches
on 12 September 2014 and 20 December 2014.
There is no evidence of any change in drift rate or, based on a 26
November Gemini image, any change in the morphology of the features
following these possible interactions, although F was seen close
to the central meridian only in the late November image.

\subsection{Unusually bright feature G}

\cite{DePater2015storm} identified several bright features in the 2014
Keck observations, one of exceptional brightness, which is labeled as
G in Fig.\ \ref{Fig:rectpro}. This appears to have faded
dramatically within a month or so.  A similarly bright cloud feature
was detected in August 2005 \citep{Sro2007bright}, but at a higher latitude (31\degx N).
It brightened and faded dramatically within a few months, and extended to similarly high
altitudes ($\sim$300 mbar level). 

\section{High latitude polar cloud features.}

\subsection{Feature morphology}\label{Sec:polemorph}

\begin{figure*}[!htb]\centering
\includegraphics[width=6.5in]{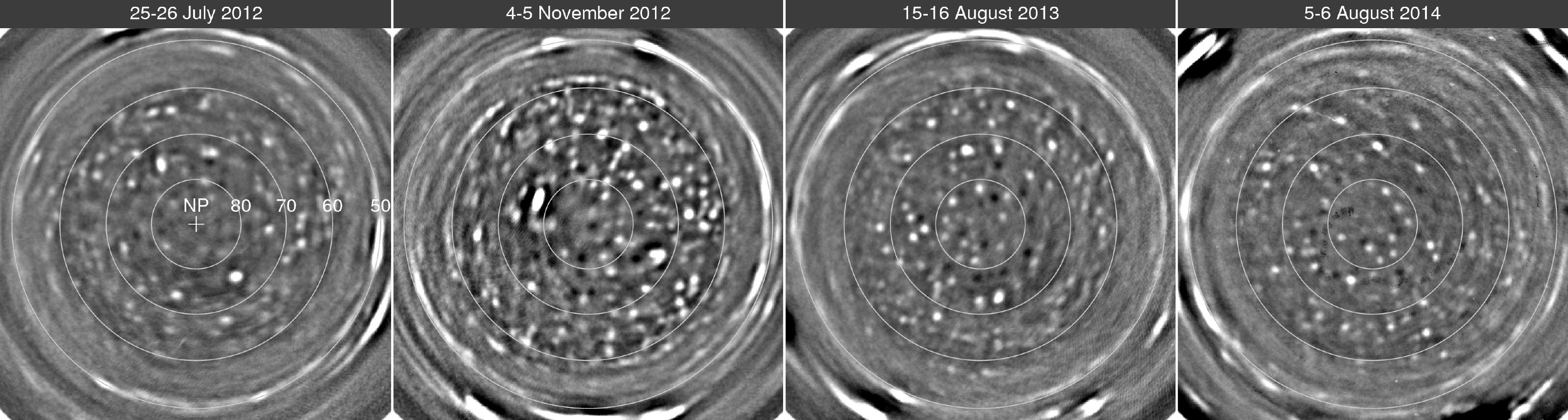}\\[0.01in]
\includegraphics[width=6.5in]{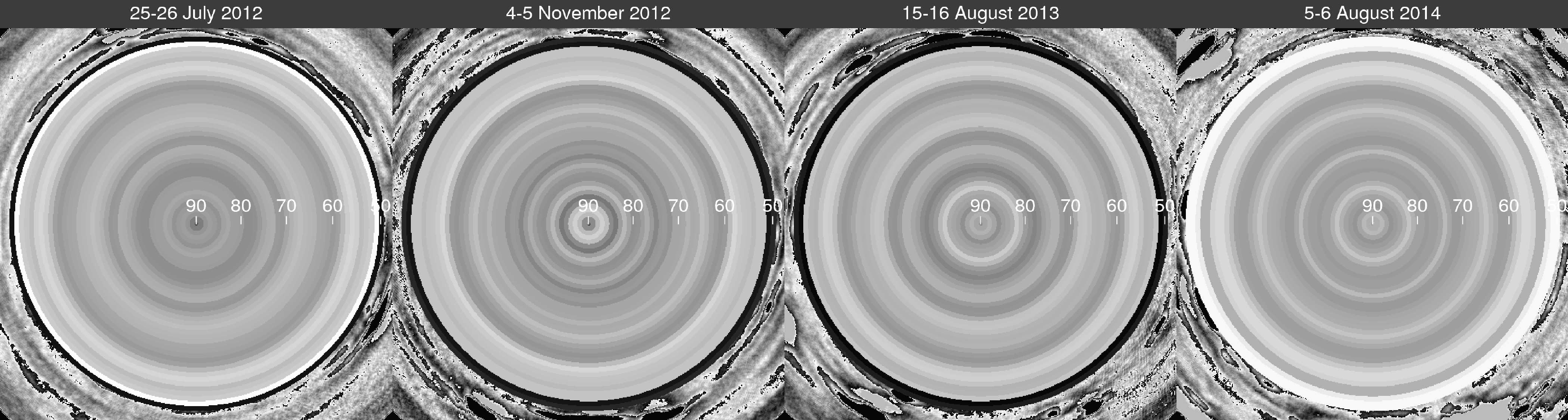}
\caption{Polar composites (top row) from 2012, 2013, and 2014, in
  which images were averaged after removing a fixed 4.1\degx/h
  rotation relative to the planet's fixed coordinate system.  The lack
  of zonal smearing at latitudes greater than 55\degx N confirm the
  existence of precise solid body rotation in this region. These
  images were high-pass filtered with a smoothing length of 25 pixels,
  compared to a displayed image size of 500 pixels.  Zonal averages of
  the composites in 1\deg bins (bottom row), showing a crudely
  consistent pattern in the location of bright features.}
\label{Fig:poleviews}
\end{figure*}

The 2007 equinox observations \citep{Sro2009eqdyn} provided the first
detection of cloud features north of the 60\degx N westward jet peak.
More features were seen in 2011 images, and it appeared that the north
polar region was peppered with small low-contrast discrete clouds
\cite{Sro2012polar}, presenting what looked like a field of
fair-weather cumulus convection on earth, but on a much larger scale.
This was not expected because such features had never been seen in the
south polar region.  From 2012 onward, the improved views of the north
polar region of Uranus have allowed us to combine observations from
successive observing nights to form a complete picture of the polar
region.  Because the entire region from about 63\degx N to at least
83\degx N moves in solid body rotation, we were able to average images
with that rotation period removed to obtain a pole-centered high S/N
view of Uranus' north polar region for each of the three years of
observation (Fig.\ \ref{Fig:poleviews}).  These show a continued
prevalence of large numbers of small bright features.

Almost all the polar bright features have similar size and shape, which is
mainly circular.  Typical  diameters are 600-800 km, which is comparable
to imaging resolution, and thus is an upper limit for the smaller
features.  There are
also less numerous but well-defined small dark spots, many of which
are well separated from bright spots and thus cannot be explained as
an artifact of high-pass filtering.  The small bright features have a
typical contrast of 1-2\% in the raw images, with a much brighter than
typical feature in the 4-5 November image reaching 10\%.  The dark
features have only a fraction of the contrast of the bright features
(as well as the opposite sign). The lifetimes of the polar features
can extend for at least 1.5 planet rotations, as quite a few of these
have been tracked over that time period (not continuously, but with a
one- rotation gap). The large number of features seen in the November
2012 data set seems to indicate the clear peak in activity. At a casual
glance these spots appear to be fairly uniformly distributed in
latitude and longitude, down to a latitude of about 55\degx N, where
there begins a transition to a morphology dominated by longitudinally
stretched streaky features often occurring within long narrow regions
of enhanced brightness with a width comparable to the diameter of the
small polar spots.  However, there is a pattern in the distribution of
features that has at least some crude year-to-year consistency, as
discussed in Section\ \ref{Sec:polepat}.

It is worth recalling, as demonstrated in Fig. 1 of
\cite{Sro2014stis}, that when the south polar region was imaged in
2003, using the same telescope, camera, and essentially similar AO
system, no discrete polar features of any kind were observed between
50\degx S and the south pole.  That observation was made in southern
hemisphere fall, while the current observations were made during
northern hemisphere spring.  One might surmise that this seasonal
difference is the cause of this striking morphological asymmetry.  A
long season of radiative cooling from the top of the atmosphere during
Uranus' northern winter might tend to produce an unstable thermal
profile that favors vertical mixing, which is likely to be most active
following equinox.  On the other hand, the long summer of heating from
above during southern hemisphere summer might stabilize the thermal
profile and inhibit vertical convection in southern polar regions in
the fall.  Although this has a plausible ring to it, there are other
indicators of vertical convection, such as depleted methane mixing
ratios in both polar regions at the same time \citep{Sro2014stis},
which argues against such seasonal modulations in at least the large
scale flow.  We are expecting that this ``convective'' activity might
continue for a while as the northern hemisphere moves into summer, but
it is plausible to expect it to eventually dissipate as summer heating
intensifies.  We also expect the formation of a polar cap cloud
\citep{Hammel2007var}, which appears to have begun
\citep{DePater2015storm}.

\subsection{Zonal-average patterns in polar clouds}\label{Sec:polepat}

\begin{figure}[!htb]\centering
\includegraphics[width=3.5in]{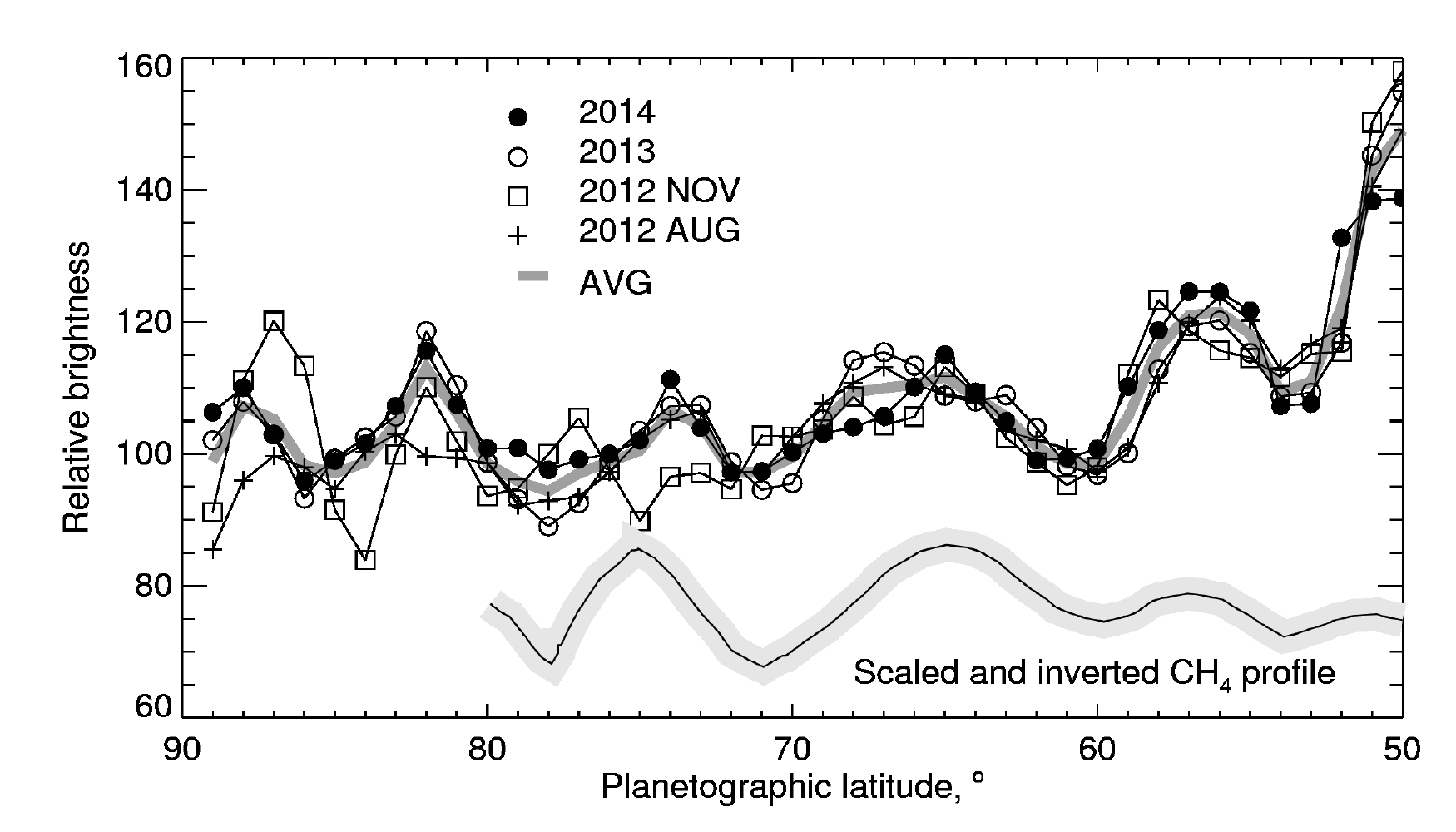}
\caption{Zonal average of relative brightness as a function of latitude for each data
set during 2012-2014 (thin lines), derived
from polar composites displayed in Fig.\ \ref{Fig:poleviews}. The average over all data
sets (giving equal weights to each year) is plotted as a thick
dark gray curve, and a scaled and inverted version of the methane profile of \cite{Sro2014stis} is plotted in the
lower part of the figure, where the variations between 60\deg and 80\deg correspond to CH$_4$ volume
mixing ratio variations of about $\pm$0.005 relative to a mean of 0.02.  As noted by \cite{Sro2014stis},
these variation might actually be caused by para fraction variations induced by local vertical
convection. }
\label{Fig:zonalav}
\end{figure}

A closer look reveals that the zonal average relative brightness
between 50\degx N and 90\degx N displays a roughly consistent pattern,
evident in the zonally averaged polar projections in
Fig.\ \ref{Fig:poleviews}, and the plot of relative brightness versus
latitude in Fig.\ \ref{Fig:zonalav}.  In the average for all three
years, the two 2012 values are averaged together to represent the
average for that year.  All three years have minima near
planetographic latitudes of 53-54\degx N, 60-61\degx N, 70-71\degx N,
and 76-80\degx N, though the 2012 November 4-5 composite is somewhat
of an anomaly, with both larger numbers of polar features and features
of greater contrast than seen in other years.  Other details regarding
positioning and widths and numbers of various bands vary from year to
year.  Between 60\degx N and 80\degx N, the August 2012 and 2013
composites have better pattern agreement with each other than either
has with the 2014 composite.  But at 80\deg N and 90\deg N the 2013
and 2014 composites agree much better with each other than with the
2012 composites.  Whether these changes from year to year are real
trends, or just due to stochastic variations is hard to evaluate
without more observations.  Certainly 2012 demonstrates considerable
variability within a single year.  The 3-year average pattern is
roughly consistent with observed high-latitude variations in the
apparent methane mixing ratio inferred from 2012 STIS observations by
\cite{Sro2014stis}, who noted the correlation with 2012 Keck
observations.  They also pointed out that these apparent mixing ratio
variations might actually be caused by para fraction variations
induced by local vertical convection.

\begin{figure}[!htb]\centering
\includegraphics[width=3.5in]{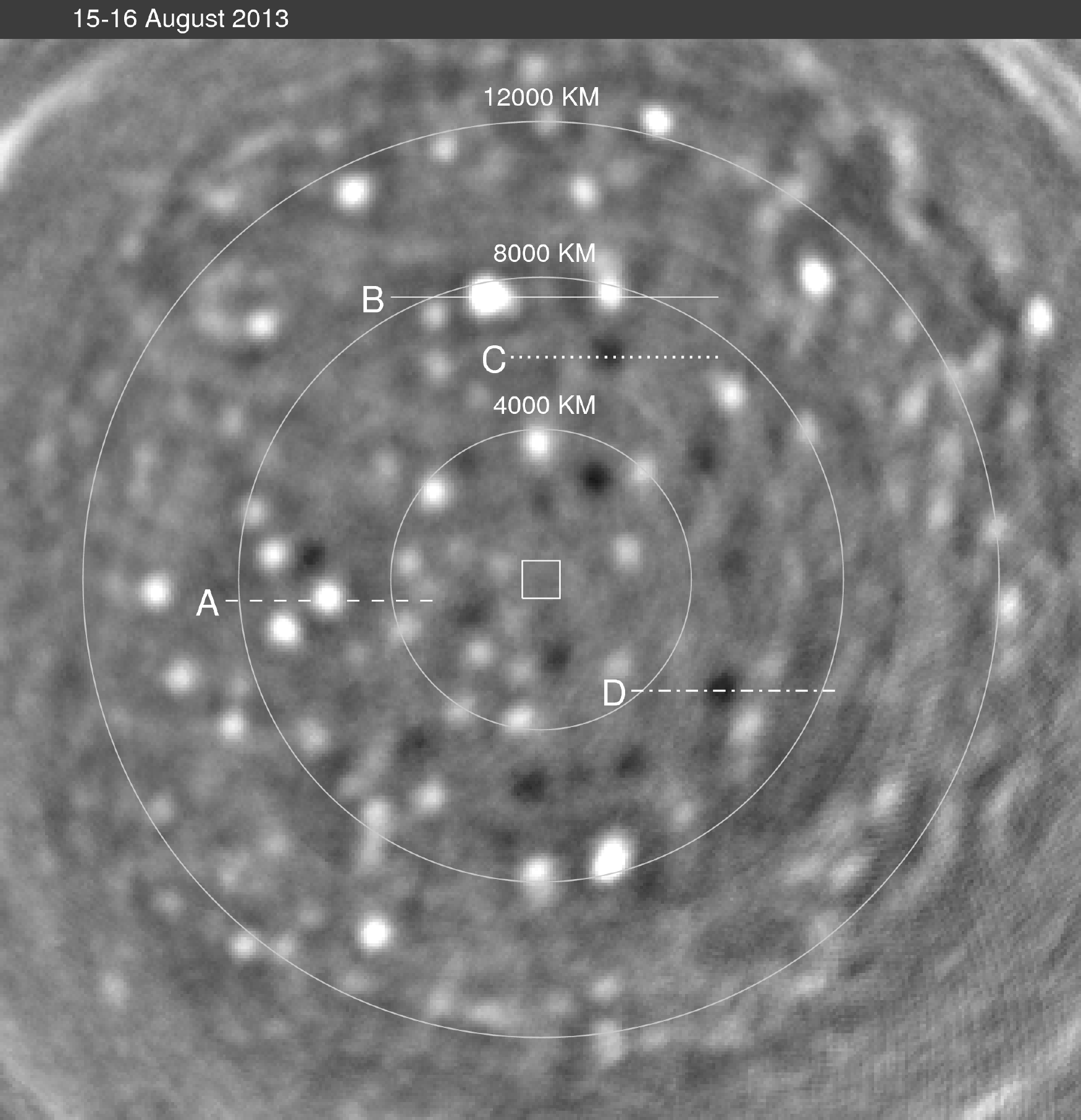}
\includegraphics[width=3.5in]{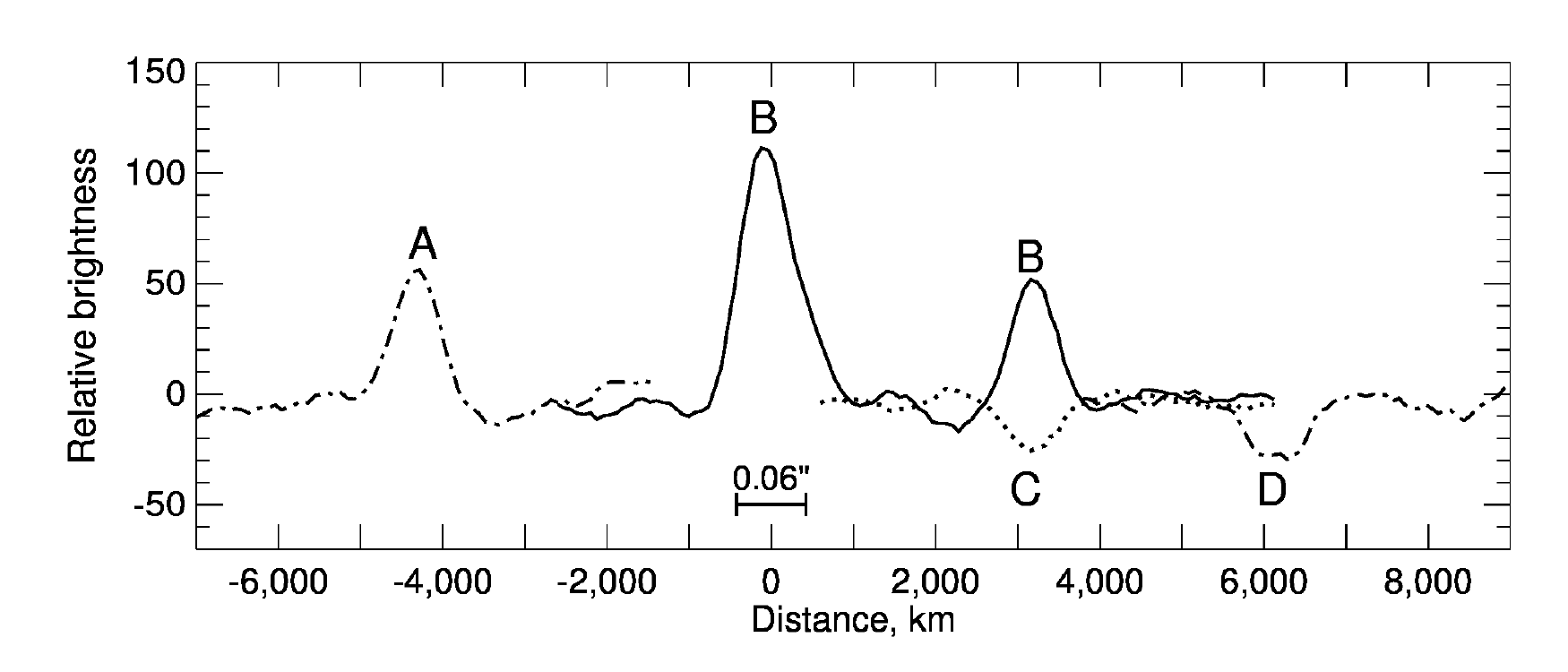}
\caption{Detailed view of polar bright and dark spots in August 2013
  (upper) and line scans through several spot features (lower). Note
  that dark spots often occur in isolation from bright features. Line
  scans indicate that the spot features are generally not resolved;
  they appear to have Gaussian cross sections and
  full-width-half-maximum values comparable to that of the imaging
  system (about 0.06$''$ in H under ideal conditions). In the upper
  panel, circles of constant distances from the pole provide a sense
  of the physical scale.  The small square centered at the pole is
  1000 km on a side. This image was also high-pass filtered as in
  Fig.\ \ref{Fig:poleviews}.  The contrast in unfiltered images was
  measured to be 2.7\% for the brighter feature along scan B and 2\%
  for the feature intersected by scan A.}
\label{Fig:spotdetail}
\end{figure}

A more detailed view of polar cloud features is presented in
Fig.\ \ref{Fig:spotdetail}, where we see that almost all features have
about the same size, which is 600 km to 800 km according to 2-D
Gaussian fits to many of the features.  Line scans through several
bright and dark spots are plotted in the bottom panel of
Fig.\ \ref{Fig:spotdetail}, showing that both bright and dark features
have FWHM values comparable to the Keck NIRC2 PSF (about 0.06$''$).
Thus, these features are generally not resolved and might actually be
considerably smaller than they appear in these images.  The dark
spots, which are possibly regions of reduced cloud opacity, perhaps
produced by downwelling motions, have lower contrast than bright
features, but are of similar apparent size. Spacing between features
is typically 1000 km to several thousand km.  The dark spots are
generally found at higher latitudes than the bright features, most
within 20\deg of the pole.  There is also another change in
distribution between regions close to the pole, where spots appear to
be randomly distributed, and regions further from the pole, where
features seem to appear more often as beads on a string, with the
``string'' in this case lying along a circle of constant latitude.

\section{Near-equatorial waves}\label{Sec:waves}

\begin{figure}[!htb]\centering
\includegraphics[width=3.5in]{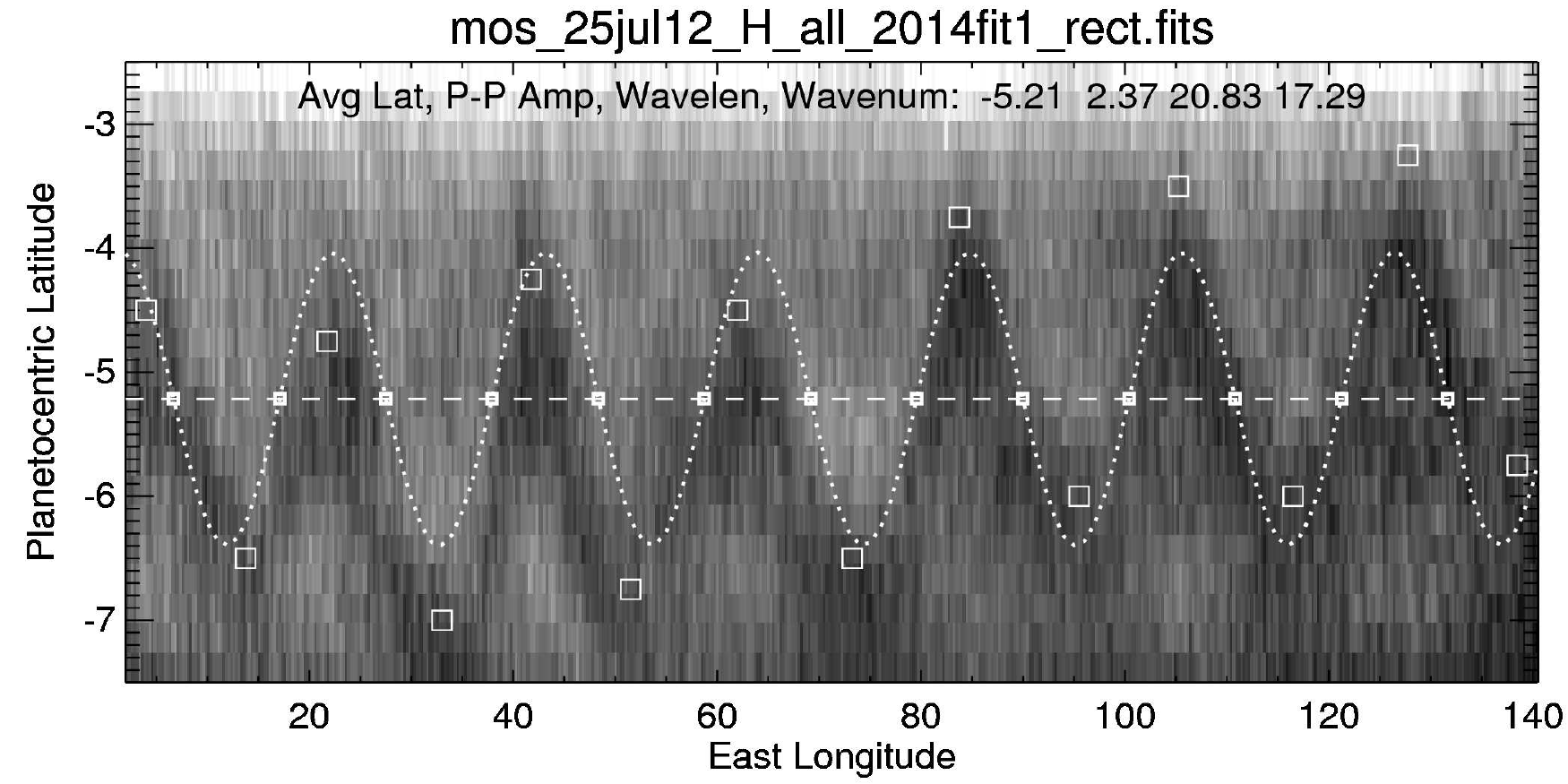}
\includegraphics[width=3.5in]{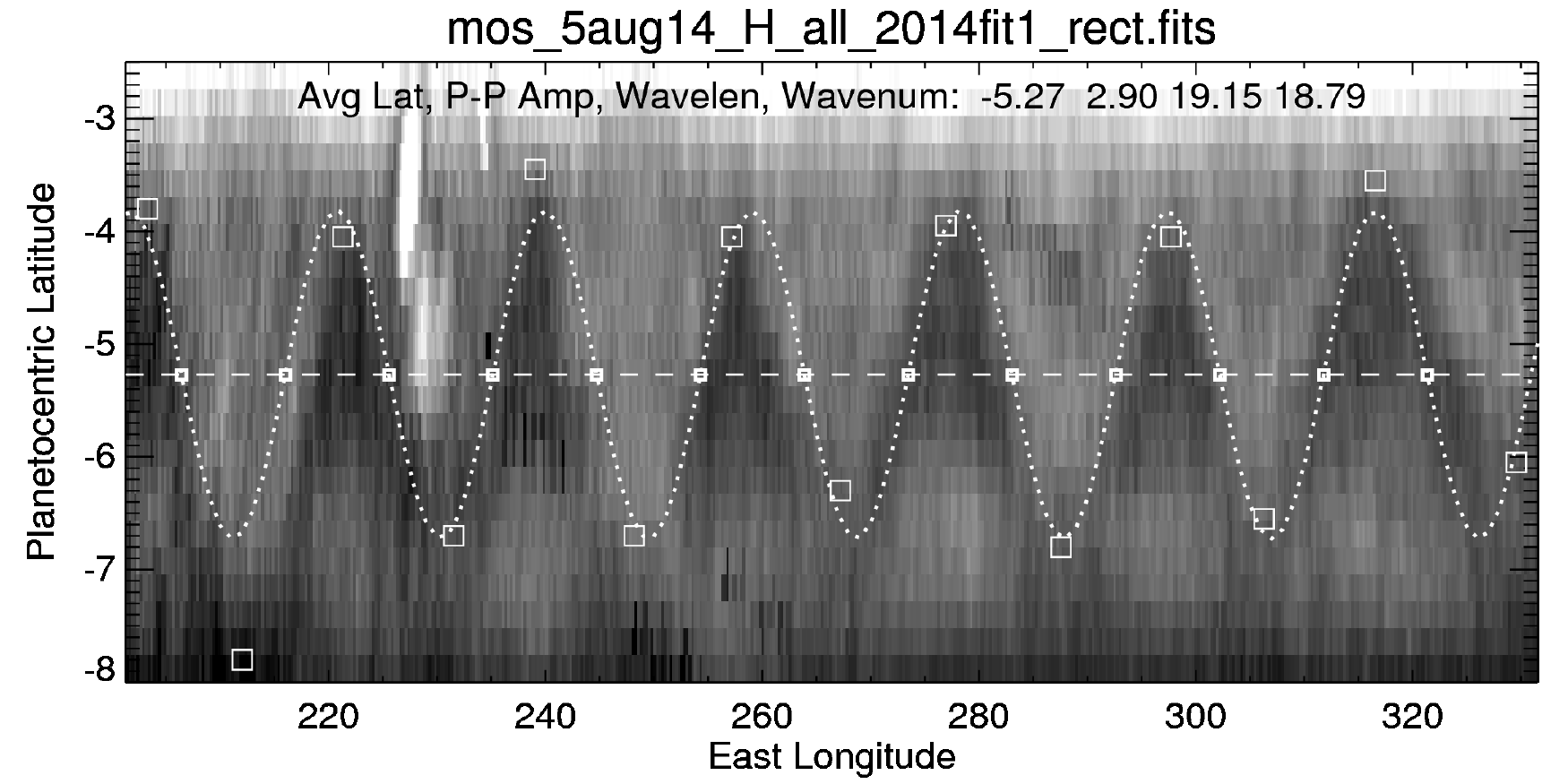}
\caption{South-equatorial waves in 2012 (top) and 2014 (bottom), with sinusoidal fits
for each to the lower boundary of the bright region above the dark ``ribbon'' feature.  For
context, see Fig.\ \ref{Fig:imsamples}.}
\label{Fig:wavefits}
\end{figure}

One of the more striking features in Uranus' atmosphere is the
scalloped wave form that was first brought to light by high-S/N
measurements in 2012.  The immediate appearance is that of two sine
waves that criss-cross each other, similar to a two strand braid
(Fiq.\ \ref{Fig:featuresEKW}).  This appears strikingly similar
to vertically varying cloud forms created by Kelvin-Helmholtz instabilities
generated by sharp vertical gradients in the horizontal winds.  A different
wave mechanism must be at work here however, perhaps Kelvin waves
or mixed internal gravity-Rossby waves.  A correct identification would
be helped by measuring its dispersion relation (phase speed versus
wavelength), which is not likely to be feasible with  current observations.  
 The lower boundary of this morphology is a sort of ribbon wave
 (Fiq.\ \ref{Fig:wavefits}), though the ribbon does not have a
 constant latitudinal width.  The better defined upper boundary of
 the ribbon, which is the lower boundary of the scalloped features, is
 somewhat sinusoidal, with a longitudinal wavelength of about 20\deg
 (corresponding to a wavenumber of 17 to 19), with only a small decrease from
 21\deg in 2012 to 19\deg in 2014, and a latitudinal
 peak-to-peak amplitude of about 2.4-2.9\degx.  The dark ribbon itself has a width
 of about 1-2\deg in latitude.  The ribbon continues around the
 planet, but the transverse wave does not; instead it seems highly
 damped on one side and well defined on the other.  This can be seen
 from images in Fig.\ \ref{Fig:imsamples}, specifically pairs taken on
 successive nights, which look at opposite sides of the planet.  For
 example, compare pairs B-D, J-L, N-P, and R-T.  In the last case,
 instead of the transverse wave, we see small discrete bright features
 in the same latitude region. These characteristics are more clearly
 evident in the rectangular maps of Fig.\ \ref{Fig:rectpro},
 especially in the 2012 maps.

Ribbon waves have also been seen in the cloud forms on Saturn, one
located at the peak of a mid-latitude eastward jet in the northern
hemisphere \citep{Sro1983JGR} and one located near the peak of an
eastward jet in the southern hemisphere
\citep{Sanchez-Lavega2000satwind}.  The latter waves had peak-to-peak amplitudes of
1-1.3\deg in latitude, and wavelengths of 5-6\deg in longitude, smaller than
the waves we observed on Uranus by a factor of two in amplitude and a
factor of four in wavelength.  The mechanism suggested to explain the
Saturnian waves is baroclinic instability \citep{Godfrey1986}.  A
viable mechanism that might explain the uranian ribbon wave remains to
be determined.

\section{Persistent latitudinal patterns in reflectivity}

\begin{figure*}[!htb]\centering
\includegraphics[width=6in]{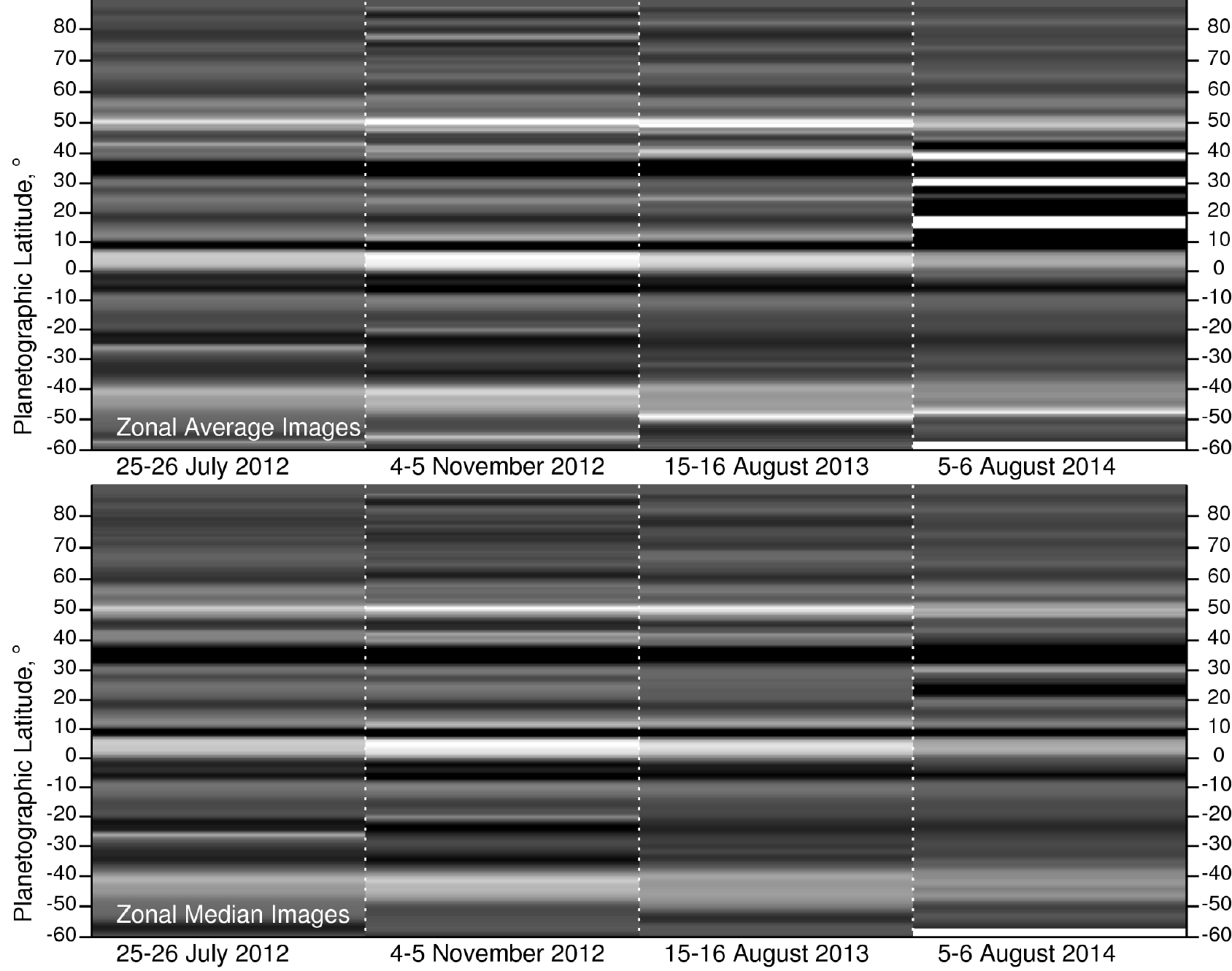}
\caption{Images of zonal mean (top) and zonal median (bottom) relative
  brightness values obtained from high-pass filtered mosaicked image
  maps displayed in Fig.\ \ref{Fig:rectpro}. The median images avoid
  contributions from discrete cloud features. In both images black and
  white correspond to approximate I/F deviations of -0.4$\times
  10^{-4}$ and +0.8$\times 10^{-4}$ respectively (the central disk I/F
  in H is about 0.01).}
\label{Fig:zonalimage}
\end{figure*}

\begin{figure*}[!htb]\centering
\includegraphics[width=5in]{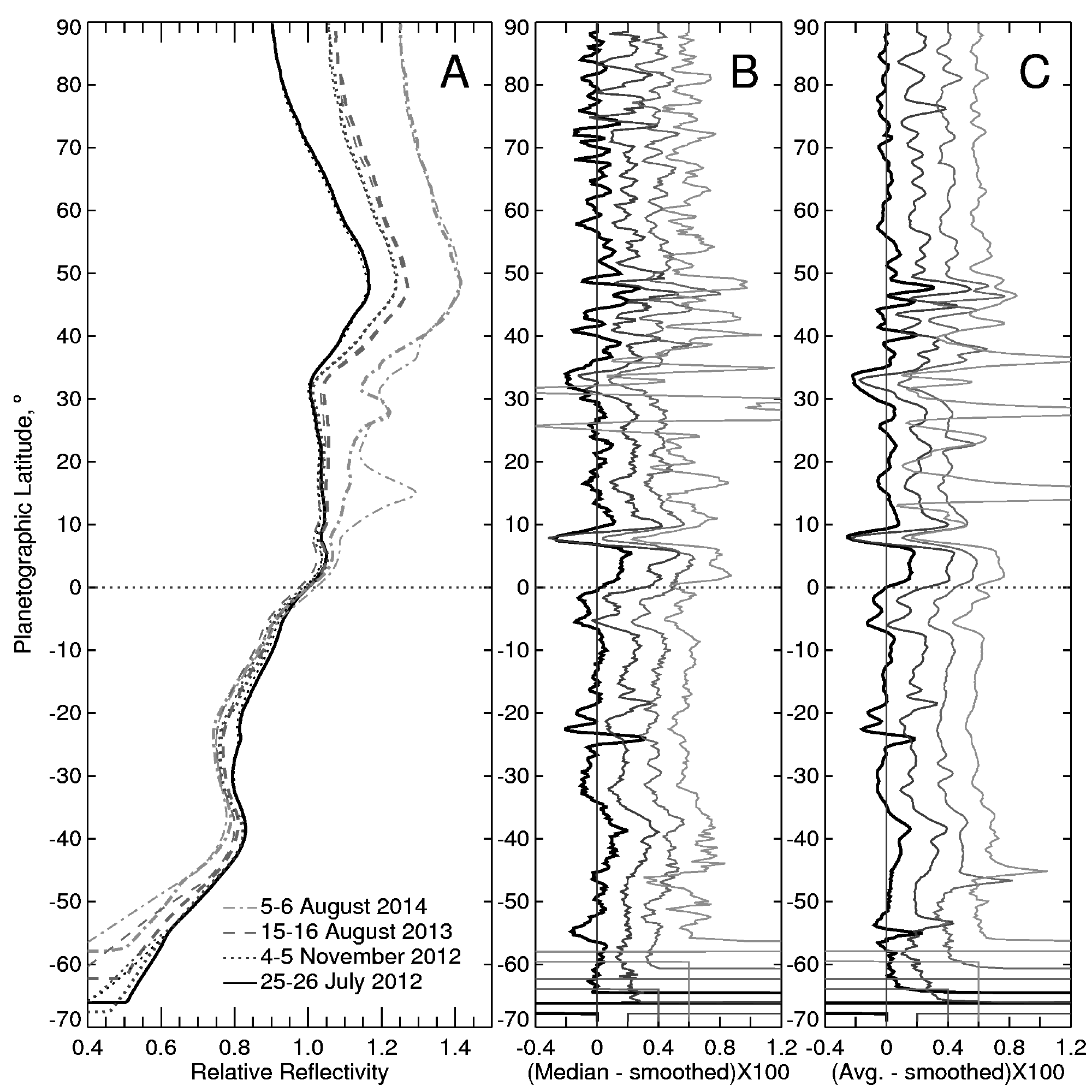}
\caption{Zonal mean and zonal median reflectivities from rectangular
  maps of mosaicked images without high pass filtering and normalized
  to 1.0 at the equator (A), and high-pass filtered deviation profiles 
 for median (B), and mean (C) values for each data set. The subtracted
smooth profiles used a 3.25\deg boxcar. In A, thinner
  lines show mean brightness (which includes discrete cloud
  contribution) and thicker lines show median brightness. In B and C
  the lightest curve is for the earliest data set (25-26 July, 2012)
  and the darkest curve is for the latest.}
\label{Fig:latprofile}
\end{figure*}

To investigate the latitudinal band patterns over the whole globe we
sampled our mosaicked images in narrow latitude bands, computing both
zonal means and zonal median values to avoid the contributions of
bright cloud features.  The results from each time period are shown in
Fig.\ \ref{Fig:zonalimage} in image form and in
Fig.\ \ref{Fig:latprofile} as normalized plots.  The zonal average,
shown in the top panel in Fig.\ \ref{Fig:zonalimage} includes the
effect of discrete cloud features, and the bright bands in the 2014
image are due to the eruption in cloud activity during that year.  The
median images (lower panel) are much less affected by discrete cloud
features and yet show many of the same persistent patterns, especially
the brighter bands near 40-50\degx S, 10-20\degx S, 0-8\degx N,
10-12\degx N, 18-31\degx N, 38-42\degx N, and 48-52\degx N.  The
exceptions are mainly for the 2014 observations where the large number
of discrete features has apparently obscured the background band
patterns in the northern hemisphere.  The plots of latitudinal
profiles in Fig.\ \ref{Fig:latprofile} make clear how subtle the
patterns really are.  These features cannot be directly discerned in
the direct averages and median profiles shown in panel A.  They only
become obvious when the profiles have their smoothed versions
subtracted, and the difference amplified by a factor of 100, as in
panels B and C.

The large scale relative changes in panel A with latitude are somewhat
distorted by the variation of effective view angle with latitude in
the mosaicked images, especially at high southern latitudes, where the
effective view angle cosines are of necessity much smaller than for
mid-latitudes and for all of the northern hemisphere.  To examine this
issue we formed a median image without remapping to a fixed time, then
sampled latitudinal brightness profiles at fixed view angle cosines.
At $\mu=0.3$ we found that in July 2012 the I/F at 50\deg N was about
25\% brighter than the equator when compared at the same view angles,
but 18\% brighter in the mosaicked images. The same comparison at
45\degx S, found a ratio of 0.95 for the sampling at equal view
angles, but 0.83 in our mosaicked image.  However, the trend of
increasing brightness with time at high northern latitudes is also
seen when sampled at constant view angle, as is the trend of darkening
with time at high southern latitudes, both with respect to equatorial
values.


\section{Cloud composition and vertical structure}

\subsection{Cloud composition}

The composition of the clouds is constrained by indirect measurements.
Spectral observations show that the main condensable in the upper
troposphere is methane.  At the
pressures and temperatures of the brighter clouds (about 1 bar and 77
Kelvin) it seems certain that they are mainly made of frozen methane
particles.  The deeper clouds, at pressures near 1.6 bars, are less
certain.  H$_2$S is a good candidate for these clouds, along with
NH$_4$SH for the even deeper layers \citep{DePater1991Icar}.

\subsection{Vertical structure}\label{Sec:vertical}

\begin{figure*}[!htb]\centering
\includegraphics[width=6.in]{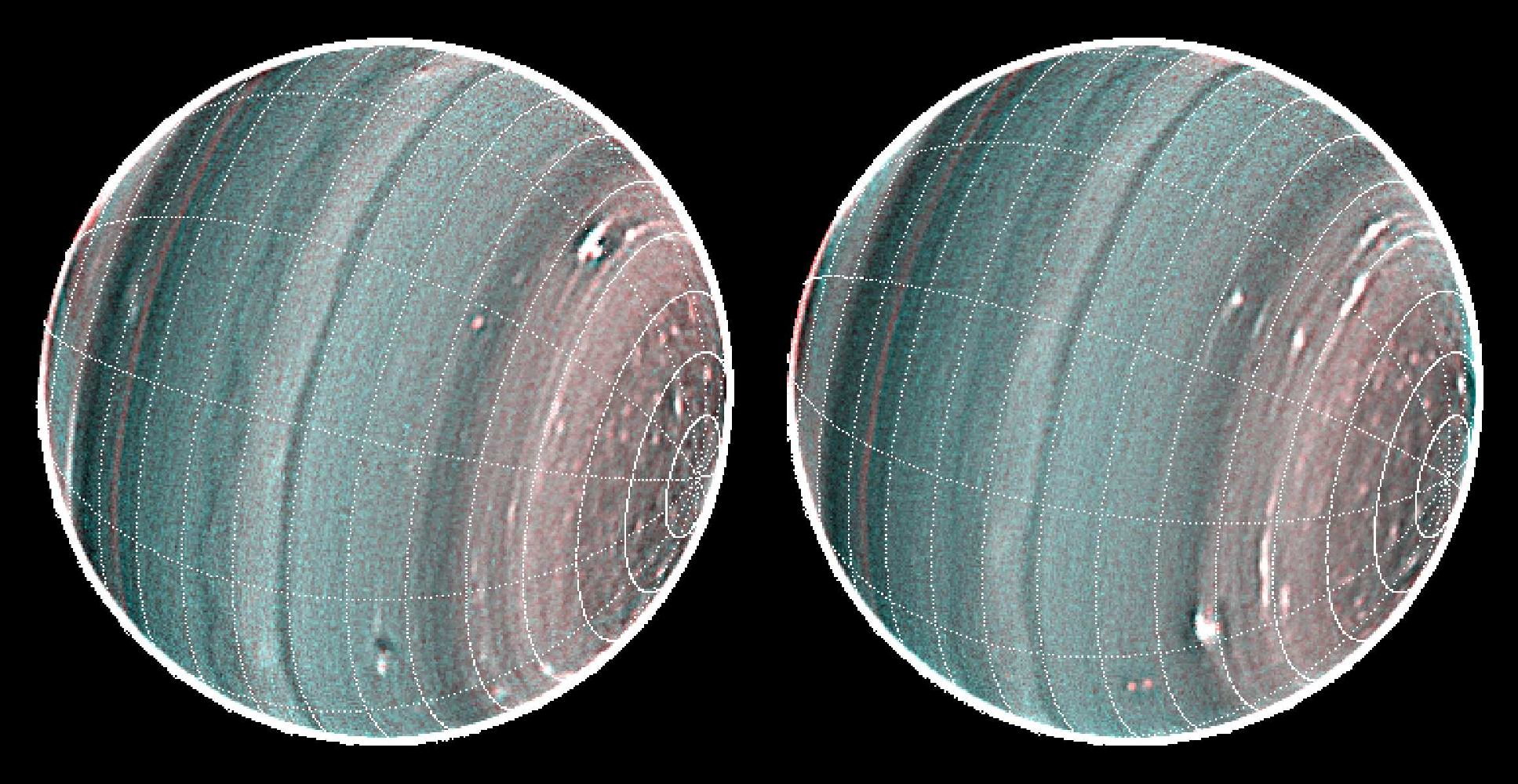}
\caption{Broad cloud altitude changes indicated by H and Hcont
  brightness differences in these 25 (L) and 26 (R) July 2012 image
  composites. Here low altitude clouds have a blue tint, high altitude
  optically thin clouds have a red tint, while clouds that are both
  high and optically thick appear white. Some of the color gradient
  between high and low latitudes may be related to the decline in the
  upper tropospheric methane mixing ratio with latitude.  These images
  have been processed to amplify the high spatial frequencies using
  I(enhanced)=I + k*(I-smoothed(I)), with k set to 35 and the
  smoothing length set to 9 pixels (0.09 arcseconds). The grid
  intervals are 10\deg in planetographic latitude and 30\deg in
  longitude.}
\label{Fig:2012color}
\end{figure*}

The spectral filters used in our high-S/N observing program had to be
very limited due to the time required to obtain high S/N ratios.
Besides our primary H filter, we also used Hcont, CH4S, and
limited K$'$ filters, to provide constraints on effective cloud
altitudes (see Fig. 3), although we cannot constrain the many parameters of a
detailed vertical structure model with these observations.  Effective
altitudes were estimated for major features seen in 2014 images by
\cite{DePater2015storm}, finding that the larger bright feature had
effective cloud top pressures from as low as 300 mb to as high as 1.2
bars.  From a similar analysis of northern high-latitude features in
2011 Keck images, \cite{Sro2012polar} found cloud altitudes of 200 mb
to 400 mb for the two brightest features, and found that the small
polar cloud features (many unresolved) were at considerably higher
pressures, most in the 1-2 bar range.  Since the latter analysis, it
has been confirmed that the methane mixing ratio at high polar
latitudes is depleted substantially in the upper troposphere relative
to low latitudes \citep{Sro2014stis}. That implies that a more complex
analysis is needed to interpret spectral constraints on cloud
altitudes in the north polar region, which we leave for future work.
For the moment, we can provide a qualitative view of relative cloud
height differences using H and Hcont images from our 2012 Keck data
set.  In Fig.\ \ref{Fig:2012color}, we display two color-composite
images in which we assign Hcont to blue and green color channels, and
H to the red color channel. Deep clouds are attenuated more in H than
in Hcont, so they appear to have a blue tint, while higher altitude
clouds of low optical depth have relatively greater brightness
increases above background in H than in Hcont, thus appearing with a
red tint, while optically thick high altitude clouds can appear
equally bright in both channels, and thus can appear white, with
appropriate display enhancements.

Among the high-altitude features, based on their visibility in K$'$
images, we find that B and C reached pressures less than one bar.  No
features reached that level in the southern hemisphere.  We suspect
that feature A may have reached that level in 2012, but lacking any
K$'$ images from that period, we need to carry out a full radiative
transfer analysis to confirm it. Its bright clouds were not visible in
K$'$ images from 2013 and 2014.  Feature F was the only feature
prominent in amateur images from 2014, which were typically taken with
a 625-nm long-pass filter (cut off by CCD response falloff).  F is not
a high-altitude feature; it is not visible in Keck K$'$ images, but
apparently is of sufficient optical thickness to provide contrast at
wavelengths with less methane absorption. At these wavelengths, the background
atmosphere becomes too bright to allow detection of optically thin
high-altitude features, even though they can be very prominent in K$'$
images.  An analysis of cloud pressures for major 2014 features by
\citep{DePater2015storm} shows that C4 (their feature 1) reached
levels of 420-720 mbar, that F (their feature 2) did not extend much
above 2 bars, while G (their feature Br) had a complex structure with
component elements reaching pressures from 300-700 mbar.

\section{Discussion: Polar circulation and static stability}

There may be a connection between the numerous small, apparently
convective, cloud features in the north polar region of Uranus, and
its circulation being solid-body over the region where this suggested
convection occurs.  If the static stability (as measured by its
Brunt-Vaisala frequency or its Rossby deformation radius $L_R$
\citep{pedlosky1982geophysical}) were large, then the planetary
east-west flow would be expected to act as if were two-dimensional and
quasi-geostrophic (QG) \citep{marcus93}. In the mid-latitudes between
10\deg and 60\deg in both hemispheres, we can show that the
mean east-west flow is well-approximated with a QG model with an $L_R
\simeq 6000$~km and with a weak potential vorticity gradient, in
accord with \citet{deng2007comparative}. A QG model does not
approximate the flow well in the region of the equatorial jet, where
the Coriolis parameter passes through zero and where the east-west
flow is likely to be more three-dimensional and driven by cellular
motions such as a Hadley cell \citep{yamazaki2005hadley}. In addition,
a QG model does not approximate the near solid-body rotation that we
observe at the north polar region of Uranus. On the other hand, if the
north polar region has low static stability with intermittent, local
convection and/or baroclinic instabilities, as suggested by the
region’s profuse clouds and by the fact that it recently passed from
winter to summer (see Section\ \ref{Sec:polemorph}), then there is reason to expect
that solid-body rotation would develop there.
   
In modeling stars, it was traditionally believed that the turbulent
mixing within convection zones made the convective regions of stars
rotate as solid bodies \citep{maeder2000evolution}. In fact, a number
of different authors have argued that in convectively stable regions
of stars with internal gravity waves or with baroclinic instabilities,
such as the Goldreich-Schubert-Fricke instability, the flow is driven
to solid-body rotation \citep{maeder2000evolution}. However,
measurements of the rotation curves in our sun, using helioseismology
(c.f., Fig.~5 in \citep{beck2000comparison}) show a more complex
picture of rotation. Solar observations, along with three-dimensional
numerical simulations \citep{miesch2006solar} in which there is mixing,
show that only polar regions within 20\deg -- 30\deg of
the poles rotate as solid bodies. Although there is differential rotation in the outermost surface of
the sun, deeper, at the boundary of
the convection zone and the radiative region (which is a nearly
spherical surface with constant pressure), there is near solid-body
rotation. (See the comparison between the surfaces of constant
rotation and the location of the convective zone boundary in Fig.~1 of
\cite{balbus2009differential}). Because we expect that clouds that we
use in determining the east-west velocities or Uranus are located near
the boundary of its convective and radiative zones, it is the
differential (or lack thereof) rotation curve at this boundary that is
relevant to us.

A simple argument by \cite{balbus2009simple} based on weak instability
and the thermal wind equation explains the solid-body rotation at the
poles and differential rotation elsewhere. His original argument
required that the weak instability be due to magnetic fields, but
later showed that {\it any} weak instability, including convective or
baroclinic, would suffice \citep{balbus2009differential}. The decrease
in the static stability at the north pole of Uranus due to the cooling
of the atmosphere during winter over the long season of darkness may
be sufficient to allow weak, new baroclinic instabilities to form or
old instabilities to strengthen and cause the north polar region to
rotate as a solid body during its winter and spring.  If this
connection between solid body rotation and weak static stability is
valid, then as the north polar region of Uranus passes into summer, we
would expect to observe a decrease in the number of clouds at the
north polar region and an increase in its differential
rotation. Future observations should be capable of testing this
hypothesis.  

The angular velocity acceleration needed to transform the polar
circulation from the solid body form seen in north polar spring to the
more complex form seen at southern hemisphere solstice might be
produced by vertical advection of angular momentum from below or from
poleward meridional motions, for which conservation of angular
momentum would tend to increase angular velocity with decreasing
distance to the pole.  Whether such a transformation actually takes
place, and how effective these mechanisms might be in producing the
speculated transformation remains to be determined.

\section{Summary and Conclusions}

We reported on the analysis of an extensive data set of high-quality AO images acquired from
Keck and Gemini observatories from 2012 through 2014.  Further enhancement of image quality
was obtained by averaging multiple images in a body fixed coordinate system, allowing the
improvement of signal to noise ratios while avoiding the smear due to planet rotation. Our
results are summarized below.

\begin{enumerate}

\item We made over 850 measurements of cloud features in high-S/N
  images from 2012, 2013, and 2014. These are heavily weighted towards
  the northern hemisphere not only because the sub-earth and sub-solar
  latitudes were in the northern hemisphere, but also because Uranus
  produced more cloud features there. The number of trackable cloud
  features in 2014 was exceptionally large compared to other
  years. Gemini observations produced far fewer trackable features than Keck
  observations, mainly due to the higher performance of the Keck AO
  system as a result of its ability to use Uranus as a wave-front
  reference.

\item These observations revealed an active polar region, with many
  small cloud features of 600-800 km, comparable to the Keck telescope's
  resolution. Most of the features are bright, but a number of dark
  features were also seen of about the same size and mainly circular
  shape. These features are found between the north pole and about
  55\degx N.  The small polar cloud features were seen in every year
  of high-S/N observations, but were most apparent in 4-5 November
  2012 observations, where they seemed to have higher contrast and
  greater numbers.  A sizable fraction of these features lived long
  enough to track on successive nights, providing an accurate
  determination of their wind speeds.

\item Our new measurements firmly established that the high-latitude
zonal winds of Uranus, between 60\degx N and at least 83\degx N, closely adhere to solid-body
rotation at a rate of 4.08$\pm$0.015\degx/h westward relative to body-fixed coordinates.  When
winds are plotted in units of m/s, the transition to solid-body rotation appears as a zonal
jet with a peak velocity of 260 m/s westward

\item We were able to fill in details of the zonal profile in the
  12\degx N - 30\degx N region, which has been under-sampled in prior
  data sets.

\item Binned wind measurements show a stair-step appearance at middle latitudes, suggesting
that the wind profile may not be an entirely smooth function of latitude.

\item We discovered a substantial difference between the near-equatorial motions of waves
and those of small discrete cloud features, with the former moving at a speed of
0.4\degx /h eastward and the latter at a speed of 0.1\degx /h eastward. This may mean 
that the equatorial mass flow is much slower than previously inferred and that the
waves move eastward faster than the zonal flow by 0.3\degx /h. 

\item Three symmetric (even order) Legendre polynomial fits were
  derived from the new wind observations. One is based on essentially
  the entire data set.  A second is based on observations including
  the wave motions but excluding the small discrete feature tracking
  that was obtained from the 2014 observations, and a third is based
  on observations that exclude the wave motion measurements.

\item We also found that the preponderance of current and past observations are
consistent with a north-south  mid-latitude asymmetry in the zonal wind profile of Uranus. Using
an asymmetric model to characterize this difference, we find that the maximum
asymmetry amplitude is  0.09\degx /h when all high accuracy observations
from Voyager onward are included, and 0.135\degx /h, when all but 2012-2104 observations
are included, with the main asymmetries peaking near $\pm$30\degx.
 It appears that the asymmetry has decreased with time. We did not include in this analysis
the enormous high-latitude asymmetry implied by the recent reanalysis of Voyager images
by \cite{Kark2015vgr}.


\item Although there are no HST or groundbased wind measurements at
  high southern latitudes (50\degx S - 90\degx S), a recent reanalysis
  of 1986 Voyager 2 Uranus observations by \cite{Kark2015vgr} has
  yielded wind results there that are very different from northern
  winds at corresponding northern latitudes.  This large north-south
  asymmetry might be seasonal.  However, only minimal changes have
  been seen in the north polar region between 2011 and 2014, and at
  middle latitudes only tiny changes have been seen since 1986.  This
  argues against a large seasonal change in polar winds, though it
  cannot be ruled out.

\item We found two types of equatorial wave features. One type, seen
  in prior observations, as well as in our more recent observations,
  are diffuse bright features a few degrees north of the equator,
  spaced about 30\deg to 40\deg apart in longitude.
The other kind of wave feature, not observed prior to these high S/N
observations, is a transverse wave marked by a dark ribbon with a
latitudinal width of about 1\deg and a longitudinal wavelength of
about 20\degx. However, the transverse wave amplitude, which is of the
order of several degrees of latitude over about half of the planet's
circumference, damps to nearly zero over the remaining longitude
range.

\item We found that zonal averages of brightness of polar cloud
  features has a latitudinal pattern in which brightness minima occur
  near 53.5\degx N, 61\degx N, 71.5\degx N, and 78\degx N.  This is a good
match to the apparent \chf mixing ratio variations inferred by \cite{Sro2014stis}.
More obvious patterns are seen at lower latitudes, but most are only visible
with high-pass filtering.

\item Zonal maps made from images acquired on successive nights in August 2012, November 2012,
  August 2013, and August 2014, show persistent patterns, and six easily distinguished
  long-lived cloud features, which we were able to track for long periods that ranged from
5 months to over two years. Two at similar latitudes are associated with dark spots, and move
with the atmospheric zonal flow close to the location of their associated dark spot
instead of following the flow at the latitude of the bright features. These features
retained their morphologies and drift rates in spite of several close interactions.
A second pair of features at similar latitudes also survived several close approaches.
Several of the long-lived features also exhibited equatorward drifts and latitudinal
oscillations.

\item Among the high-altitude features, based on their visibility in
  K$'$ images, we find that B and C reached pressures less than one
  bar, and thus are likely at least partly composed of methane ice.
  No features reached that level in the southern hemisphere.  We
  suspect that Feature A reached that level in 2012, but lacking any
  K$'$ images from that period we would need a full radiative
  transfer analysis to confirm it.

\item There is a correlation between the region of polar ``convective''
cloud forms and the region of solid body rotation, both extending
from about 60\deg N to at least 83\deg N.  There are dynamical
reasons to expect that these might be related, and that when
the convection subsides, the winds might also change.

\end{enumerate}

As Uranus moves towards its 2030 northern hemisphere summer solstice, a large
seasonal shift in wind speeds should occur if the recent results of
\cite{Kark2015vgr} for the southern hemisphere indicate a seasonal
asymmetry.  What also might happen is that the cloud features that we
currently use to track motions in this region either disappear or
become obscured by an overlying haze, a result suggested by the fact
that we have never seen near-IR cloud features in the south polar
region.  We can only hope that any seasonal changes will become
observable before the tracers we use to observe them disappear.

\section*{Acknowledgments}

LAS, PMF, and HBH acknowledge support from NASA's Planetary
Astronomy Program (Grant NNX13AH65G for LAS and PMF). LAS and PMF also
acknowledge NASA Keck observing support (JPL Grant 1485335).  We thank
staff at the W. M. Keck Observatory, which is made possible by the
generous financial support of the W. M. Keck Foundation.  We thank
those of Hawaiian ancestry on whose sacred mountain we are privileged
to be guests. Without their generous hospitality none of our
groundbased observations would have been possible.  We also thank
staff at the Gemini observatory, which is operated by the Association
of Universities for Research in Astronomy, Inc., under a cooperative
agreement with the NSF on behalf of the Gemini partnership: the
National Science Foundation (United States), the National Research
Council (Canada), CONICYT (Chile), the Australian Research Council
(Australia), Ministério da Ciência, Tecnologia e Inovação (Brazil) and
Ministerio de Ciencia, Tecnología e Innovación Productiva
(Argentina). We thank Erich Karkoschka for providing his tabulated
correlation results in advance of publication.


\begin{thebibliography}{41}
\expandafter\ifx\csname natexlab\endcsname\relax\def\natexlab#1{#1}\fi
\expandafter\ifx\csname url\endcsname\relax
  \def\url#1{\texttt{#1}}\fi
\expandafter\ifx\csname urlprefix\endcsname\relax\def\urlprefix{URL }\fi

\bibitem[{{Alexander}(1965)}]{Alexander1965}
{Alexander}, A.~F.~O., 1965. {The planet Uranus; a history of observation,
  theory, and discovery}. New York, American Elsevier Pub.~Co., 1965.

\bibitem[{{Allison} et~al.(1991){Allison}, {Beebe}, {Conrath}, {Hinson}, and
  {Ingersoll}}]{Allison1991uranbook}
{Allison}, M., {Beebe}, R.~F., {Conrath}, B.~J., {Hinson}, D.~P., {Ingersoll},
  A.~P., 1991. {Uranus atmospheric dynamics and circulation}. In: Bergstralh,
  J.~T., Miner, E.~D., Matthews, M.~S. (Eds.), {Uranus}. {University of
  Arizona, Tucson}, pp. 253--295.

\bibitem[{{Archinal} et~al.(2011){Archinal}, {AHearn}, {Bowell}, {Conrad},
  {Consolmagno}, {Courtin}, {Fukushima}, {Hestroffer}, {Hilton}, {Krasinsky},
  {Neumann}, {Oberst}, {Seidelmann}, {Stooke}, {Tholen}, {Thomas}, and
  {Williams}}]{Archinal2011}
{Archinal}, B.~A., {AHearn}, M.~F., {Bowell}, E., {Conrad}, A., {Consolmagno},
  G.~J., {Courtin}, R., {Fukushima}, T., {Hestroffer}, D., {Hilton}, J.~L.,
  {Krasinsky}, G.~A., {Neumann}, G., {Oberst}, J., {Seidelmann}, P.~K.,
  {Stooke}, P., {Tholen}, D.~J., {Thomas}, P.~C., {Williams}, I.~P., 2011.
  {Report of the IAU Working Group on Cartographic Coordinates and Rotational
  Elements: 2009}. Cel. Mech. \& Dyn. Astr. 109, 101--135.

\bibitem[{Balbus(2009)}]{balbus2009simple}
Balbus, S.~A., 2009. A simple model for solar isorotational contours. Monthly
  Notices of the Royal Astronomical Society 395~(4), 2056--2064.

\bibitem[{Balbus et~al.(2009)Balbus, Bonart, Latter, and
  Weiss}]{balbus2009differential}
Balbus, S.~A., Bonart, J., Latter, H.~N., Weiss, N.~O., 2009. Differential
  rotation and convection in the sun. Monthly Notices of the Royal Astronomical
  Society 400~(1), 176--182.

\bibitem[{Beck(2000)}]{beck2000comparison}
Beck, J.~G., 2000. A comparison of differential rotation measurements--(invited
  review). Solar Physics 191~(1), 47--70.

\bibitem[{{de Pater} et~al.(1991){de Pater}, {Romani}, and
  {Atreya}}]{DePater1991Icar}
{de Pater}, I., {Romani}, P.~N., {Atreya}, S.~K., 1991. {Possible microwave
  absorption by H2S gas in Uranus' and Neptune's atmospheres}. Icarus 91,
  220--233.

\bibitem[{{de Pater} et~al.(2011){de Pater}, {Sromovsky}, {Hammel}, {Fry},
  {LeBeau}, {Rages}, {Showalter}, and {Matthews}}]{DePater2011}
{de Pater}, I., {Sromovsky}, L., {Hammel}, H.~B., {Fry}, P.~M., {LeBeau},
  R.~P., {Rages}, K.~A., {Showalter}, M.~R., {Matthews}, K., 2011.
  {Post-equinox Observations of Uranus: Berg's Evolution, vertical structure,
  and track towards the equator}. Icarus 215, 332--345.

\bibitem[{{de Pater} et~al.(2015){de Pater}, {Sromovsky}, {Fry}, , {Hammel},
  {Baranec}, and {Sayanagi}}]{DePater2015storm}
{de Pater}, I., {Sromovsky}, L.~A., {Fry}, P.~M., , {Hammel}, H.~B., {Baranec},
  C., {Sayanagi}, K., 2015. {Record-breaking storm activity on Uranus in 2014}.
  Icarus 252, 121--128.

\bibitem[{Deng and LeBeau(2007)}]{deng2007comparative}
Deng, X., LeBeau, R.~P., 2007. Comparative cfd simulations of the dark spots of
  uranus and neptune. In: 18th AIAA Computational Fluid Dynamics Conference,
  Miami, FL. pp. 25--28.

\bibitem[{{Fry} et~al.(2012{\natexlab{a}}){Fry}, {Sromovsky}, {de Pater},
  {Hammel}, and {Rages}}]{Fry2012}
{Fry}, P.~M., {Sromovsky}, L.~A., {de Pater}, I., {Hammel}, H.~B., {Rages},
  K.~A., 2012{\natexlab{a}}. {Detection and Tracking of Subtle Cloud Features
  on Uranus}. Astron. J. 143, 150--161.

\bibitem[{{Fry} et~al.(2012{\natexlab{b}}){Fry}, {Sromovsky}, {Rages},
  {Hammel}, and {de Pater}}]{Fry2012DPS}
{Fry}, P.~M., {Sromovsky}, L.~A., {Rages}, K.~A., {Hammel}, H.~B., {de Pater},
  I., 2012{\natexlab{b}}. {Uranus High Signal-to-noise Ratio Near-IR Imaging:
  Recent Results}. In: AAS/Division for Planetary Sciences Meeting Abstracts.
  Vol.~44 of AAS/Division for Planetary Sciences Meeting Abstracts. p.
  \#412.20.

\bibitem[{{Godfrey} and {Moore}(1986)}]{Godfrey1986}
{Godfrey}, D.~A., {Moore}, V., 1986. {The Saturnian ribbon feature - A
  baroclinically unstable model}. Icarus 68, 313--343.

\bibitem[{{Hammel} et~al.(2005){Hammel}, {de Pater}, {Gibbard}, {Lockwood}, and
  {Rages}}]{Hammel2005winds}
{Hammel}, H.~B., {de Pater}, I., {Gibbard}, S., {Lockwood}, G.~W., {Rages}, K.,
  2005. {Uranus in 2003: Zonal winds, banded structure, and discrete features}.
  Icarus 175, 534--545.

\bibitem[{{Hammel} and {Lockwood}(2007)}]{Hammel2007var}
{Hammel}, H.~B., {Lockwood}, G.~W., 2007. {Long-term atmospheric variability on
  Uranus and Neptune}. Icarus 186, 291--301.

\bibitem[{{Hammel} et~al.(2001){Hammel}, {Rages}, {Lockwood}, {Karkoschka}, and
  {de Pater}}]{Hammel2001Icar}
{Hammel}, H.~B., {Rages}, K., {Lockwood}, G.~W., {Karkoschka}, E., {de Pater},
  I., 2001. {New Measurements of the Winds of Uranus}. Icarus 153, 229--235.

\bibitem[{{Hammel} et~al.(2009){Hammel}, {Sromovsky}, {Fry}, {Rages},
  {Showalter}, {de Pater}, {van Dam}, {Lebeau}, and {Deng}}]{Hammel2009Icar}
{Hammel}, H.~B., {Sromovsky}, L.~A., {Fry}, P.~M., {Rages}, K., {Showalter},
  M., {de Pater}, I., {van Dam}, M.~A., {Lebeau}, R.~P., {Deng}, X., 2009. {The
  Dark Spot in the atmosphere of Uranus in 2006: Discovery, description, and
  dynamical simulations}. Icarus 201, 257--271.

\bibitem[{{Karkoschka}(1998)}]{Kark1998Sci}
{Karkoschka}, E., 1998. {Clouds of High Contrast on Uranus}. Science 280,
  570--572.

\bibitem[{{Karkoschka}(2015)}]{Kark2015vgr}
{Karkoschka}, E., 2015. {Uranus' southern circulation revealed by Voyager 2:
  Unique characteristics}. Icarus 250, 294--307.

\bibitem[{{Lindal} et~al.(1987){Lindal}, {Lyons}, {Sweetnam}, {Eshleman}, and
  {Hinson}}]{Lindal1987}
{Lindal}, G.~F., {Lyons}, J.~R., {Sweetnam}, D.~N., {Eshleman}, V.~R.,
  {Hinson}, D.~P., 1987. {The atmosphere of Uranus - Results of radio
  occultation measurements with Voyager 2}. J. Geophys. Res. 92, 14987--15001.

\bibitem[{Maeder and Meynet(2000)}]{maeder2000evolution}
Maeder, A., Meynet, G., 2000. The evolution of rotating stars. Annual Review of
  Astronomy and Astrophysics 38, 143--190.

\bibitem[{Marcus(1993)}]{marcus93}
Marcus, P., 1993. Jupiter's {G}reat {R}ed {S}pot and other vortices. Annual
  Review of Astronomy and Astrophysics 31, 523--573.

\bibitem[{{Meeus}(1997)}]{Meeus1997}
{Meeus}, J., 1997. {Equinoxes and solstices on Uranus and Neptune}. Journal of
  the British Astronomical Association 107, 332.

\bibitem[{Miesch et~al.(2006)Miesch, Brun, and Toomre}]{miesch2006solar}
Miesch, M.~S., Brun, A.~S., Toomre, J., 2006. Solar differential rotation
  influenced by latitudinal entropy variations in the tachocline. The
  Astrophysical Journal 641~(1), 618.

\bibitem[{Pedlosky(1982)}]{pedlosky1982geophysical}
Pedlosky, J., 1982. Geophysical fluid dynamics. New York and Berlin,
  Springer-Verlag, 1982. 636 p. 1.

\bibitem[{{Porco} et~al.(2005){Porco}, {Baker}, {Barbara}, {Beurle}, {Brahic},
  {Burns}, {Charnoz}, {Cooper}, {Dawson}, {Del Genio}, {Denk}, {Dones},
  {Dyudina}, {Evans}, {Giese}, {Grazier}, {Helfenstein}, {Ingersoll},
  {Jacobson}, {Johnson}, {McEwen}, {Murray}, {Neukum}, {Owen}, {Perry},
  {Roatsch}, {Spitale}, {Squyres}, {Thomas}, {Tiscareno}, {Turtle}, {Vasavada},
  {Veverka}, {Wagner}, and {West}}]{Porco2005}
{Porco}, C.~C., {Baker}, E., {Barbara}, J., {Beurle}, K., {Brahic}, A.,
  {Burns}, J.~A., {Charnoz}, S., {Cooper}, N., {Dawson}, D.~D., {Del Genio},
  A.~D., {Denk}, T., {Dones}, L., {Dyudina}, U., {Evans}, M.~W., {Giese}, B.,
  {Grazier}, K., {Helfenstein}, P., {Ingersoll}, A.~P., {Jacobson}, R.~A.,
  {Johnson}, T.~V., {McEwen}, A., {Murray}, C.~D., {Neukum}, G., {Owen}, W.~M.,
  {Perry}, J., {Roatsch}, T., {Spitale}, J., {Squyres}, S., {Thomas}, P.,
  {Tiscareno}, M., {Turtle}, E., {Vasavada}, A.~R., {Veverka}, J., {Wagner},
  R., {West}, R., 2005. {Cassini Imaging Science: Initial Results on Saturn's
  Atmosphere}. Science 307, 1243--1247.

\bibitem[{{Sanchez-Lavega} et~al.(2000){Sanchez-Lavega}, {Rojas}, and
  {Sada}}]{Sanchez-Lavega2000satwind}
{Sanchez-Lavega}, A., {Rojas}, J.~F., {Sada}, P.~V., 2000. {Saturn's Zonal
  Winds at Cloud Level}. Icarus 147, 405--420.

\bibitem[{{Sayanagi} and {al.}(2015)}]{Sayanagi2015too}
{Sayanagi}, K.~M., {al.}, e., 2015. {Evolution of bright storms on Uranus during 2014-2015 observed
  by HST and ground-based telescopes}.
  Icarus, in preparation.

\bibitem[{{Smith} et~al.(1986){Smith}, {Soderblom}, {Beebe}, {Bliss}, {Brown},
  {Collins}, {Boyce}, {Briggs}, {Brahic}, {Cuzzi}, {Morrison}, and
  {co-authors}}]{SmithBA1986}
{Smith}, B.~A., {Soderblom}, L.~A., {Beebe}, R., {Bliss}, D., {Brown}, R.~H.,
  {Collins}, S.~A., {Boyce}, J.~M., {Briggs}, G.~A., {Brahic}, A., {Cuzzi},
  J.~N., {Morrison}, D., {co-authors}, 1986. {Voyager 2 in the Uranian system -
  Imaging science results}. Science 233, 43--64.

\bibitem[{{Smith}(1986)}]{Smith1986}
{Smith}, P.~H., 1986. {The vertical structure of the Jovian atmosphere}. Icarus
  65, 264--279.

\bibitem[{{Sromovsky} and {Fry}(2005)}]{Sro2005dyn}
{Sromovsky}, L.~A., {Fry}, P.~M., 2005. {Dynamics of cloud features on Uranus}.
  Icarus 179, 459--484.

\bibitem[{{Sromovsky} et~al.(2002){Sromovsky}, {Fry}, and
  {Baines}}]{Sro2002spots}
{Sromovsky}, L.~A., {Fry}, P.~M., {Baines}, K.~H., 2002. {The Unusual Dynamics
  of Northern Dark Spots on Neptune}. Icarus 156, 16--36.

\bibitem[{{Sromovsky} et~al.(2012{\natexlab{a}}){Sromovsky}, {Fry}, {Hammel}, ,
  {de Pater}, {Rages}, {Showalter}, {Merline}, {Tamblyn}, {Neyman}, {Margot},
  {Fang}, {Colas}, {Dauvergne}, { G\a'omez-Forrellad}, { Hueso},
  {S\a'anchez-Lavega}, and {Stallard}}]{Sro2012bs}
{Sromovsky}, L.~A., {Fry}, P.~M., {Hammel}, H.~B., , {de Pater}, I., {Rages},
  K.~A., {Showalter}, M.~R., {Merline}, W.~J., {Tamblyn}, P., {Neyman}, C.,
  {Margot}, J.-L., {Fang}, J., {Colas}, F., {Dauvergne}, J.-L., {
  G\a'omez-Forrellad}, J.~M., { Hueso}, R., {S\a'anchez-Lavega}, A.,
  {Stallard}, T., 2012{\natexlab{a}}. {Episodic bright and dark spots on
  Uranus}. Icarus 220, 6--22.

\bibitem[{{Sromovsky} et~al.(2009){Sromovsky}, {Fry}, {Hammel}, {Ahue}, {de
  Pater}, {Rages}, {Showalter}, and {van Dam}}]{Sro2009eqdyn}
{Sromovsky}, L.~A., {Fry}, P.~M., {Hammel}, H.~B., {Ahue}, W.~M., {de Pater},
  I., {Rages}, K.~A., {Showalter}, M.~R., {van Dam}, M.~A., 2009. {Uranus at
  equinox: Cloud morphology and dynamics}. Icarus 203, 265--286.

\bibitem[{{Sromovsky} et~al.(2012{\natexlab{b}}){Sromovsky}, {Fry}, {Hammel},
  {de Pater}, and {Rages}}]{Sro2012dps}
{Sromovsky}, L.~A., {Fry}, P.~M., {Hammel}, H.~B., {de Pater}, I., {Rages},
  K.~A., 2012{\natexlab{b}}. {First Views of North Polar Clouds and Circulation
  on Uranus}. In: AAS/Division for Planetary Sciences Meeting Abstracts.
  Vol.~44. p. Abstract \#504.01.

\bibitem[{{Sromovsky} et~al.(2012{\natexlab{c}}){Sromovsky}, {Fry}, {Hammel},
  {de Pater}, and {Rages}}]{Sro2012polar}
{Sromovsky}, L.~A., {Fry}, P.~M., {Hammel}, H.~B., {de Pater}, I., {Rages},
  K.~A., 2012{\natexlab{c}}. {Post-equinox dynamics and polar cloud structure
  on Uranus}. Icarus 220, 694--712.

\bibitem[{{Sromovsky} et~al.(2007){Sromovsky}, {Fry}, {Hammel}, {de Pater},
  {Rages}, and {Showalter}}]{Sro2007bright}
{Sromovsky}, L.~A., {Fry}, P.~M., {Hammel}, H.~B., {de Pater}, I., {Rages},
  K.~A., {Showalter}, M.~R., 2007. {Dynamics, Evolution, and Structure of
  Uranus' Brightest Cloud Feature}. Icarus 192, 558--575.

\bibitem[{{Sromovsky} et~al.(2014){Sromovsky}, {Karkoschka}, {Fry}, {Hammel},
  {de Pater}, and {Rages}}]{Sro2014stis}
{Sromovsky}, L.~A., {Karkoschka}, E., {Fry}, P.~M., {Hammel}, H.~B., {de
  Pater}, I., {Rages}, K.~A., 2014. {Methane depletions in both polar regions
  of Uraus inferred from HST/STIS and Keck/NIRC2}. Icarus 238, 137--155.

\bibitem[{{Sromovsky} et~al.(1993){Sromovsky}, {Limaye}, and
  {Fry}}]{Sro1993Icar}
{Sromovsky}, L.~A., {Limaye}, S.~S., {Fry}, P.~M., 1993. {Dynamics of Neptune's
  Major Cloud Features}. Icarus 105, 110--141.

\bibitem[{{Sromovsky} et~al.(1983){Sromovsky}, {Revercomb}, {Krauss}, and
  {Suomi}}]{Sro1983JGR}
{Sromovsky}, L.~A., {Revercomb}, H.~E., {Krauss}, R.~J., {Suomi}, V.~E., 1983.
  {Voyager 2 observations of Saturn's northern mid-latitude cloud features -
  Morphology, motions, and evolution}. \jgr 88~(17), 8650--8666.

\bibitem[{Yamazaki et~al.(2005)Yamazaki, Read, and Skeet}]{yamazaki2005hadley}
Yamazaki, Y., Read, P., Skeet, D., 2005. Hadley circulations and kelvin
  wave-driven equatorial jets in the atmospheres of jupiter and saturn.
  Planetary and Space Science 53~(5), 508--525.

\end{thebibliography}

\end{document}